\documentclass[twocolumn,floatfix]{aastex631}
\usepackage{hyperref}

\usepackage{amsmath}
\usepackage{float,soul}
\let\cleardoublepage=\clearpage

\newcommand\N[1]{{\color{blue} \bf #1}} %N8

\begin{document}

\title{A Multiwavelength Autopsy of the Interacting IIn Supernova 2020ywx: Tracing its Progenitor Mass-Loss History for 100 Years before Death}

\newcommand{\UCB}{\affiliation{Department of Astronomy, University of California, Berkeley, CA 94720-3411, USA}}
\newcommand{\NRAO}{\affiliation{National Radio Astronomy Observatory,
520 Edgemont Rd, Charlottesville VA 22903, USA}}
\newcommand{\NCRA}{\affiliation{National Centre for Radio Astrophysics,
TIFR, Ganeshkhind, Pune 411007, India}}
\newcommand{\UVA}{\affiliation{Department of Astronomy, University of Virginia,
 Charlottesville VA 22904-4325, USA}}
\newcommand{\UA}{\affiliation{Steward Observatory, University of Arizona, 933 North Cherry Avenue, Tucson, AZ 85721-0065, USA}}
\newcommand{\GeminiNorth}{\affiliation{Gemini Observatory, 670 North A`ohoku Place, Hilo, HI 96720-2700, USA}}
\newcommand{\UCSD}{\affiliation{Department of Astronomy \& Astrophysics, University of California, San Diego, 9500 Gilman Drive, MC 0424, La Jolla, CA 92093-0424, USA}}
\newcommand{\Goddard}{\affiliation{Astrophysics Science Division, NASA Goddard Space Flight Center, Mail Code 661, Greenbelt, MD 20771, USA}}
\newcommand{\Catalyst}{\altaffiliation{LSSTC Catalyst Fellow}}
\newcommand{\LCO}{\affiliation{Las Cumbres Observatory, 6740 Cortona Drive, Suite 102, Goleta, CA 93117-5575, USA}}
\newcommand{\UCSB}{\affiliation{Department of Physics, University of California, Santa Barbara, CA 93106-9530, USA}}

\author[0009-0004-7268-7283]{Raphael Baer-way}\UVA \NRAO
\author[0000-0002-0786-7307]{Poonam Chandra} \NRAO \NCRA
\author[0000-0001-7132-0333]{Maryam Modjaz} \UVA
\author[0000-0001-8367-7591]{Sahana Kumar} \UVA
\author[0000-0002-7472-1279]{Craig Pellegrino} \UVA
\author[0000-0002-9117-7244]{Roger Chevalier} \UVA
\author[0000-0002-7627-4839]{Adrian Crawford} \UVA
\author[0000-0002-9820-679X]{Arkaprabha Sarangi}\affiliation{DARK, Niels Bohr Institute, University of Copenhagen, Jagtvej 155A, 2200 Copenhagen, Denmark}
\affiliation{Indian Institute of Astrophysics, 100 Feet Rd, Koramangala, Bengaluru, Karnataka 560034, India}
\author[0000-0001-5510-2424]{Nathan Smith}
\UA
\author[0000-0003-2611-7269]{Keiichi Maeda}\affiliation{Department of Astronomy, Kyoto University, Kitashirakawa-Oiwake-cho, Sakyo-ku, Kyoto 606-8502, Japan}
\author[0000-0002-8070-5400]{A.J. Nayana}\UCB
%\author[0000-0002-3356-5855]{Alak Ray}\affiliation{Homi Bhabha Centre for Science Education, TIFR, Mumbai 400088, India}
\author[0000-0003-3460-0103]{Alexei~V.~Filippenko}
\UCB
\author[0000-0003-0123-0062]{Jennifer E. Andrews}
\GeminiNorth
\author[0000-0001-7090-4898]{Iair Arcavi}\affiliation{School of Physics and Astronomy, Tel Aviv University, Tel Aviv 69978, Israel}
\author[0000-0002-4924-444X]{K. Azalee Bostroem}
\UA \Catalyst
\author[0000-0001-5955-2502]{Thomas G.~Brink}
\UCB
%\affiliation{Draper-Wood-Robertson Specialist in Astronomy}
\author[0000-0002-7937-6371]{Yize Dong}\affiliation{Center for Astrophysics | Harvard \& Smithsonian, 60 Garden Street, Cambridge, MA 02138-1516, USA}
\author[0000-0002-4661-7001]{Vikram Dwarkadas}\affiliation{Department of Astronomy and Astrophysics, University of Chicago, 5640 S. Ellis Ave., ERC 569, Chicago, IL 60637}
\author[0000-0003-4914-5625]{Joseph R. Farah}\UCSB \LCO
%\author[0000-0003-3460-0103]{Alexei~V.~Filippenko}
%\UCB
%\author[0000-0002-9017-3567]{Anna Y.Q. Ho}\affiliation{Department of Astronomy, Cornell University, Ithaca, NY 14853, USA}
\author[0000-0003-4253-656X]{D. Andrew Howell}\UCSB \LCO
\author[0000-0002-1125-9187]{Daichi Hiramatsu}
\affiliation{Center for Astrophysics | Harvard \& Smithsonian, 60 Garden Street, Cambridge, MA 02138-1516, USA}
\affiliation{The NSF AI Institute for Artificial Intelligence and Fundamental Interactions, USA}
\author[0000-0002-0832-2974]{Griffin Hosseinzadeh}
\UCSD
\author[0000-0001-5807-7893]{Curtis McCully}\UCSB \LCO
\author[0000-0002-7015-3446]{Nicolas Meza}\affiliation{Department of Physics and Astronomy, University of California, 1 Shields Avenue, Davis, CA 95616-5270, USA}
\author[0000-0001-9570-0584]{Megan Newsome}\UCSB \LCO
\author[0000-0003-0209-9246]{Estefania Padilla Gonzalez}\UCSB \LCO
\author[0000-0002-0744-0047]{Jeniveve Pearson}
\UA
\author[0000-0003-4102-380X]{David J. Sand}
\UA
\author[0000-0002-4022-1874]{Manisha Shrestha}
\UA

\author[0000-0003-0794-5982
]{Giacomo Terreran}\UCSB \LCO
\author[0000-0001-8818-0795]{Stefano Valenti}\affiliation{Department of Physics and Astronomy, University of California, 1 Shields Avenue, Davis, CA 95616-5270, USA}
\author[0000-0003-2732-4956]{Samuel Wyatt}
\Goddard
\author[	
0000-0002-6535-8500]{Yi Yang}\UCB

\author[0000-0002-2636-6508]{WeiKang Zheng}
\UCB
%\affiliation{Bengier-Winslow-Eustace Specialist in Astronomy}

%\author[0000-0002-7937-6371]{Yize Dong}\affiliation{Department of Physics and Astronomy, University of California, Davis, 1 Shields Avenue, Davis, CA 95616-5270, USA}
%\author[0000-0003-1637-267X]{Kuntal Misra}
%\affiliation{Aryabhatta Research Institute of Observational Sciences, Manora Peak 263001, India}



\begin{abstract}
While the subclass of interacting supernovae with narrow hydrogen emission lines (SNe IIn) consists of some of the longest-lasting and brightest SNe ever discovered, their progenitors are still not well understood. Investigating SNe IIn as they emit across the electromagnetic spectrum is the most robust way to understand the progenitor evolution before the explosion. This work presents X-ray, optical, infrared, and radio observations of the strongly interacting Type IIn SN 2020ywx covering a period $>1200$ days after discovery. Through multiwavelength modeling, we find that the progenitor of 2020ywx was losing mass at $\sim10^{-2}$--$10^{-3} 
\mathrm{\,M_{\odot}\,yr^{-1}}$ for at least 100\,yrs pre-explosion using the circumstellar medium (CSM) speed of $120$\,km\,s$^{-1}$ measured from our optical and NIR spectra. Despite the similar magnitude of mass loss measured in different wavelength ranges, we find discrepancies between the X-ray and optical/radio-derived mass-loss evolution, which suggest asymmetries in the CSM. Furthermore, we find evidence for dust formation due to the combination of a growing blueshift in optical emission lines and near-infrared continuum emission which we fit with blackbodies at $\sim$ 1000 K. Based on the observed elevated mass-loss over more than 100 years and the configuration of the CSM inferred from the multiwavelength observations, we invoke binary interaction as the most plausible mechanism to explain the overall mass-loss evolution. SN~2020ywx is thus a case that may support the growing observational consensus that SNe IIn mass loss is explained by binary interaction.

\end{abstract}


\section{Introduction} \label{sec:intro}
Interacting supernovae (SNe) are defined by significant interaction between the exploding star's ejecta and the dense surrounding circumstellar matter (CSM) expelled during the late stages of the progenitor star's life. This interaction generates shocks which create sustained emission across the electromagnetic spectrum from the radio to the X-rays \citep{ Chevalier_17,Chugai_Danziger}. Most interacting SNe are Type IIn supernovae (SNe IIn; \citealt{Smith_2016}), with \citet{Schlegel_1990} being the first to note and classify the defining narrow hydrogen emission lines in SNe IIn (for a full overview of SN subtypes see  \citealt{Filippenko_1997},  \citealt{Galyam_2017}, and \citealt{Modjaz_2019}). While SNe IIn constitute between $\sim 5$\% \citep{Cold23-SNIInrate-PTFZTF} to 9$\%$ \citep{Smith_2011} of core-collapse SNe (with many evolving subtypes \citep{Yesmin_2021ukt}), the progenitor class that gives rise to these objects is relatively unconstrained. 
\par
The most significant piece of evidence used to make deductions about SN IIn progenitors is the star's pre-explosion mass-loss rate. The mass-loss rates measured for SNe IIn range from $\sim 10^{-4}$ to $10^{0}\,\rm{M_{\odot}\,yr^{-1}}$ \citep{Taddia_2013}, which pushes the limit for single-star wind-driven mass loss well past the breaking point of $\sim 10^{-4}\, \mathrm{M_{\odot}\,yr^{-1}}$ \citep{so:06}. %\citep{Woosley_1993}.
Luminous blue variables (LBVs) are the only observed class of star that could produce such high (and relatively fast at $> 100$\,km\,s$^{-1}$) mass-loss rates \citep{Smith_2014,so:06,Smith_2016,Taddia_2013}, although it is now suspected that many LBVs originate in binaries (see, e.g., \citealt{Aghakhanloo_2023,st:15}).  While LBVs can match the measured mass-loss rates, it is unclear why a star would explode directly after the LBV phase, when traditional single-star models with strong mass loss expect LBVs to enter a Wolf-Rayet phase (for potentially $> 10,000$\,yr before the explosion, depending on initial mass) where it would lose nearly all of its hydrogen \citep{heger03}. %\citep{Chevalier_17}. 
Nevertheless, there has been direct observational evidence of LBVs exploding as SNe IIn (e.g., SN 2005gl, \citealt{GalYam2007}; SN 2009ip, \citealt{Mauerhan_2012}). 

\par This is not to say that every case of high mass loss is associated with an LBV progenitor. \citet{Reguitti_2024} found from a limited sample that only $\sim 30$\% of SNe IIn had pre-SN, LBV-like outbursts.  Wave-driven instabilities, pulsational pair instabilities, and binary interaction are other potential mechanisms that may produce the requisite mass loss for SNe IIn \citep{woosley17,Smith_2007,Taddia_2013,sa:14,Quataert_2012,wf21}. However, for nearly all SNe~IIn, the predictions of the pulsational pair instability do not match observed spectral properties \citep{ws22,Smith_2015da}, and the very brief $\sim$1 yr timescale for wave-driven mass loss \citep{Quataert_2012,wf21} falls far short of the sustained high mass loss needed for decades to centuries in most SNe~IIn \citep{Smith_2015da}.  This favors violent binary interaction as a primary agent for the strong pre-SN mass loss \citep{sa:14}. In particular, recent work investigating spectropolarimetry of SNe IIn \citep{Bilinksi} has suggested that some kind of binary interaction leading to asymmetric CSM may be the leading cause of pre-explosion mass loss in SNe IIn.

\par CSM interaction in SNe is associated with dust formation as well. Infrared (IR) spectra of SNe IIn have revealed strong blackbody emission, providing robust confirmation for the presence of dust \citep{Fox_2011}. However, the origins of the dust have been contested. While some have posited the presence of IR echoes or pre-existing dust for various SNe \citep{Fransson_2010jl,  Andrews_2010jl,Tartaglia_2020}, others have suggested that new formation of dust in the ejecta or the dense shell between shocks is more likely, as evidenced by the evolving blueshift seen in many optical SN IIn spectra \citep{Sarangi_2018,Gall_2010jl,Smith_2012,smith20,Smith_2015da}. This phenomenon is also seen clearly in some SNe~Ibn, like SN 2006jc \citep{smith08}. There have been separate arguments for pre-existing and newly forming dust even for the same SNe such as SN 2010jl \citep{Fransson_2010jl, Gall_2010jl,Maeda_2010jl}. Despite the differing arguments about the location of the dust, it is difficult to dispute that dust forms in the dense shell of many SNe IIn based on progressively more observational evidence \citep{Gall_2010jl,Smith_2015da}. 

\par To better probe the shock physics and the structure within the CSM and ejecta, radio and X-ray emission can provide a more detailed picture. Interacting SNe are the only SN subtype expected to produce long-lasting radio emission in the form of synchrotron radiation from the accelerated electrons in the forward shock (the shock moving outwards into the CSM) \citep{Chevalier_1982}. While any SNe IIn should produce radio emission, it has only been seen in a fraction of observed SNe IIn. The first prototypical SNe IIn (SNe 1988Z, 1978K, 1986J) were bright and well observed in the radio \citep{VanDyk_1988Z, Ryder}, while VLBI observations provided great spatial detail for non-SNe IIn like SN 1993J and SN 2011dh \citep{Bietenholz_2012}. Despite this early SN IIn treasure trove, in the following three decades, fewer than 100 SNe have been detected at radio wavelengths \citep{Bietenholz_2021}. 
\par Nevertheless, extensive modeling and analysis has been done to paint the theoretical picture of these dense radio-producing shocks. \citet{Chandra_2018} and \citet{Chevalier_17} give a broad overview of the different types of absorption expected (synchrotron self-absorption or free-free absorption). An understanding of the absorption mechanism from observational data is significant as it allows for an estimate of the pre-explosion mass-loss rate.
\par 
In the X-rays, the general picture in most interacting SNe is that the heating from a collisionless shock will generate thermal photons hot enough to emit X-rays \citep{Chevalier_17,Margalit_2022}. Other nonthermal mechanisms are possible, such as inverse Compton scattering \citep{Chevalier_17}, and there has been a diversity in the emission seen from X-ray-luminous SNe over the last three decades \citep{Dwarkadas_2012}.  X-ray observations of interacting SNe and SNe in general are critical owing to the wealth of information they reveal about element abundances, the ejecta, and CSM density structure as well as the presence of ionized/neutral lines which provide information about the density and state of the shock \citep{Dwarkadas_16}.

\par 
Thus, the X-ray, radio, and optical results can be combined to paint an overall picture of the pre-SN evolution from hundreds to thousands of years pre-explosion up to the explosion date itself.
 This is what we seek to do in this work for SN 2020ywx, an object for which we obtained extensive radio, near-IR (NIR), optical, and X-ray data. SN 2020ywx was first discovered on 4 November 2020 (MJD 59157.64; all dates in this paper given in UTC) by ATLAS in the o band \citep{Tonry_atlas, atlas_server}. The SN was at $\alpha = 11^{h}53^{m}26^{s}.20$ and $\delta = 10^\circ 53' 47.29''$ in host galaxy WISEA J115326.45+105347.2 \citep{2020ywx_disc} at a distance of 96\,Mpc (derived from the central redshift of the host). While the host redshift is $z = 0.0217$, we find from our higher resolution spectra that the H\,II region near the SNe shows emission lines (which should be centered at 0 velocity) that are redshifted by $\sim 60$\,km\,s$^{-1}$. We thus adopt a redshift $z = 0.0219$ for this SN, assuming that galactic rotation causes the 60\,km\,s$^{-1}$ offset from the host redshift. We assume an uncertainty of 10\% in the distance given that we have no alternative measurements. Using the \citet{2011_reddening} dust-map calibrations, we find a Galactic line-of-sight extinction $E(B-V)=0.024$\,mag. Given there was no Na\,I\,D absorption visible in any optical spectra taken of the SN, we estimate this to be the total extent of the reddening (i.e., no reddening from the host galaxy) and apply it to all optical light curves and spectra.  
\par Following the initial discovery, classification was done by \citet{2020ywx_class} using a spectrum taken with the EFOSC2-NTT instrument on the European Southern Observatory New Technology Telescope. SN 2020ywx was identified as an SN IIn owing to the striking narrow lines. With the last nondetection 106 days earlier on 21 July 2020, the SN was found after having been behind the Sun post-peak and long past the explosion. Given the heterogeneity among SN IIn light curves \citep{Taddia_2013,Ransome_2024,Baerway2024} and the relative homogeneity of the spectra \citep{Filippenko_1997,Ransome_2024} (essentially unchanging for $> 100$ days), there is no simple way to model SN 2020ywx in context with other SNe IIn to constrain the explosion date. We thus take the explosion date as the midpoint between the last nondetection and the first detection, on 13 September 2020 (MJD 59105), % We thus estimate this as the date of explosion 
with an uncertainty of $\pm 53$ days. A similar approach has been taken for other SNe \citep{Anderson-2014} --- with the caveat that the nondetections and detections are usually much closer in time. We note that any differences in derived explosion date will not impact our results significantly given that the optical, radio, and X-ray data are generally taken $> 300$ days post-discovery/explosion. 
\begin{figure*}[htb!]
\gridline{\fig{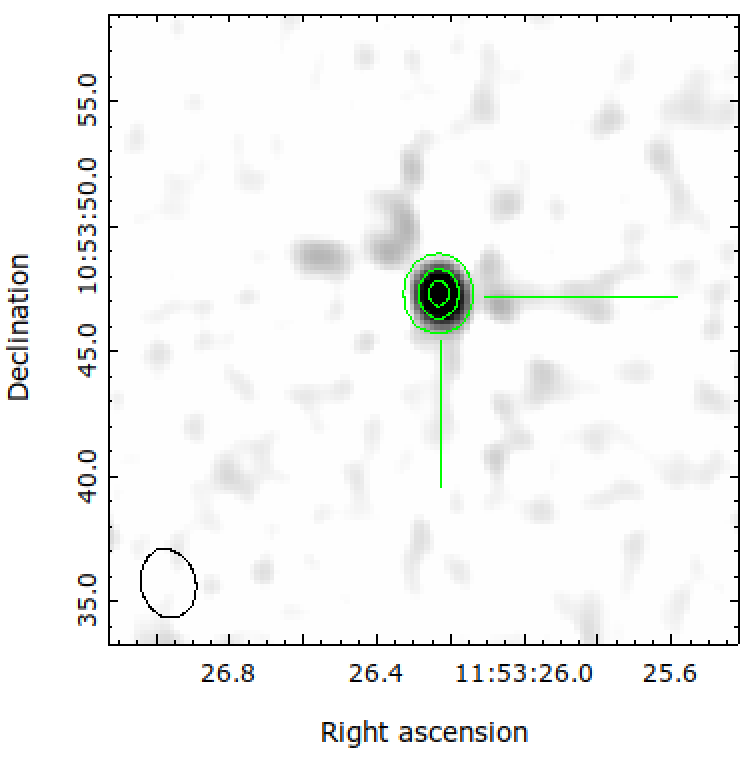}{0.35\textwidth}{ Radio 5 GHz VLA image 11/2021}\hspace{-5.0 cm}
         \fig{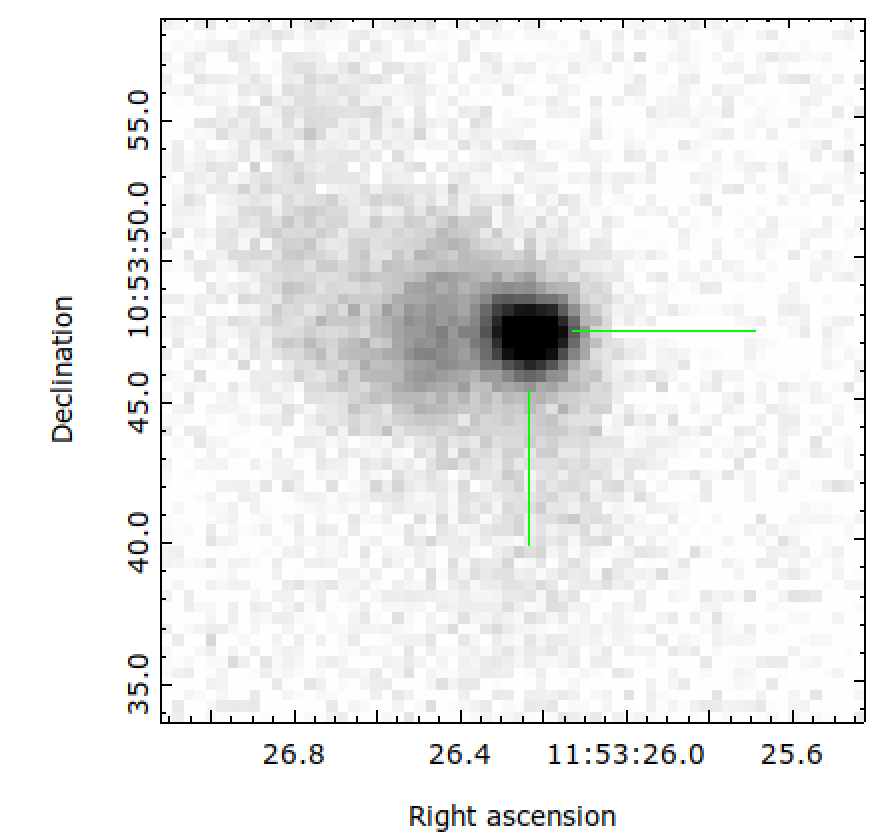}{0.37\textwidth}{LCO Optical r-band image 12/2021}}
\vspace{-0.5 cm}
\gridline{\fig{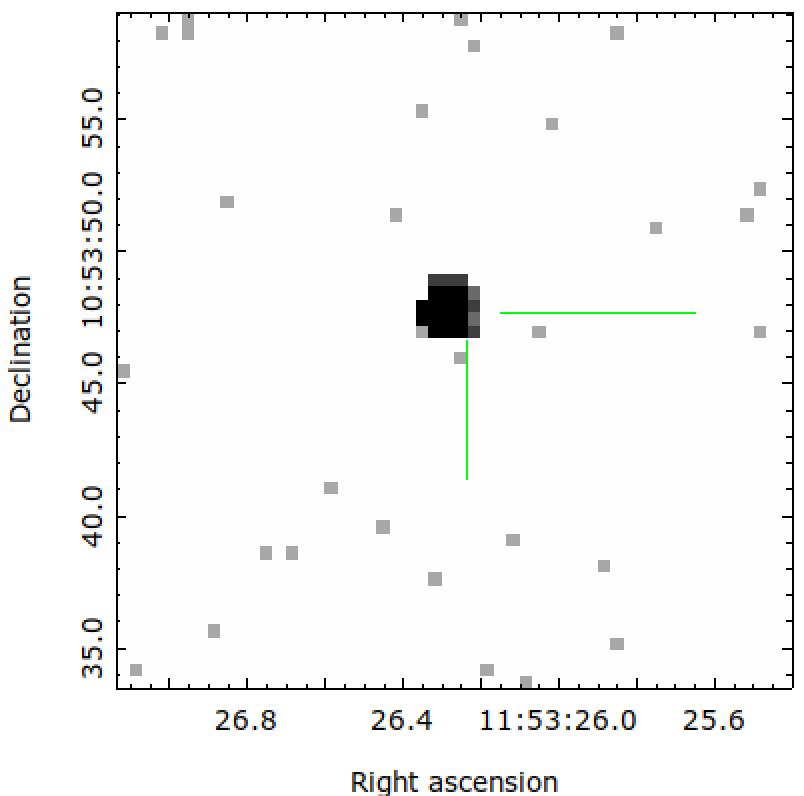}{0.35\textwidth}{\textit{Chandra} ACIS-S image 11/2021}}
\caption{Radio, optical, and X-ray images of SN 2020ywx all taken $\sim 500$ days post-explosion. The crosshairs point at the reported location of the SN \citep{2020ywx_disc}. North is up and east is to the left. We emphasize the lack of emission from the host galaxy at radio and X-ray wavelengths. The beam size is shown in the radio image at 5\,GHz in the lower-left corner. We additionally show 0.5 mJy contours in the radio image. We note that the host galaxy is not detected at all radio frequencies, including down to 0.4 GHz. }
\label{fig:Overall_images}
\end{figure*} 

\par
 The layout of this paper is as follows. In \S~\ref{sec:Data}, we describe the data taken and the associated reductions. \S~\ref{sec:Analysis} describes the fitting/modeling of our data. We interpret the results in \S~\ref{sec:Interpretations} in the context of all the fitting and modeling.  In \S~\ref{sec:Disc}, we discuss the combined results to understand the overall picture of the SN progenitor evolution and mass-loss mechanism. We summarize our work on SN 2020ywx in \S~\ref{sec:Conclusion} and emphasize key takeaways from this object in the context of SNe IIn as a class.
 
\section{Data Reduction}\label{sec:Data}
We obtained extensive X-ray (\S~\ref{sec:Xraydata}), optical/NIR (\S~\ref{sec:Optdata}), and radio (\S~\ref{sec:Radio}) data of SN 2020ywx over $> 3$ yr. 
\subsection{X-Ray}\label{sec:Xraydata}

As detailed in Table \ref{table:xray_log} (all data tables in the appendix), we obtained four epochs of \textit{Chandra} data from 231  to 1219 days post-explosion of SN 2020ywx \dataset[(DOI]{https://doi.org/10.25574/cdc.355} here) as well as a set of four short exposure \textit{Swift}-XRT observations at 180 days post-explosion. The first two {\it Chandra} epochs were taken under an approved TOO proposal (PI P. Chandra) in Cycle 22 and the last two epochs in Cycle 24 (with one set of 2 exposures which we combined from March 2023) as part of program GO-24500411 (PI P. Chandra).  Data were taken with the ACIS-S instrument and grating NONE in VFAINT mode. The host galaxy is not X-ray bright (as confirmed by matching the optical coordinates of the host to our X-ray images --- see Figure \ref{fig:Overall_images}), and thus it was straightforward to obtain a spectrum using CIAO's \citep{Ciao_2006} \texttt{specextract}.
We fit thermal plasma models (details of choosing the model are laid out in \S~\ref{sec:Analysis}) to the reduced \textit{Chandra} data using HEASARC's \texttt{xspec} v12.14.0. The initial fitting was done using $\chi^2$ statistics, and final estimates/errors were found using $\chi^2$ estimates as priors and Goodman-Weare MCMC with 50,000 chains and a 10,000 step burn-in with 20 walkers. Uncertainties are reported as 1$\sigma$ estimates from the posterior chains. Fluxes at individual energies were found using the \texttt{PIMMS}\footnote{https://cxc.harvard.edu/toolkit/pimms.jsp} toolkit as \texttt{xspec} is unable to obtain flux density at specific energies. 

The four \textit{Swift} observations, all from March 2021 (see Table \ref{table:xray_log}), were combined using the \textit{Swift} online data-reduction tools \citep{Swift_tools}. Although the signal-to-noise ratio (S/N) is low, we are able to construct a spectrum and use the parameters from the \textit{Chandra} data around the same epoch ($\sim 200$ days) to estimate the flux combining all \textit{Swift} data.

\subsection{Optical/NIR}\label{sec:Optdata}

In the optical, follow-up observations of SN 2020ywx started soon after discovery. Zwicky Transient Facility (ZTF) \citep{ztf_overview} $g/r$-band and LCO (Las Cumbres Observatory) $gri$-band photometry  followed the initial ATLAS detection within weeks and the first Las Cumbres Observatory (LCO) spectrum was taken 1 month after discovery. All LCO data are obtained as part of the Global Supernova Project (GSP). LCO photometric data were reduced using \texttt{lcogtsnpipe} \citep{Valenti_LCOpipe}, which calculates point-spread-function (PSF) magnitudes after finding zero-points and color terms \citep{Steson_daophot}. We converted the $gri$-band values to AB magnitudes \citep{Oke_Gunn} using SDSS catalogs \citep{Smith_AB}. 
\par 
As the SN was relatively close to its host galaxy ($\sim 0.5''$ away), host subtraction was performed for the optical photometry. We used SDSS (Sloan Digital Sky Survey) pre-SN host images to perform subtraction in the $g$, $r$, and $i$ bands for the LCO data, using the HOTPANTS \citep{HOTPANTS} algorithm for the subtraction. ZTF \citep{ztf_overview} data were obtained for 500 days post-explosion in the $g$ and $r$ bands \citep{ztf_reduction}.
\par 
Further optical observations over the years post-explosion were taken by other telescopes such as the Keck 10\,m, the Shane 3\,m telescope at Lick Observatory, the Bok telescope, and the MMT telescope.
In total in the optical, there were 12 LCO spectra, 1 LRIS Keck spectrum, 2 Lick Kast spectra, 3 Bok telescope B\&C spectra, and 6 MMT spectra taken of SN 2020ywx.
\par
LCO spectra were obtained on both the 2.0\,m Faulkes Telescope North at Haleakala Observatory and the Faulkes Telescope South at Siding Spring Observatory with the FLOYDS robotic spectrographs. The wavelength coverage is 3500--10,000\,\AA  and the resolution is $R \approx 1000$. 
Reduction of the LCO spectra was performed using the \texttt{floydspec}\footnote{\url{https://github.com/svalenti/FLOYDS_pipeline/}} pipeline. The Lick/Kast spectra were taken across the red and blue channels in the range 3500--8000\,\AA\ \citep{Kast_1993} and were reduced in the standard manner, obtained with the long slit at or near the parallactic angle to prevent differential light loss due to atmospheric dispersion \citep{Filippenko_1982}. The Keck LRIS \citep{LRIS_Keck} spectrum obtained on 29 April 2022 at slightly higher resolution $R \approx 4000$ (in the range 3100--6890\,\AA) was also reduced using standard IRAF \citep{iraf_86,iraf_93} techniques. MMT data were obtained with the Binospec \citep{Binospec_2019} instrument in single-object mode and the blue-channel instrument. Binospec data were reduced using a dedicated  pipeline \footnote{\url{https://bitbucket.org/chil_sai/binospec/wiki/Home}}. Further MMT data were taken with the Blue Channel spectrograph \citep{Angel_79}, which were reduced using IRAF. The Bok spectra were obtained using the Boller and Chivens spectrograph on the 2.3\,m Bok Telescope at Steward Observatory, and were reduced again using standard IRAF routines. A summary of the optical spectra is given in Table \ref{table:optical_log}. 
\par 
Two of the NIR spectra of SN 2020ywx were obtained using the Magellan Folded-Port InfraRed Echelle (FIRE) instrument on the Baade telescope, while another was obtained using Keck's Near-InfraRed Echellette Spectrometer \citep[NIRES;][$R \approx 5000$]{Wilson04}. The FIRE data were reduced using FIREhose with xtellcorr for telluric corrections. The Keck NIRES observation was reduced using \verb|Spextool| \citep{Cushing04}. \verb|Spextool| performs flat fielding and wavelength calibration, followed by flux calibration of the combined spectra using an A0V star, at a similar airmass than the target, for proper telluric correction.  
\subsection{Radio}\label{sec:Radio}
We obtained 3 epochs of Karl G. Jansky Very Large Array (VLA)  
 data of SN 2020ywx from 400--1200 days post-explosion, along with 10 epochs of Giant Meterwave Radio Telescope (GMRT) data across bands 3, 4, and 5 (250--500\,MHz, 550--850\,MHz, and 1000--1460\,MHz). Table \ref{table:Radio_log} in the Appendix lists all radio observations of SN 2020ywx used in this work. All of the VLA epochs had observations in the S (2--4\,GHz), C (4--8\,GHz), X (8--12\,GHz), Ku (12--18\,GHz), K (18--26.5\,GHz), and Ka (26.5-40\,GHz) bands.
The observations were made along with the flux calibrator 3C286 and phase calibrator J1120+1420 for both GMRT and VLA observations. The flux calibrator was used to calibrate the absolute flux and was also used as the bandpass calibrator.
\par
The VLA data were analyzed using the standard CASA tools running the most recent version of CASA-VLA v6.5.4 \citep{Casa_desc}. The data in each band were split as best as possible into symmetric subsets of channels. We then used \textit{tclean} to clean the overall image after calibration and fit a Gaussian at the SN position to measure the flux using CASA's \textit{imfit}.  Self-calibration was performed on certain datasets that had large phase errors, in particular the S-band datasets and the March 2023 dataset in the Ku band (12-18\,GHz). We ran the CASA calibration pipeline on all datasets to check against our own reductions and found that the results were consistent (within 1$\sigma$). 
\par 
GMRT reductions were also done using CASA v6.6.1. We performed an initial run through the GMRT pipeline \citep{GMRT_pipeline} followed by fine-tuned phase self-calibration using \textit{tclean},\textit{gaincal}, and \textit{applycal}. We added 10\% uncertainties in quadrature to GMRT and VLA measurements as the errors reported by CASA task \textit{imfit} may be underestimates \citep{Chandra_Kanekar_2017}.  We combined the GMRT and VLA data to make full wide-band radio spectra at three epochs from 0.4 to 40\,GHz.

\section{Data Analysis}\label{sec:Analysis}

\subsection{X-Ray Data Analysis}
\begin{figure*}[htbp!]
\gridline{\fig{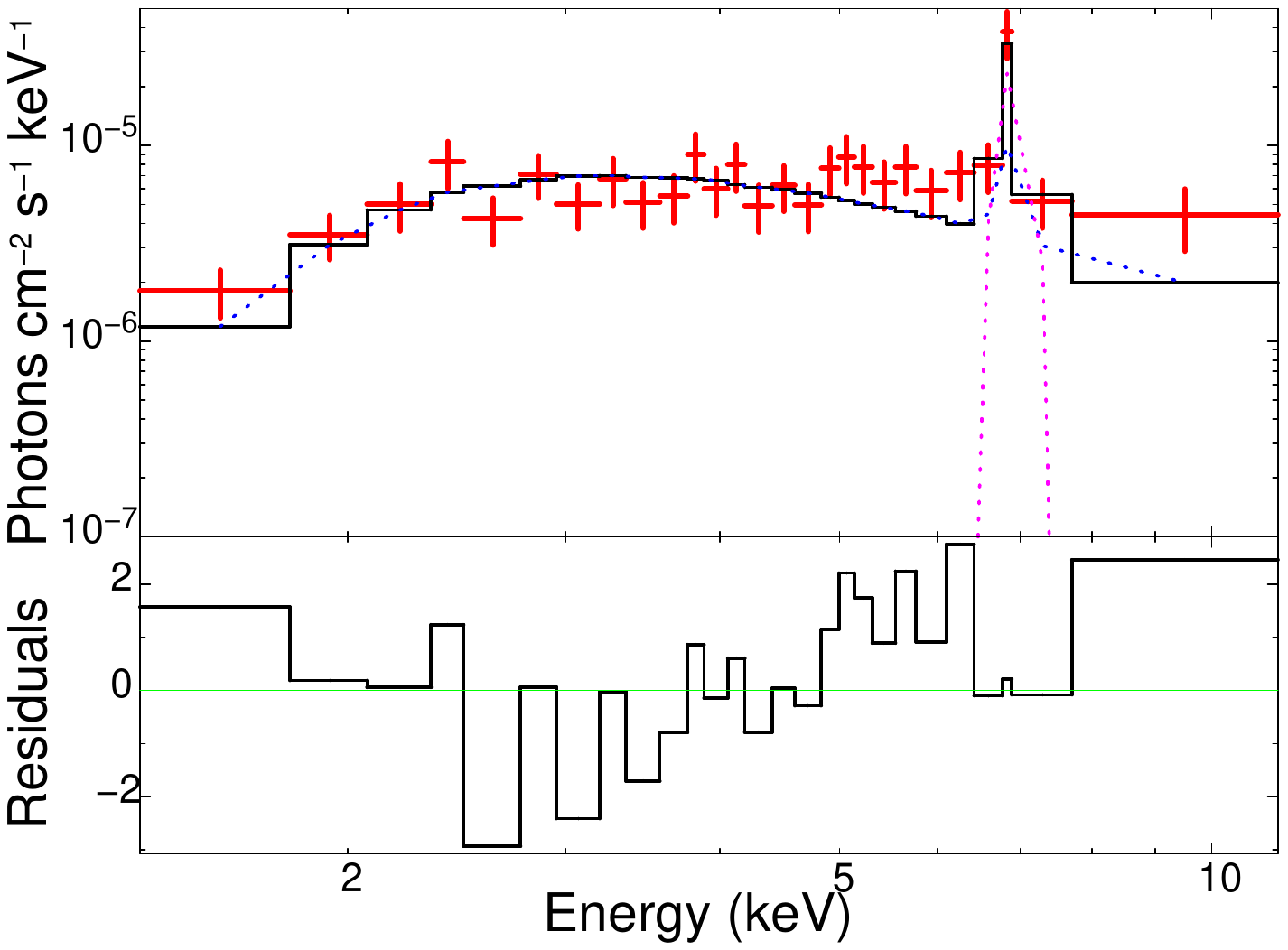}{0.4\textwidth}{ March 2021 (230 days post-explosion)}
          \fig{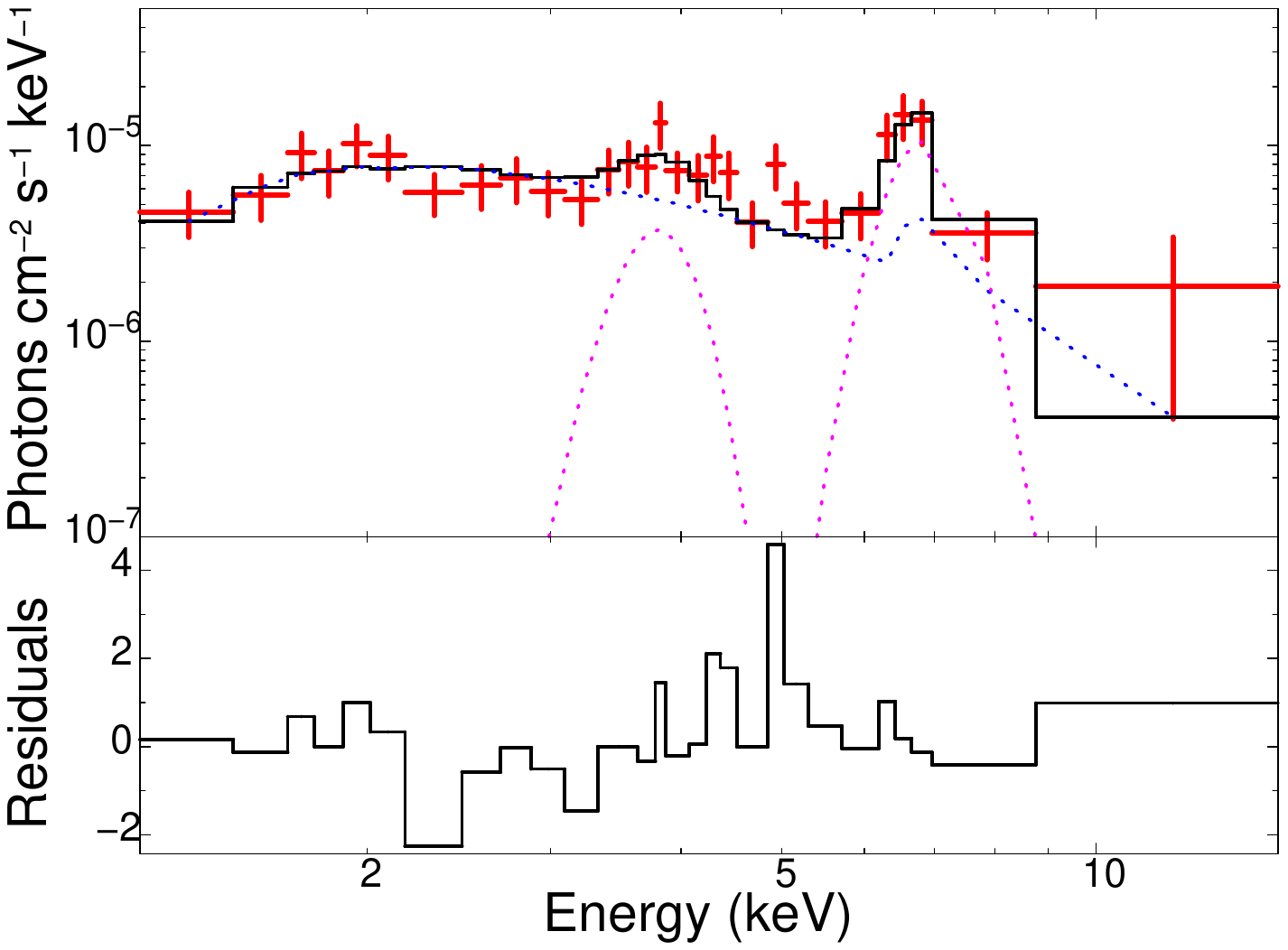}{0.4\textwidth}{November 2021 (445 days post-explosion)}}
\gridline{\fig{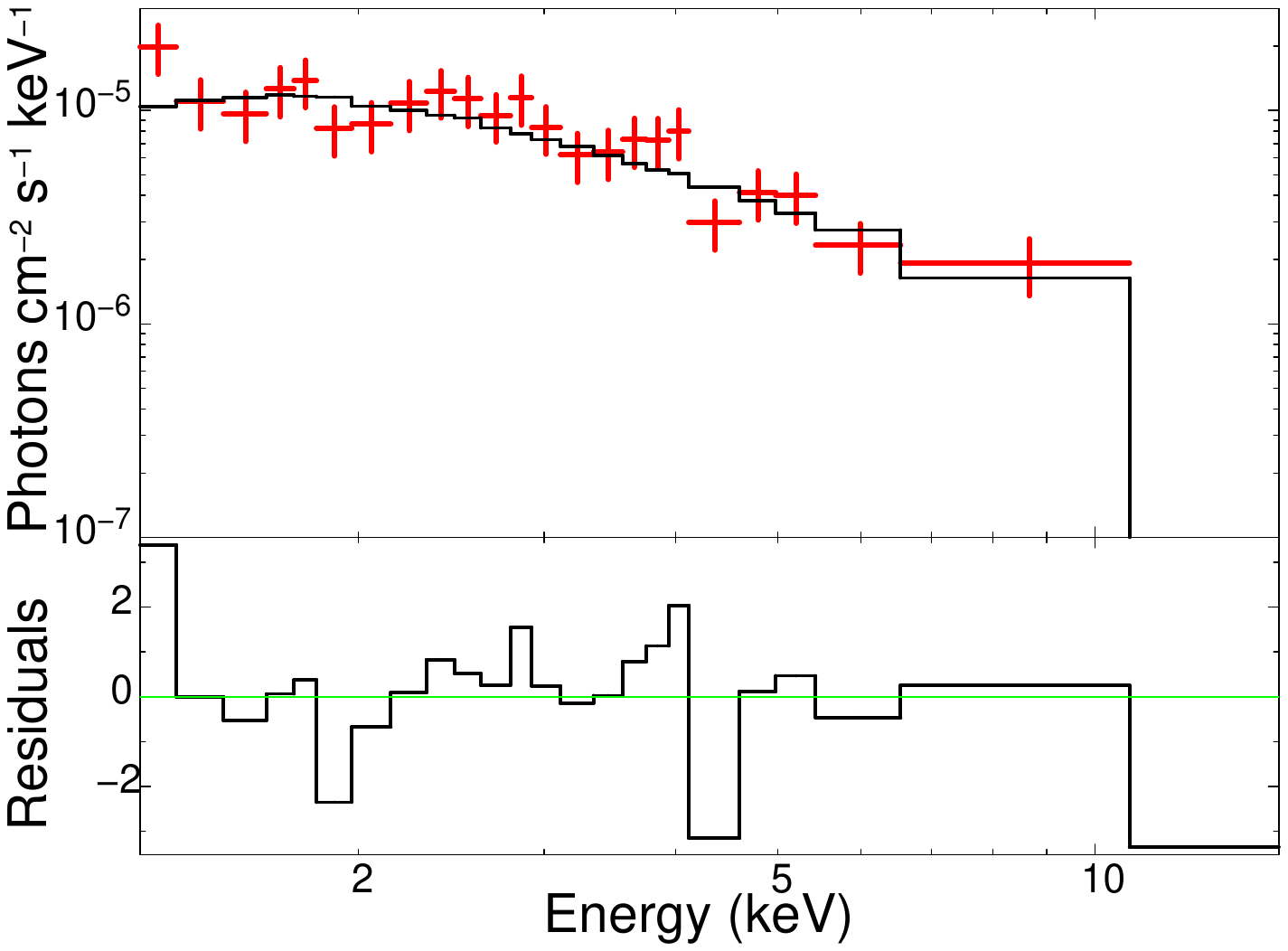}{0.4\textwidth}{March 2023 (921 days post-explosion)}
          \fig{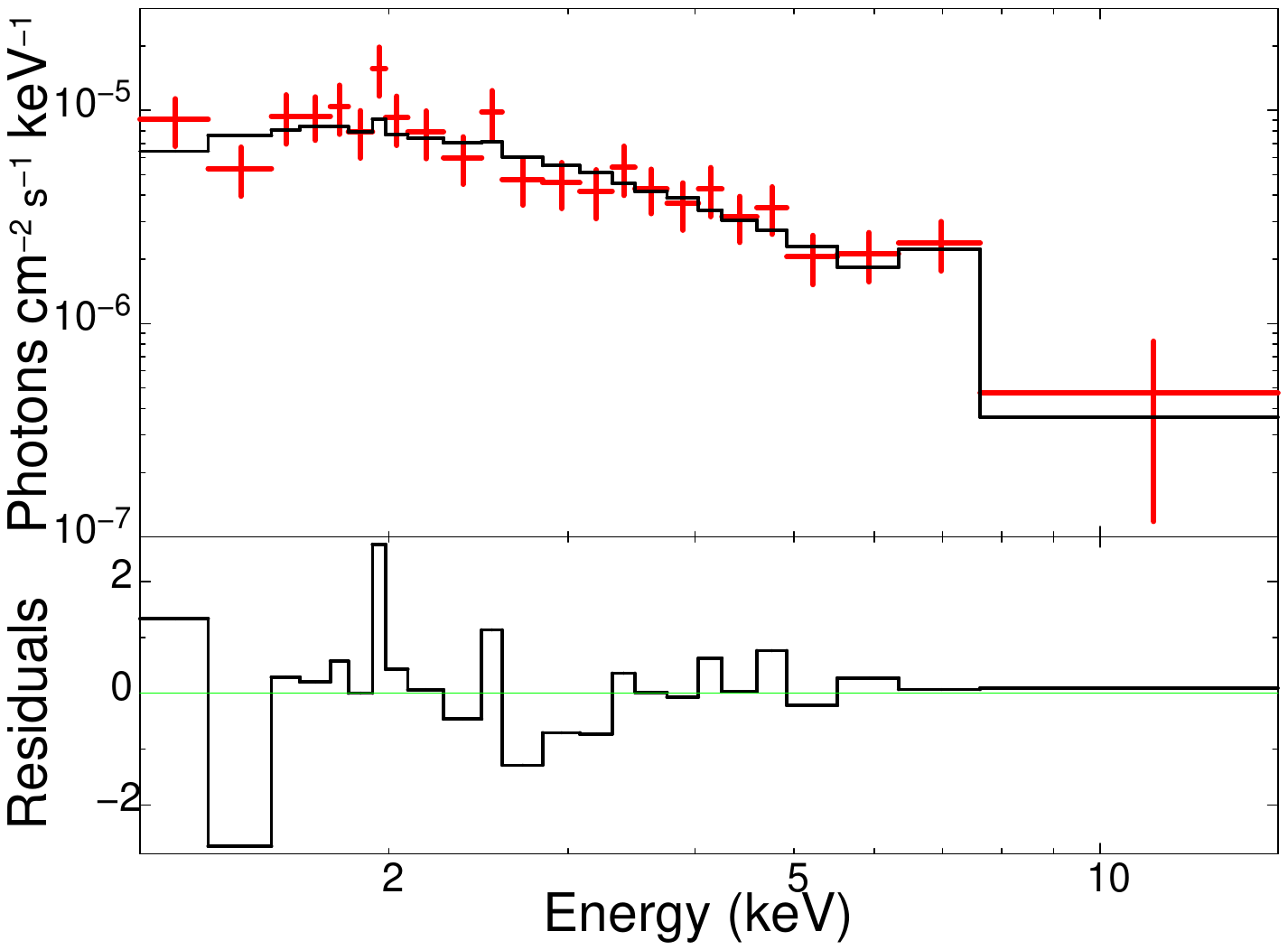}{0.4\textwidth}{January 2024 (1245 days post-explosion)}}
\caption{The four \textit{Chandra} X-ray spectra of SN 2020ywx with their fits (thermalized plasma models with associated Gaussians). The models are denoted with dashed lines in cases where there are multi-component models (at the first two epochs). The residuals are plotted as sign (data-model) $\times$ $\Delta \chi^2$. For details on the best-fit parameters, see Table \ref{table:xray_results}. We note the high temperature seen in all spectra and the declining presence of the 6.7-6.9\,keV ionized iron lines as well as the presence of the 4.0\,keV Ca line in the second epoch.}
\label{fig:Xrays}
\end{figure*} 

\begin{deluxetable*}{ccccccc}
\tablecolumns{7}
\tablewidth{2pt}
\tablecaption{ X-Ray Fit Details \label{table:xray_results}}
\tablehead{
\colhead{Epoch}  & \colhead{Model} & \colhead{$\chi^2/\nu$} & \colhead{$N_{\rm H}(10^{22}\rm {\mathrm{cm^{-2}}})$} & 
\colhead{Temperature} & \colhead{0.2--10\,keV Abs. Flux} & \colhead{Unabs. 0.2--10\,keV Flux} \\
\colhead{} & \colhead{} & \colhead{} & \colhead{} & 
\colhead{(keV)} & \colhead{($\mathrm{\frac{ergs}{cm^2 \, \text{s}}}$)} & \colhead{($\mathrm{\frac{ergs}{cm^2 \, \text{s}}}$)}}

\startdata
2021-05-01 & Thermal Plasma & 1.34 &$4.37_{-0.80}^{+1.11}$ & 20 (frozen) & $3.61_{-0.42}^{+0.40}\times 10^{-13}$ & $6.17_{-0.81}^{+0.90}\times 10^{-13}$ \\ --& Gaussian & -- & -- & 6.8 (frozen $\mathrm{\sigma}=0.1$)&  -- & $6.28_{-3.09}^{+2.89}\times 10^{-14}$  \\
2021-11-10  &Thermal Plasma & 0.96 & $1.59_{-0.39}^{+0.47}$ & 15 (frozen)& $3.78_{-0.42}^{+0.43} \times 10^{-13}$ & $5.05_{-0.59}^{+0.62} \times 10^{-13}$ \\ --
&Gaussian & -- &-- & 6.8($\sigma=0.45$ (frozen)&  -- & $1.33_{-1.11}^{+0.41}\times 10^{-13} $ \\ -- & Gaussian &--&-- & 3.8($\sigma=0.3$ (frozen))&--&$1.94_{-2.7}^{+1.10}\times 10^{-14}$\\ 2023-03-23&Thermal Plasma &1.08 &$0.76_{-0.21}^{+0.10}$& $13.598_{-1.62}^{+30.315}$&$2.49_{-0.09}^{+0.21}\times 10^{-13}$&$3.43_{-0.43}^{+0.14}\times 10^{-13}$\\ 2024-01-15&Thermal Plasma&0.84&$0.77_{-0.26}^{+0.26}$&$10.38_{-2.74}^{+32.37}$&$1.83_{-0.12}^{+0.11}\times 10^{-13}$& $2.48_{-0.31}^{+0.09}\times 10^{-13}$\\
\enddata 
%\vspace{-0.8cm}
\tablecomments{X-Ray modeling details. Some of the flux errors on the 2021 data are given by $\chi^2$ 1$\sigma$ contours rather than chain estimation owing to unbounded lower values from the MCMC chains. }
\end{deluxetable*}
\vspace{-1.0 cm}
With the X-Ray data of SN 2020ywx reduced, we performed fits to the reduced data to determine the best-fitting model. The best fit (as measured by $\chi^2$) was found by fitting each \textit{Chandra} X-ray spectrum of SN 2020ywx with a thermal plasma model (xspec's \textit{apec}) at Solar abundance (having no way to accurately constrain the host metallicity) with additional Gaussians associated with line emission. This best-fitting model is expected as the X-ray emission originates from shock heating of the particles by the collisionless shock, especially at later times.  The fits are shown with the data in Figure \ref{fig:Xrays}. The details of each fit are provided in Table \ref{table:xray_results}.

\begin{figure}[htbp!]

\gridline{\fig{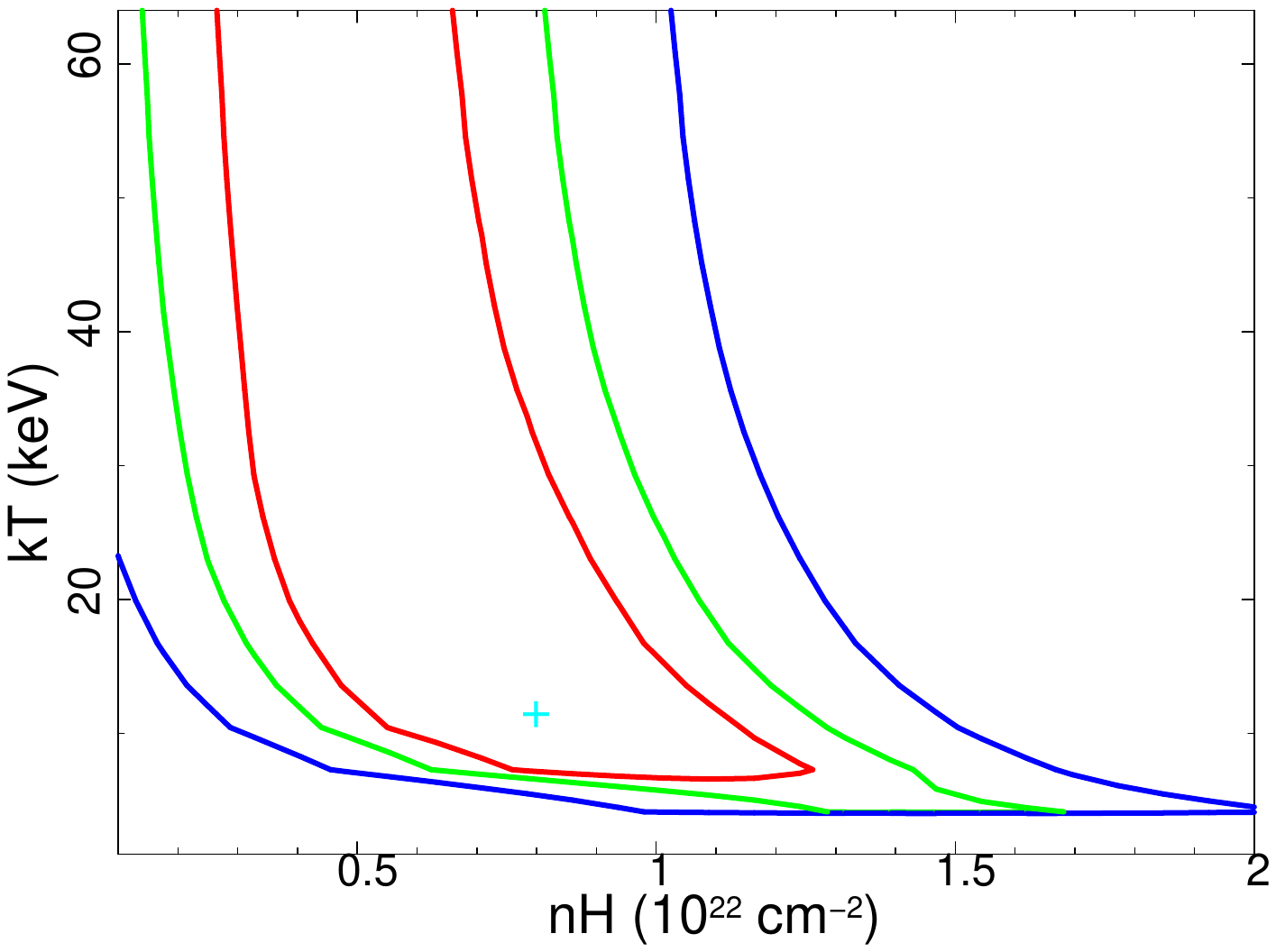}{0.4\textwidth}{}}
\vspace{-1.25 cm}
\gridline{\fig{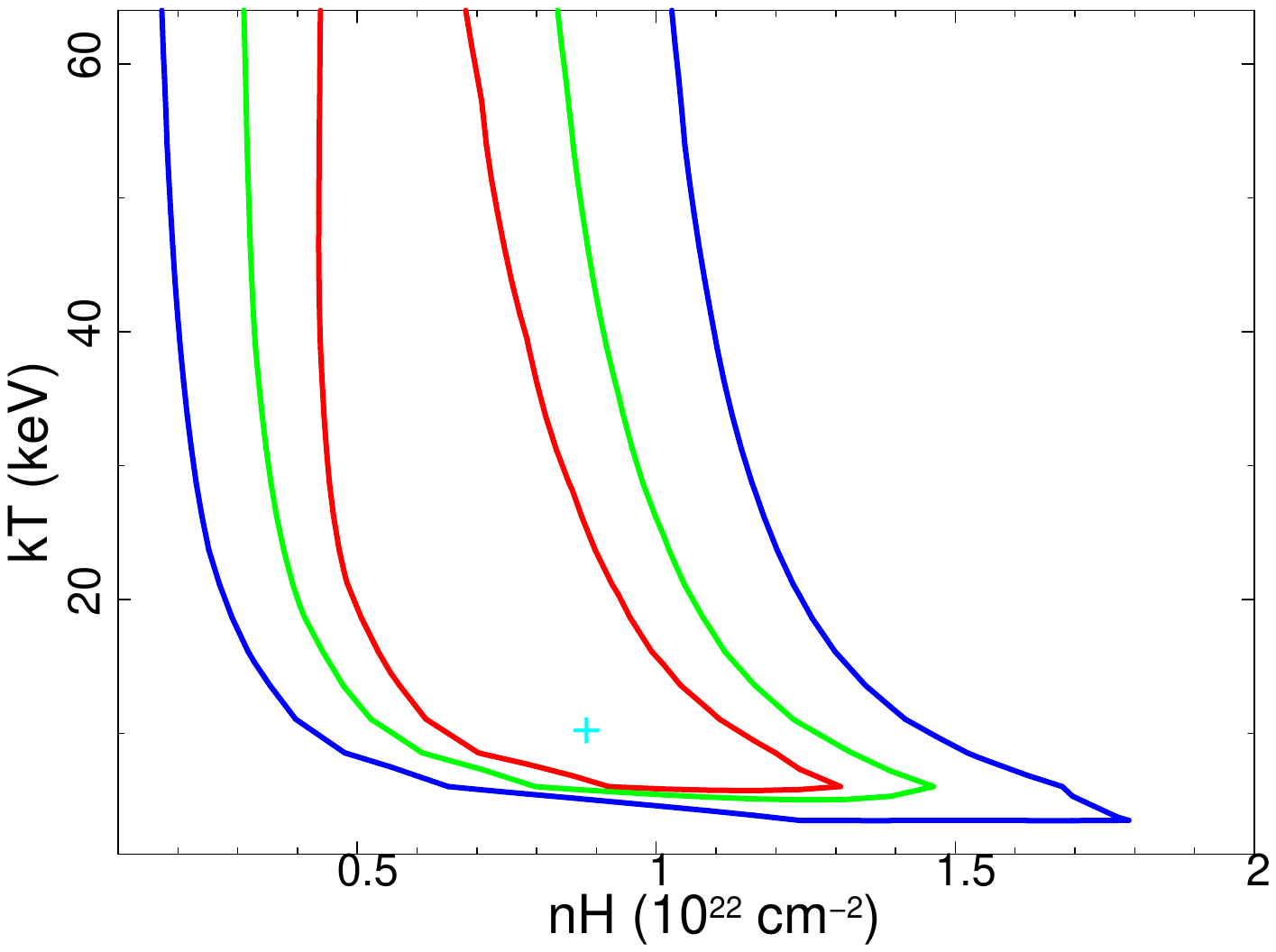}{0.4\textwidth}{}}
\vspace{-1.25 cm}
\caption{Temperature vs. column density contours from the two epochs in March 2023 and January 2024. We obtain a rough estimate for the temperature through these temperature fits.}
\label{fig:Xray_temp}

\end{figure}
\begin{figure}[htbp!]
\gridline{\fig{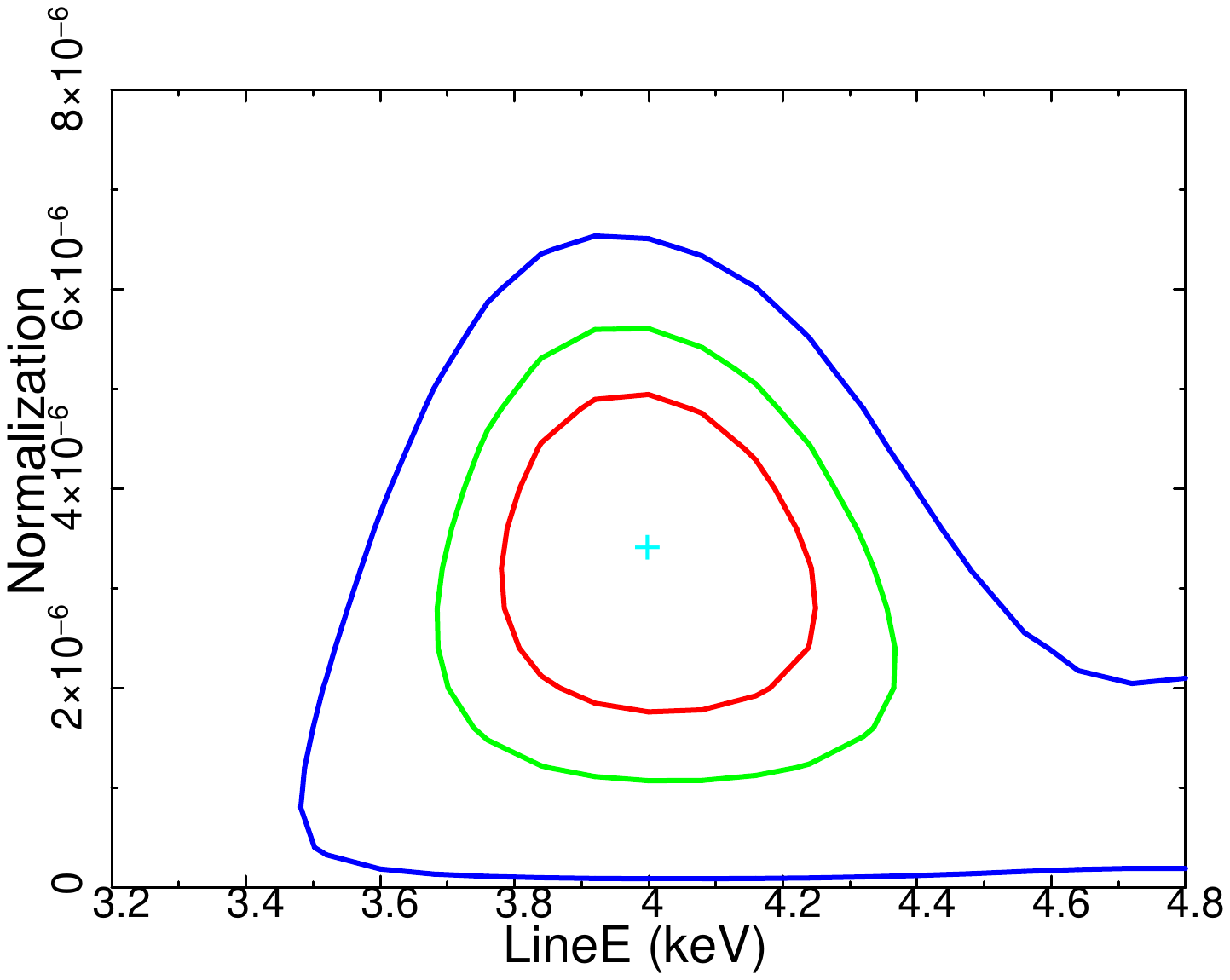}
{0.4\textwidth}{Detection significance of 3.8\,keV calcium line.}}
\vspace{-1.0 cm}
\gridline{\fig{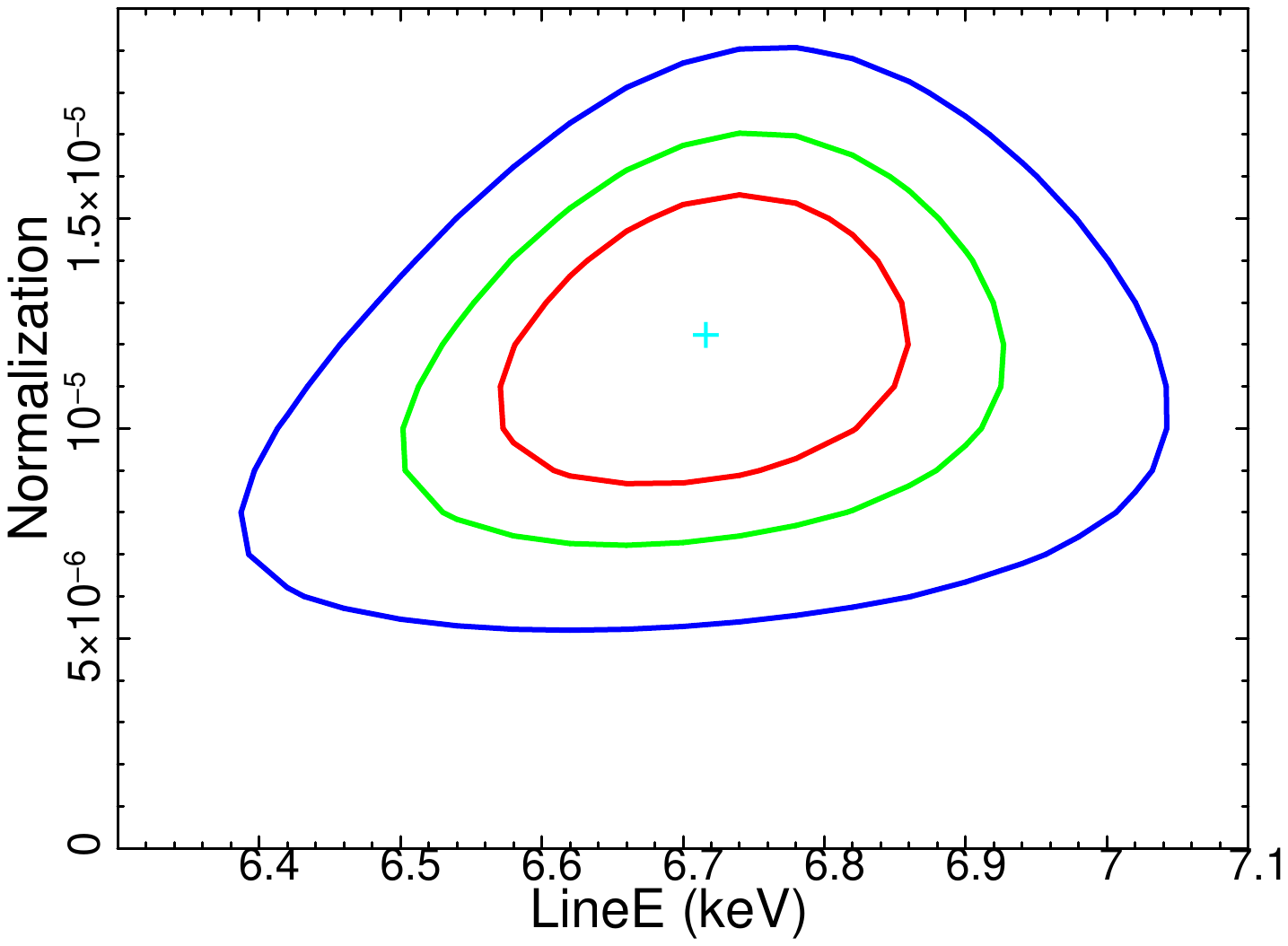}{0.4\textwidth}{Detection significance of 6.8 keV Iron line.}}
\caption{Contours of the detection significance of the two emission lines at $\sim 3.9$ and 6.8\,keV in the November 2021 X-ray spectrum of SN 2020ywx. The contours represent 1$\sigma$, 2$\sigma$, and 3$\sigma$ contours from $\chi^2$ analysis.}
\label{fig:Gaussian_confidence}
\end{figure}

\par
SNe IIn are expected to create very hot emission ($> 20$\,keV) at early times stemming from the forward shock \citep{Chevalier_17}. It is thus not surprising that the temperature is unconstrained at $> 10$\,keV over the first two epochs owing to \textit{Chandra's} sensitivity from 0.2--10\,keV. We found a temperature closer to 10\,keV in the final two epochs (see Figure \ref{fig:Xray_temp}), and thus allowed the parameter to vary, albeit with large error bars. We use these later constraints on the temperature in order to make an approximation for the temperature in the first two epochs of 20 and 15\,keV. This estimate is based on the expected power-law evolution $t^{-2/(n-2)}$ \citep{Dwarkadas_2012} of the shock-heated gas temperature, where $n$ is the ejecta density gradient exponent ($\rho_{\rm ej} \approx r^{-n}$) which we constrain at $\sim 6$ through our X-ray and radio modeling as described in \S~\ref{sec:Interpretations}.  While the ejecta density profile may be changing over time, given that the X-ray emission is probing the forward shock as also detailed in \S~\ref{sec:Interpretations} we have no way to probe the density profile through the reverse shock as was done for SN 1993J \citep{Kundu_2019}. We thus assume a non-changing ejecta density profile.
\par 
There is a distinct line at 6.5--7.0\,keV in the first two spectra. This is most likely a combination of the 6.7 and 6.9\,keV ionized Fe lines. Although the line is quite strong at our first two epochs, it faded in strength and was not detected at the later epochs. This line is confirmed at a 3$\sigma$ detection level in both early epochs as shown in Figure \ref{fig:Gaussian_confidence}. We interpret the varying width of the feature as a result of the low spectral resolution, not a true broadening. This line makes up $> 10$\% of the flux at both epochs. We froze the width of the iron-line Gaussians as otherwise the fits did not converge.  We also find an additional Gaussian in the second epoch at 3.8\,keV (the inclusion of which significantly improved the fit from a $\chi_{v}^2$ of 1.4 to 1) potentially due to an ionized line of calcium (rest energy 4.0\,keV).
\par 
Over the last two epochs, there is no line emission at a significant enough level to be detected. Given that the column density fell by a factor of 2 between epochs and that the iron line had flux $1.38 \times 10^{-13}\, \mathrm{erg\,cm^{-2}\,s^{-1}}$ at the second epoch, we would expect that the iron line would still be detectable at $> 5 \times 10^{-14}\, \mathrm{erg\,cm^{-2}\,s^{-1}}$ given the sensitivity limits of \textit{Chandra}. This nondetection is thus somewhat surprising and we discuss its implications  in \S~\ref{sec:Interpretations}. 

\par 
For the \textit{Swift} data from March 2021, we use the model values obtained from the first epoch of \textit{Chandra} data from 2 May 2021 to obtain an estimate for the 0.2--10\,keV flux, $6.01_{-0.91}^{+0.89} \times 10^{-13}\, \mathrm{erg\,cm^{-2}\,s^{-1}}$.
We combine this with the \textit{Chandra} flux values measured using our models to find the overall 0.2--10\,keV flux evolution. However, we note that given the high temperatures measured at early times, we are not capturing the full X-ray flux in our data. We thus opt to use the best-fit \textit{xspec} models to simulate the total flux over a range 0.2--50\,keV to find the ``overall'' X-ray luminosity evolution as best as possible.  We simulate the 0.2--50\,keV flux using PIMMS at each \textit{Chandra} epoch. Errors are extrapolated based on the errors on the 0.2--10\,keV unabsorbed flux values. When fitting to our 0.2--50\,keV  light curve with 10,000 MCMC chains, we find a power law that goes as $t^{-0.77}$.

\begin{figure}[htbp!]
    \centering
    \includegraphics[width=8 cm,height=6 cm]{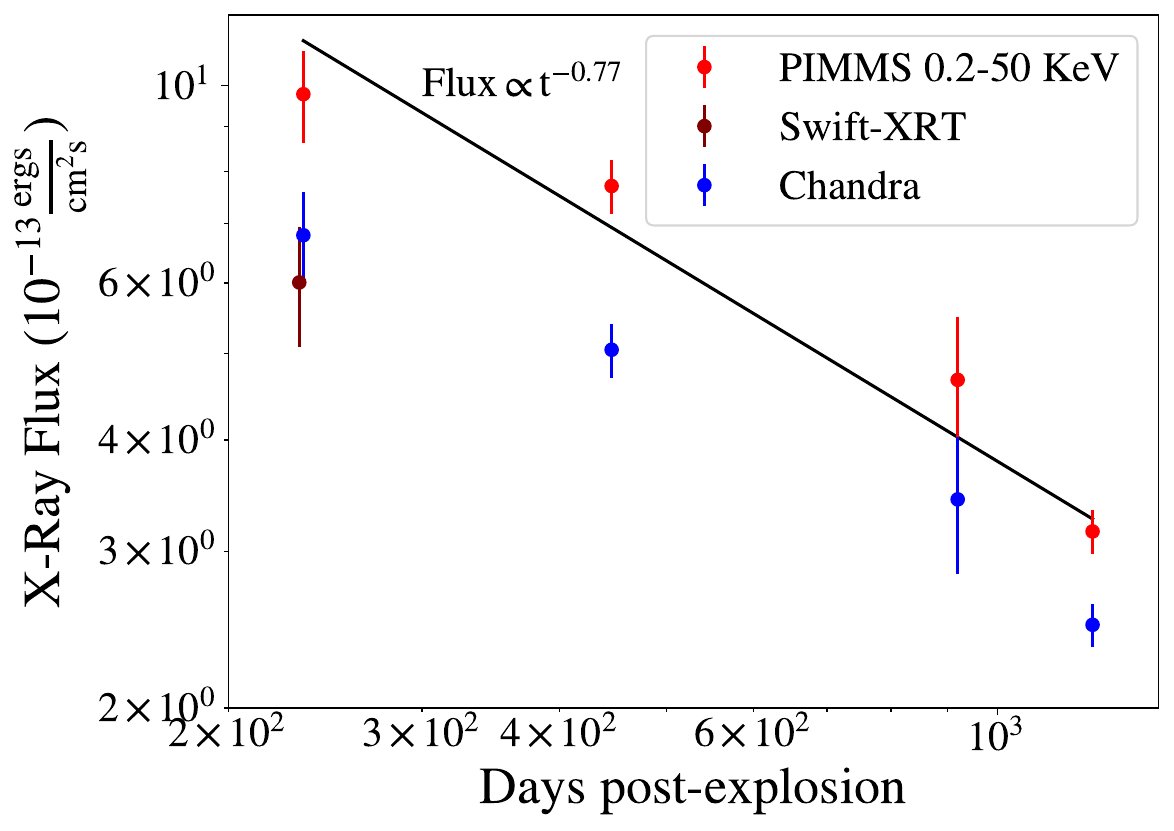}
    \caption{The fit to the 0.2--50\,keV X-ray flux of SN 2020ywx over time with a power law. We show  the measured 0.2--10\,keV flux as well as the simulated 0.2--50\,keV evolution to which we fit, finding a power law that declines as $t^{-0.77}$. This suggests a shallow $s \approx 1.85$ CSM density profile as we detail in our X-ray analysis section. }
    \label{fig:Fit_xrays}
\end{figure}

The fit to the flux is shown in Figure \ref{fig:Fit_xrays}. We expand on these results (in particular the luminosity evolution and the presence/disappearance of the iron lines) and what they mean for the SN in terms of the mass-loss rate and density gradients in the X-ray-emitting plasma in \S~\ref{sec:Interpretations}.

%\vspace{-0.5cm}

\subsection{Optical/IR}

\begin{figure*}
    \centering
    \includegraphics[width=14 cm, height =10 cm]{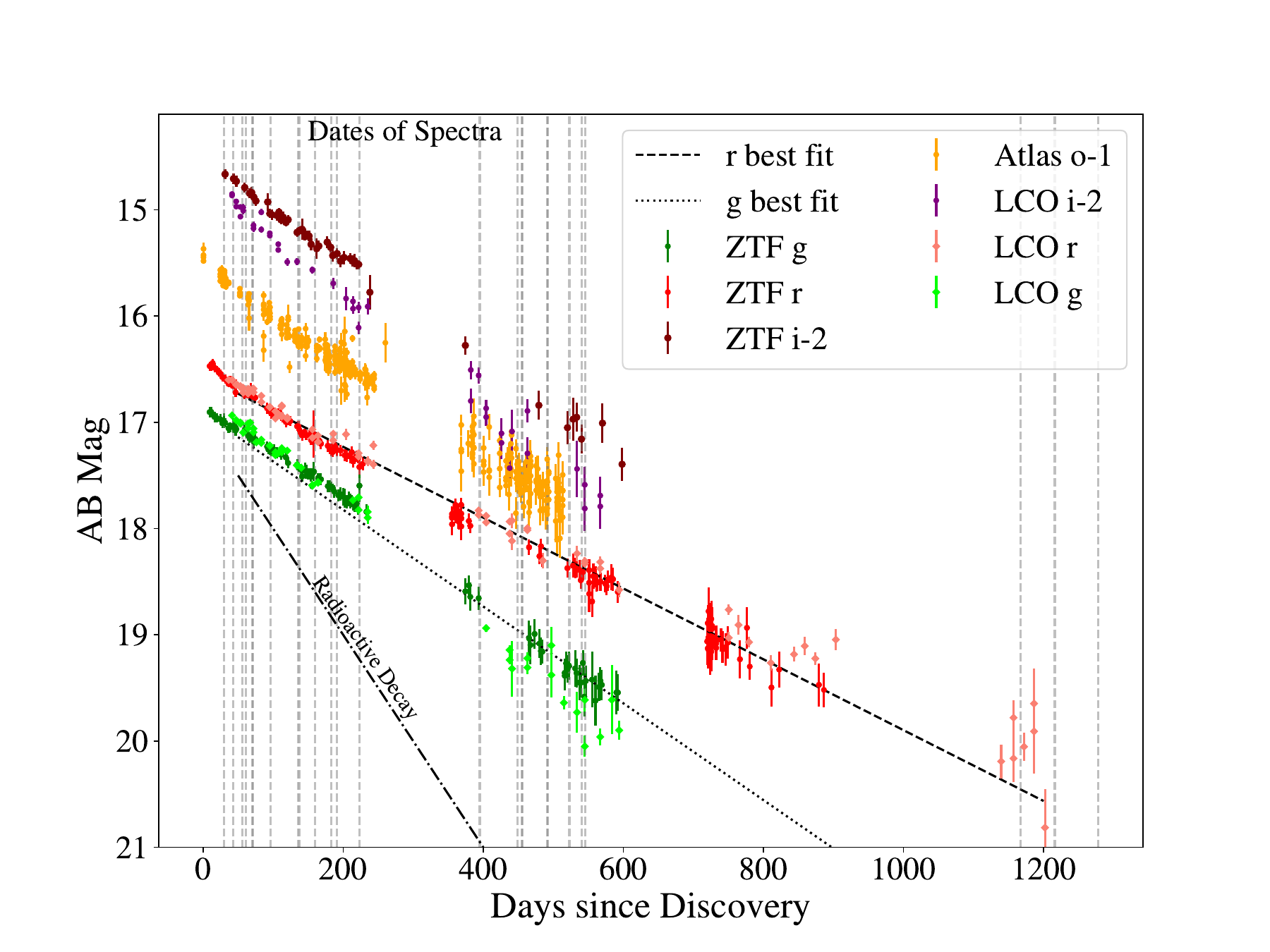}
    \caption{The host-subtracted and extinction-corrected optical light curves of SN 2020ywx in LCO $gri$, ZTF $gri$, and ATLAS $o$ bands. The LCO $i$-band results differ from those of ZTF owing to varying wavelength coverage in the two bands. The dip in LCO i band magnitude around 400 days is somewhat difficult to diagnose-it could be due to the declining Calcium NIR presence or due to instrumentation issues.  The dashed and dotted line indicates the best linear fit to the ZTF (as it is a bigger telescope) $g$ and $r$ data at 0.0046\,mag\,day$^{-1}$ and 0.0033\,mag\,day$^{-1}$, respectively.  The dash-dotted line indicates the expected radioactive decay, which the light curves clearly do not follow as their decline is more shallow. The dashed vertical lines represent the times at which optical spectra were taken.}
    \label{fig:opt_LC}
\end{figure*}

The optical light curve of SN 2020ywx covers a period of 1200 days (Figure \ref{fig:opt_LC}) and is striking in the remarkably constant linear decline for $> 1000$ days post-explosion. We fit the light curve with a linear function using MCMC fitting and find a decay rate of $0.00333_{-0.000016}^{+0.000015}$\,mag\,day$^{-1}$ in the ZTF $r$ band and $0.00455_{-0.00003}^{+0.00003}$\,mag\,day$^{-1}$ in the ZTF $g$ band. The difference in decline in the two bands is likely due to the strong H$
\alpha$ emission contributing to the $r$-band measurements. Given the lack of full coverage even into the lower-wavelength optical range (no $u$ band), and with no photometry in the ultraviolet or IR, we do not attempt measurements of the bolometric luminosity or any blackbody parameters; we do not have early-time data to constrain an inner radius or temperature of the CSM.
\par

What is most striking about the optical spectra is the similarity at each epoch. The first spectrum was taken at 83 days post-explosion. We thus missed potential early-time spectra in which the emission lines may have been better described by Lorentzians owing to electron scattering around the photosphere, as has been seen for SNe IIn such as SN 2015da and SN 2010jl \citep{Smith_2015da,Smith_2012,Zhang_2010jl}. 

\par 
In general, however, the early-time spectra look quite similar to those of prototypical SNe IIn at $\sim 100$ days. There is a blue pseudo-continuum which is likely due to a combination of many blended emission lines as a consequence of interaction seen in many subtypes of interacting SNe (e.g., SNe Ibn, IIn, and Ia-CSM; \citealt{Foley_2006jc}; \citealt{Fox_2015}). This blue pseudocontinuum fades, and the spectra become flat (no strong continuum) and line-dominated at later times.

\begin{figure*}[htbp!]
    \centering
    \includegraphics[width=12 cm, height=18 cm]{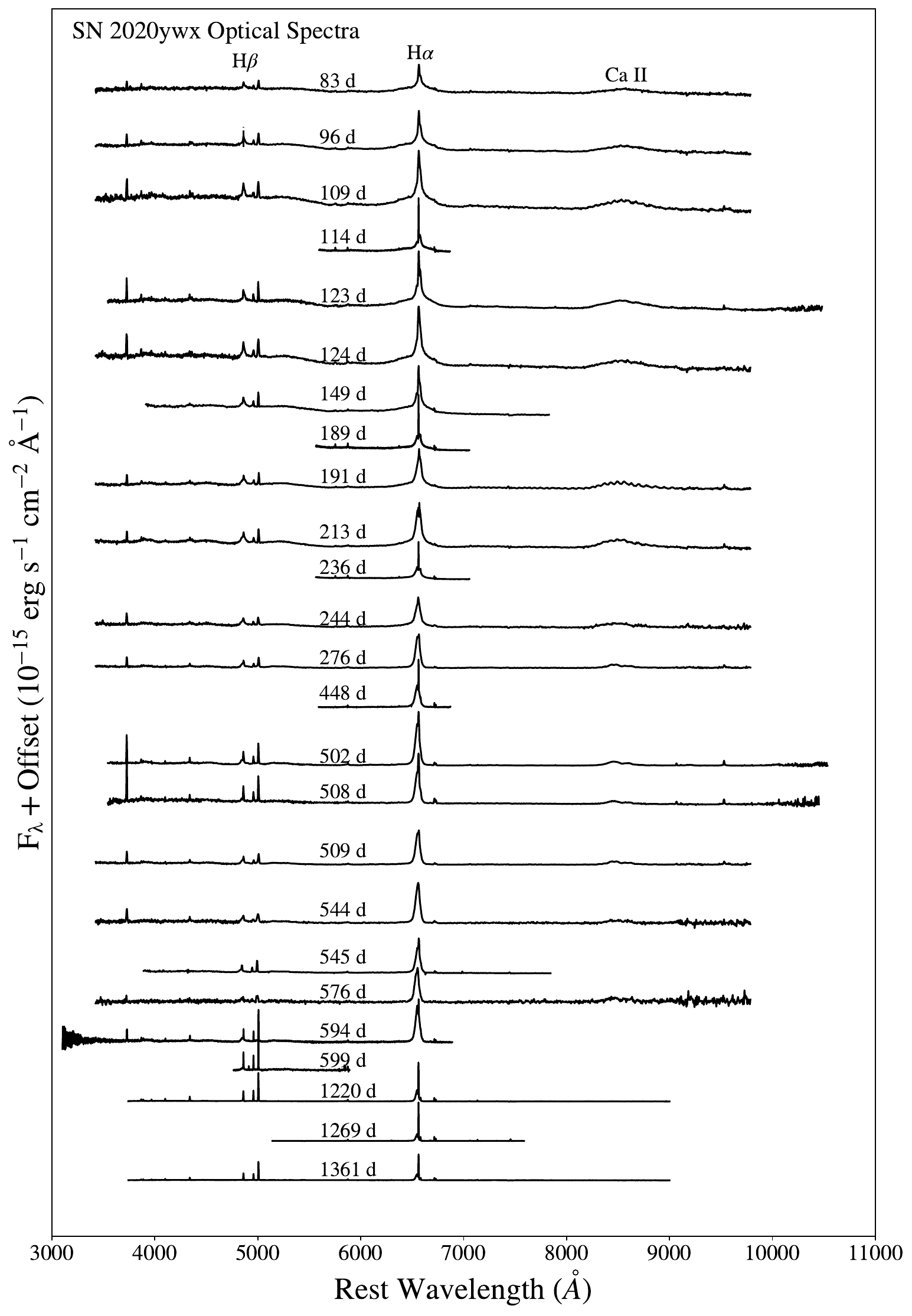}
    \caption{Extinction-corrected optical spectra of SN 2020ywx across all epochs. The H$\alpha$ emission line is most prominent, but there is additionally significant H$\beta$. We also note the declining presence of the Ca II NIR triplet.}
    \label{fig:opt_spectra}
\end{figure*}

The spectra shown in Figure \ref{fig:opt_spectra} are notable in the very strong H$\alpha$ emission combined with other hydrogen lines, a strong Ca\,II NIR triplet from the unshocked ejecta (suggested by the very high speed $\sim 10,000$\,km\,s$^{-1}$ and by the line fading over time), some H\,II region lines ([O\,III], [N\,II], [S\,II]), and narrow helium lines. Given the different resolutions of our spectra, we must be careful when interpreting the strength or absence of lines, but given that we have spectra ranging from low ($R \approx 600$) to medium ($R \approx 5000$) resolution, we are still able to immediately ascertain certain aspects of the emission. The spectra are dominated by narrow/intermediate-width hydrogen and helium emission lines coming from the unshocked CSM shed in later phases of the progenitor's life. Additionally, there is no Na\,I\,D absorption which would indicate reddening within the host galaxy, suggesting that most of the reddening is due to Milky Way absorption along the line of sight (as we assume for the optical analyses). 

\begin{figure*}[htbp!]
    \centering
    \includegraphics[width=12cm, height=10 cm]{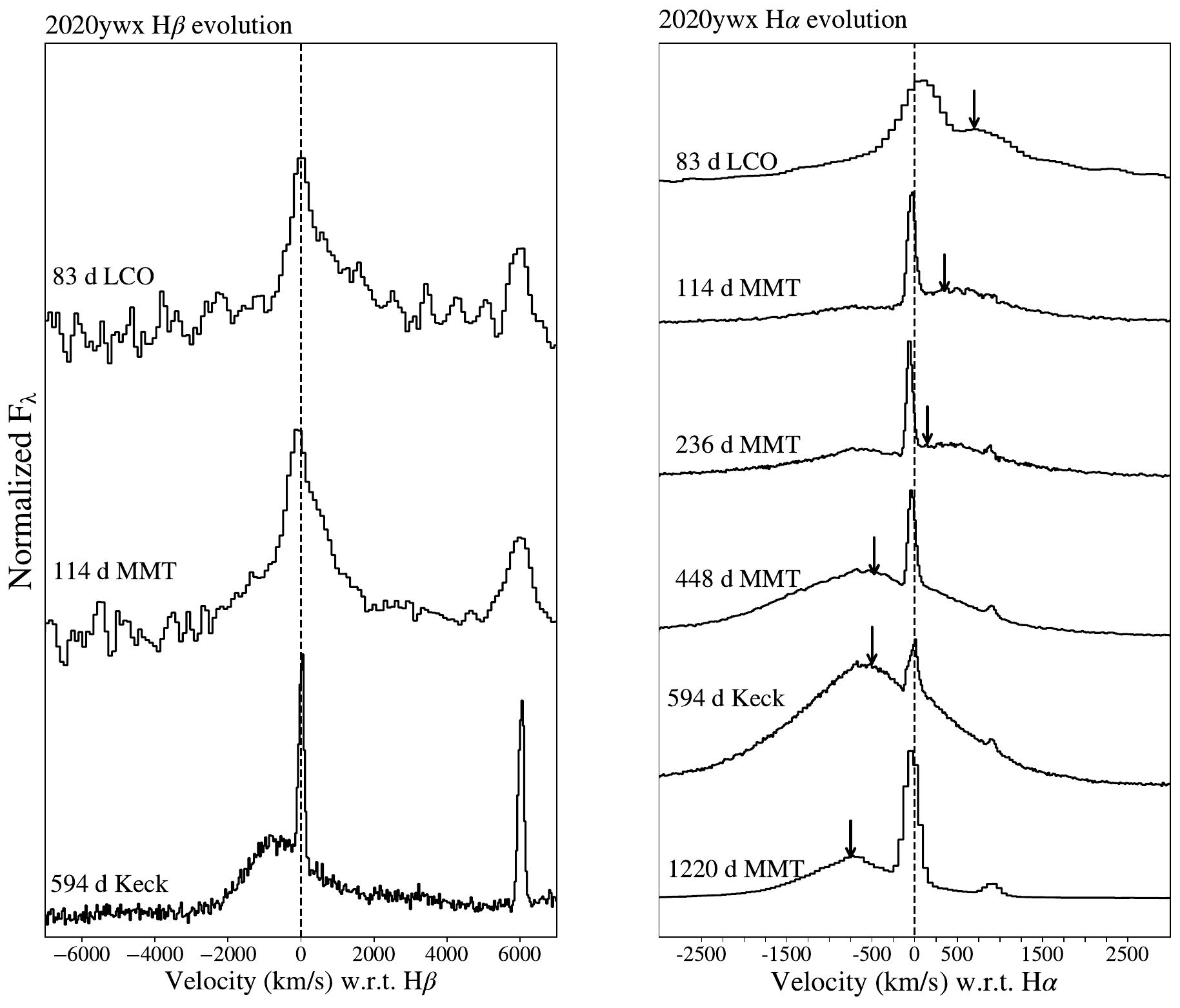}
    %add arrows for centroid velocities
    %add tick marks for velocity
    \caption{H$\alpha$ and H$\beta$ profiles across selected epochs for SN 2020ywx. The increasing blueshift of the intermediate component for H$\alpha$ and H$\beta$ (starting at 114 days at positive velocities and then going towards negative velocities, at 1220 days at $-$690 km/s) is visible. See figure \ref{fig:Blueshift} for the measured full evolution of the central velocity. We add arrows at the fitted centroid points to guide the eye to the shifting central velocity. We normalize to the H$\alpha$ peak narrow-line flux. The differing flux levels at different epochs are likely due to relative flux calibration since we normalize only to the peak value of the H$\alpha$ flux and have not used photometry to do absolute flux-calibration for this plot. }
    \label{fig:Halpha}
\end{figure*}

\par We focus mainly on the H$\alpha$ lines for our optical spectral analysis given the wealth of information they provide. The evolution of the H$\alpha$ and H$\beta$ lines is shown in Figure \ref{fig:Halpha}. It is clear that the H$\alpha$ lines in many of the spectra are contaminated by an H\,II region (revealed by the H$\alpha$ and [N\,II] $\lambda\lambda$6548, 6584 lines), which is unsurprising considering the proximity of the SN to the center of the host galaxy. This is a consistent issue for SNe IIn and has caused misidentification of SNe IIn in the past \citep{Ransome_2021}. Unfortunately, there was no spectrum taken of the host pre-SN to subtract out the narrow emission lines from the H\,II region. Thus, we perform separate fits to the H\,II region lines where they are prominent in the spectra (considering that the location of the slit will directly affect how much of the H\,II region we capture).  Another notable detail in the spectra is that the broad/intermediate components are not symmetric about zero velocity, even in the first epoch. This suggests asymmetry starting at early times and disfavors the electron-scattering hypothesis for the origin of the lines as these would be expected to be centered at zero velocity. 

\begin{figure*}
    \centering
    \includegraphics[width=15 cm, height= 15 cm]{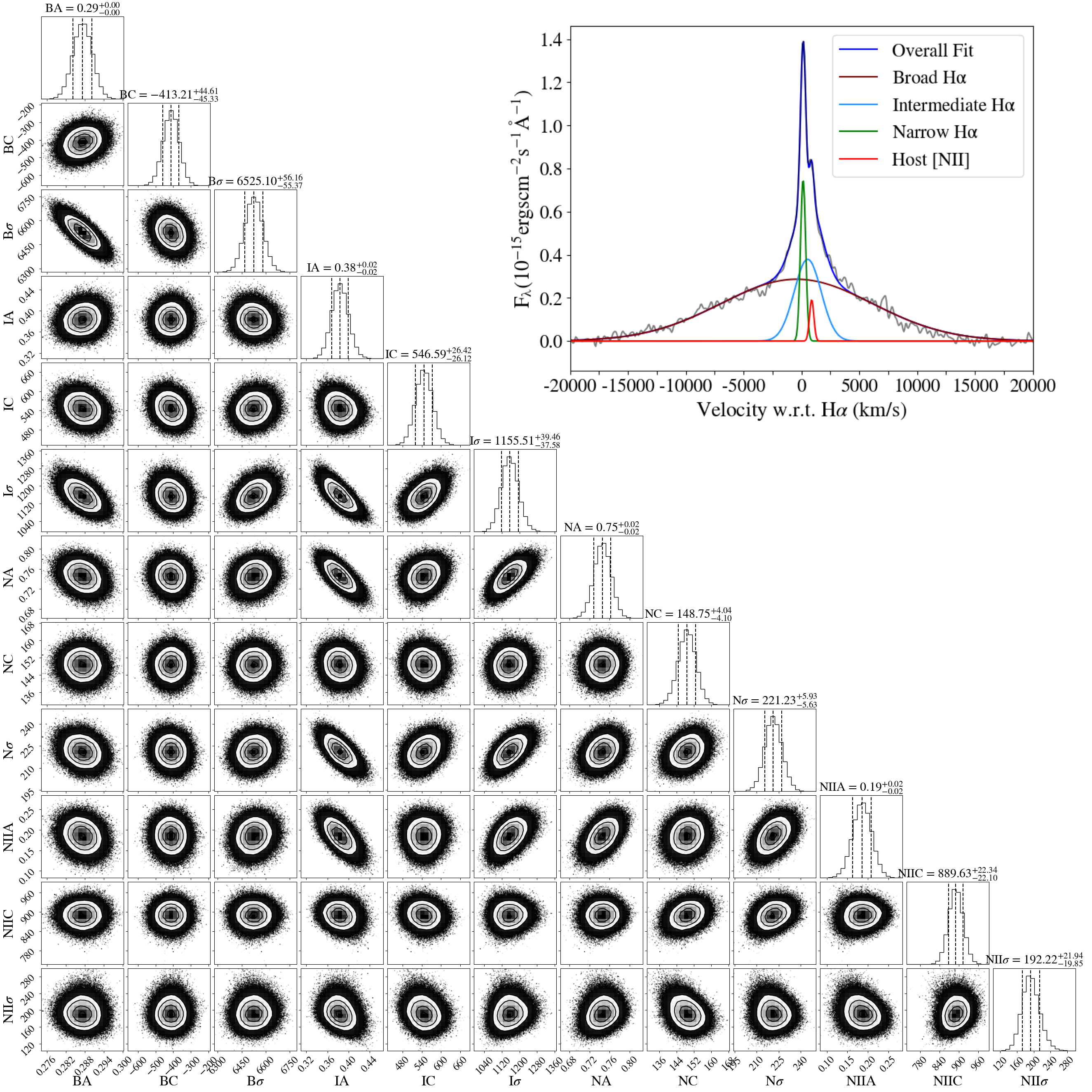}

    \caption{MCMC posteriors and the fitting (with H\,II region contamination from [N\,II]) to the first optical spectrum of SN 2020ywx (83 days post-explosion). B, I, N, and NII refer to the broad, narrow, intermediate, and [N\,II] components of the model. A and C refer to amplitude and central velocity, respectively.}
    \label{fig:MCMC_fits}
\end{figure*}

\par
We fit the continuum in our spectra with multi-order polynomials (orders vary depending on the spectrum) using \texttt{astropy} specutils and subtracted it for our line fitting. To capture the H$\alpha$ emission, we attempted Gaussian, Lorentzian, and combined Voigt fits as the emission lines are not particularly boxy, in contrast with other interacting SNe which show boxy profiles as a consequence of emission from the thin shell between the unshocked CSM and shocked ejecta \citep{Dessart_2022}. We calculated uncertainties in the fluxes when necessary using a Fourier-smoothed filter, filtering out the high-frequency noise and then taking a rolling standard deviation over a window similar to the size of a typical emission feature around 40\,\AA, similar to the procedure of \citet{Liu_2016}. Our methodology agreed within 60--140\% of the calculated uncertainty arrays for the spectra that had errors.  We performed Monte Carlo Markov Chain (MCMC) fitting on all spectra.  We use the \texttt{emcee} package \citep{Foreman_2013}, checking for convergence via the stretch statistic.
\par
\begin{deluxetable*}{cccccc}
\tabletypesize{\footnotesize}
\tablecolumns{6}
\tablewidth{0pt}
\tablecaption{Optical multicomponent $\mathrm{H\alpha}$ fitting across epochs \label{table:optresults}}
\tablehead{
\colhead{Days since explosion} & 
\colhead{Broad FWHM (km/s)  } & 
\colhead{Narrow FWHM } & 
\colhead{Intermediate FWHM } & 
\colhead{[N\,II] 6584 \AA\ FWHM} & 
\colhead{[N\,II] 6548 \AA\ FWHM}
} 

\startdata
83   & $15360_{-132}^{+133}$ & $520_{-14}^{+14}$ & $2720_{-89}^{+94}$ & $450_{-46}^{+50}$ & -- \\
189  & $13390_{-29}^{+30}$   & $80_{-1}^{+1}$    & $2630_{-6}^{+6}$   & $80_{-3}^{+3}$    & $280_{-11}^{+11}$ \\
594  & $4750_{-77}^{+80}$    & $110_{-1}^{+1}$   & $1960_{-7}^{+7}$   & $210_{-24}^{+16}$ & -- \\
1269 & --                    & $100_{-4}^{+4}$   & $1510_{-121}^{+132}$ & $260_{-117}^{+192}$ & -- \\
\enddata
\tablecomments{We provide fitting results which show the general trend across epochs. The results are rounded from the exact fitting results to 3 significant figures to reflect the accuracy of the measurements. The blending with [N\,II] lines is hard to disentangle through fitting in the lower-resolution spectra. The width of the narrow [N\,II] components is set by the resolution of the instrument. All velocities are given in km\,s$^{-1}$.}
\end{deluxetable*}

\vspace{-0.8 cm} 
The Gaussian fits were significantly better, providing a better fit both by visual inspection and by $\chi^2$ (by a factor of $> 2$). We thus proceeded with MCMC Gaussian fitting.  We fit the three hydrogen components directly, not making any assumptions beyond the Gaussian shape about the width/central velocity of the components. In the spectra in which [N\,II] lines (at 6548 and 6584\,\AA) were apparent, we fit for these [N\,II] lines additionally as best as possible. In all of the low-resolution spectra ($R \lesssim 1000$), the [N\,II] line at 6548\,\AA\ was too blended with the hydrogen components to constrain on its own. 
We thus note that our fits contain some minimal contamination from the [N\,II] line on the blue side of H$\alpha$, but we do not expect this to be a significant issue as this line contains $<1$\% of the flux of  H$\alpha$ when resolved in some of our MMT spectra.
\par
\begin{figure}[h!]
    \centering
    \includegraphics[width=8cm, height= 6cm]{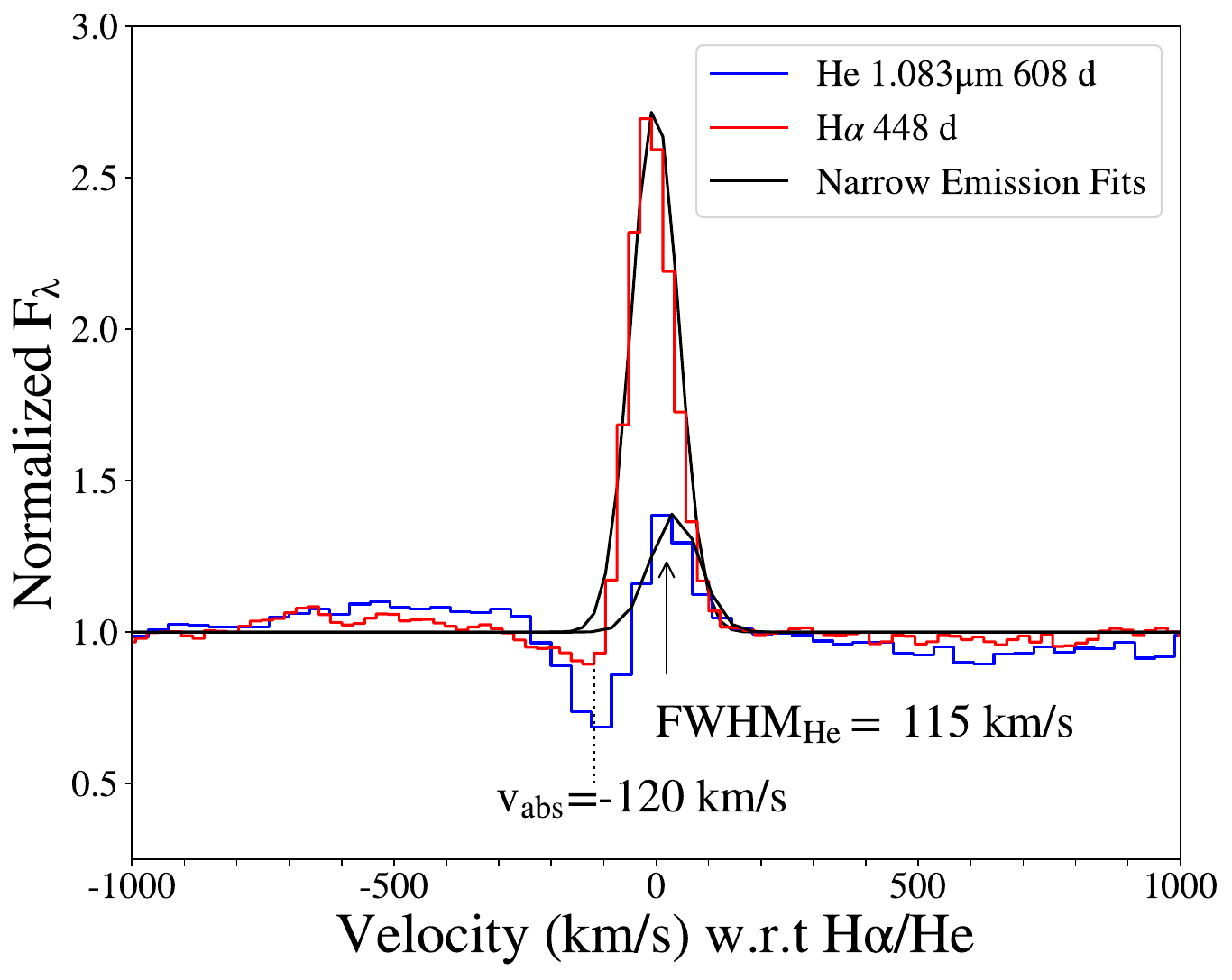}
    %add FWHM=
    \caption{P~Cygni profiles in optical H$\alpha$ (MMT spectrum at 448 days) and NIR 1.083\,$\mu$m helium (NIRES spectrum at 608 days) at two epochs 150 days apart. The spectra are normalized to the narrow features. We take the wind speed as 120\,km\,s$^{-1}$ from the absorption velocity measured in the optical spectrum and the near infrared spectrum taken 150 days apart. This speed is also consistent with all other P~Cygni profiles measured for Hydrogen in optical spectra at other epochs.}
    \label{fig:CSM}
\end{figure}

We also find a P~Cygni profile in H$\alpha$  (improving $\chi^2$ of the fit significantly) in the medium-resolution MMT spectra. 
The absorption trough of these P~Cygni profiles is $\sim -100$\,km\,s$^{-1}$ across a 300 day gap in the first 500 days post-explosion. We are able to constrain the absorption velocity (which we take as the CSM speed, since the absorption trough should exactly trace the expanding unshocked CSM \citep{Chevalier_17}) by finding the velocity at the  minimum of the absorption trough and determining uncertainties from the resolution of the spectrum (as it was clear which resolution element contained the minimum). We find $\mathrm{v_{abs}=v_{CSM}=120 \pm 22\,km\,s^{-1}}$ in the MMT spectrum at day 448. This CSM speed is relevant for our analysis across wavelengths. The fit to the H$\alpha$ profile with P~Cygni absorption is shown in Figure \ref{fig:CSM}. The presence of the absorption component confirms that part of the narrow hydrogen emission comes from the unshocked CSM; an H\,II region would not create a P~Cygni profile but the outward-moving CSM would.
\par
An example of our fits to one of our spectra with the posterior chains is plotted in Figure \ref{fig:MCMC_fits}. A portion of the fitting results that show the general evolution are given in Table \ref{table:optresults}. The upper and lower error bars are 1$\sigma$ from the MCMC posterior distributions.  We interpret the evolution and widths of the line components in the optical portion of \S~\ref{sec:Interpretations}.

\par The three NIR spectra of SN 2020ywx (see Table \ref{table:optical_log}) have similarities with the optical spectra in that they are marked by narrow and intermediate-width emission lines. We fit to hydrogen and helium lines again using MCMC methods. Of particular note is the dramatic P~Cygni profile in the helium 1.083\,$\mu m$ line in the NIRES February 2022 spectrum. We measure a full width at half-maximum intensity (FWHM) of this emission line of $115_{-26}^{+21}$\,km\,s$^{-1}$. This is quite close to the speed as measured from the absorption trough of the H$\alpha$ P~Cygni line. We show the normalized fits to this helium profile as well as H$\alpha$ in Figure \ref{fig:CSM}. Given this measurement, we assume that the CSM speed is not changing over time, given that we find an almost identical value at epochs 150 days apart. 

\par
\begin{figure}[!ht]
    \centering
    \includegraphics[width=8 cm, height= 7cm]{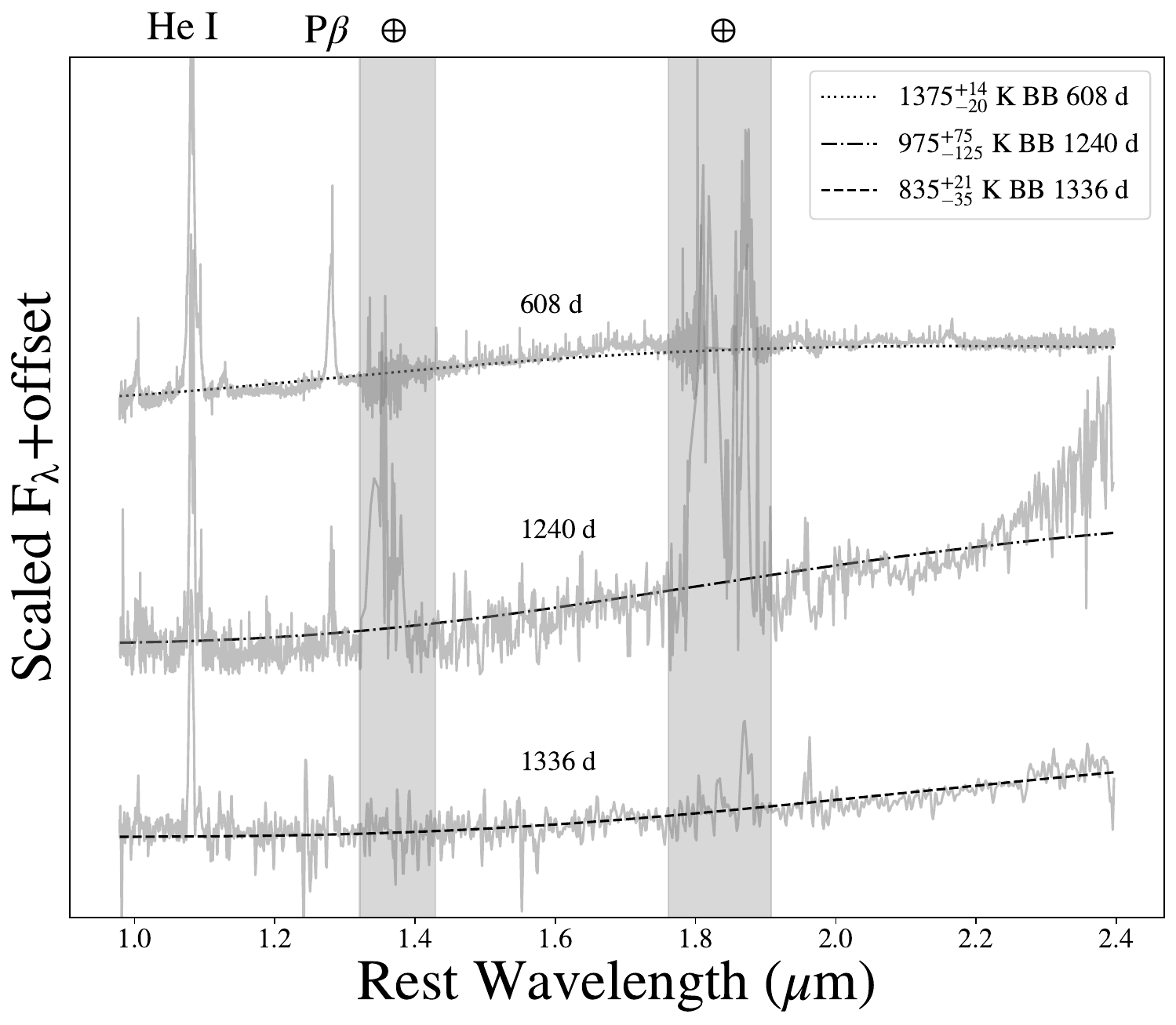}
    \caption{The NIR spectra of SN 2020ywx with accompanying cold blackbody fits. We note that at later times the continuum is redder than at earlier times, suggesting an emerging dust component. Telluric regions are shaded in gray and denoted by the $\oplus$ symbol. The prominent HeI and Paschen lines are denoted. The redward spike in continuum flux in the epoch at 1240 days is likely due to the instrument, not a physical spike in continuum flux.}
    \label{fig:IR_tot}
\end{figure}

 
Our NIR spectra are shown in Figure \ref{fig:IR_tot}. Beyond the emission lines, the continuum of the NIR spectra is interesting to consider. Over time, the continuum becomes redder and does not decline out to longer wavelengths in any of our spectra. After subtracting the optical contribution to the continuum at each epoch, we perform MCMC fitting of a blackbody to the continuum of the spectra after masking the telluric contamination and emission lines using \textit{specutils}. At each NIR epoch, we subtracted the optical contribution to the NIR continuum by fitting a blackbody to the multi-order polynomial fit to the optical continuum. At the first NIR epoch at 608 days, after subtracting an 8000\,K optical blackbody, we find a NIR blackbody at $1375_{-20}^{+14}$\,K. At the second IR epoch, the temperature of the NIR blackbody after subtracting an optical blackbody at 7300\,K now (from the MMT spectrum taken in January 2024) is $975_{-175}^{+25}$\,K. At the final epoch, we subtract a 6800\,K optical blackbody and find an IR blackbody temperature of $835_{-35}^{+21}$\,K.  Figure \ref{fig:IR_tot} shows the fits to the NIR spectra. We interpret the implications of these NIR results and what they mean for potential dust formation in \S~\ref{sec:Interpretations}.

\subsection{Radio}

\begin{deluxetable*}{cccccccccc}
\tablecaption{Best-fit parameters from radio modeling (\S~\ref{sec:Analysis}). 10\% uncertainties added in quadrature to the errorbars of the data due to well-known underestimating of radio errors from CASA's \textit{imfit} routine as described in \S~\ref{sec:Analysis}.}

\label{table:chi_radio}
\tablehead{
\colhead{Model} & \colhead{$\chi_{\nu}^2$} & \colhead{$K_{1}$} & \colhead{$K_{2}$} & \colhead{$K_{3}$} & \colhead{$\alpha$} & \colhead{$\beta$} & \colhead{$\beta'$} & \colhead{$\delta$} & \colhead{$\delta'$}
}
\startdata
SSA & 2.11 & $50.1^{+2.49}_{-2.4}$ & $0.04^{+0.002}_{-0.002}$ & N/A & $0.88^{+0.02}_{-0.02}$ & $0.51^{+0.03}_{-0.03}$ & $2.37^{+0.08}_{-0.08}$ & N/A & N/A \\
Int FFA & 1.30 & $2.30_{-0.07}^{+0.08}$ & N/A & $0.34^{+0.03}_{-0.03}$ & $1.01_{-0.03}^{+0.03}$ & $0.66_{-0.04}^{+0.04}$ & N/A & N/A & $1.85^{+0.07}_{-0.07}$ \\
Ext FFA & 2.10 & $2.92_{-0.06}^{+0.07}$ & $0.45^{+0.01}_{-0.01}$ & N/A & $1.11_{-0.02}^{+0.02}$ & $0.75_{-0.04}^{+0.04}$ & N/A & $0.93^{+0.04}_{-0.04}$ & N/A \\
Int+Ext FFA & 1.51 & $1.43^{+0.04}_{-0.04}$ & $0.58^{+0.06}_{-0.05}$ & $0.35^{+0.04}_{-0.03}$ & $1.14_{+0.03}^{-0.03}$ & $0.88^{+0.04}_{-0.04}$ & N/A & $1.53_{+0.05}^{-0.02}$ & $1.71^{+0.10}_{-0.11}$
\enddata
\end{deluxetable*}

\vspace{-0.5 cm}

In interacting SNe, radio emission arises as nonthermal synchrotron emission from relativistically accelerated electrons in the forward shock. This emission occurs in a region exterior to the SN photosphere in the forward-shock region \citep{Chevalier_1982, Weiler_1986}. However, it is expected to be absorbed either by synchrotron self-absorption (SSA) or free-free absorption (FFA). The varying optical depth across frequency creates the two regions in the spectrum of a radio SN: the optically thin declining emission and optically thick rising emission. Lower frequency emission is highly attenuated at early times; but as the optical depth decreases with time, eventually the entire radio spectrum becomes optically thin. 
\par We assume a self-similar synchrotron model, as detailed by \citet{Chevalier_1982}, where the shock radius evolves as $r\propto t^m$, with $m$ related to the ejecta density gradient by $m=(n-3)/(n-s)$. Here, $s$ gives the density structure of the CSM as $\rho_{\rm{CSM}}\propto r^{-s}$ and $n$ is given by the ejecta density profile as  \begin{equation}
\mathrm{\rho_{ej}}\propto \rho_{0}\left(\frac{t}{t_{0}}\right)^{-3}\left(\frac{v}{v_{0}}\right)^{-n}\, . 
\end{equation} 
We expect $s$ to be $\sim 2$ for a wind-like constant mass-loss CSM, but examine this assumption given that many SNe IIn have shown different density profiles \citep{Chandra_2018}. The absorption effects can be parametrized through the temporal and spectral evolution of the flux. For FFA, meaning that the synchrotron emission is absorbed by external (to the shock-CSM interaction) electrons and ions, we have \citep{Chandra_2020}
\begin{equation}
F(\nu,t)=K_{1}\left(\frac{\nu}{5 \hspace{0.1 cm} \mathrm{GHz}}\right)^{-\alpha}\left(\frac{t}{\rm{1000 \hspace{0.1 cm} days}}\right)^{-\beta}e^{-\mathrm{\tau_{FFA}(\nu,t)}} ,
\end{equation}
where $\alpha=(p-1)/2$, $p$ is the nonthermal electron energy index, and all $K$ are constants for fitting.
The FFA optical depth is given by $$\tau_{\rm{FFA}}(\nu,t)=K_{2}\left(\frac{\nu}{5\hspace{0.1 cm} \mathrm{GHz}}\right)^{-2.1}\left(\frac{t}{1000 \hspace{0.1 cm}{\rm days}}\right)^{-\delta}.$$ 
Here, $\delta$ describes the optical depth and is related to the density gradients by $\delta = m(2s-1)$. If the synchrotron emission is being absorbed by synchrotron electrons themselves (SSA), \citet{Weiler_1986} give the functional form as 
\begin{equation}
    F(\nu,t)=K_{1}\left(\frac{\nu}{5 \hspace{0.1 cm} \mathrm{GHz}}\right)^{2.5}\left(\frac{t}{\rm{ 1000 \hspace{0.1 cm}days}}\right)^{-\beta'}(1-e^{-\rm{\tau_{SSA}(\nu,t)}}),
\end{equation} 
where the SSA optical depth is given by $$\tau_{\rm{SSA}}=K'(\frac{\nu}{\mathrm{5 \hspace{0.1 cm} GHz}})^{(-\alpha-2.5)}(\frac{t}{\rm{1000 \hspace{0.1 cm} days}})^{(-\beta'+\beta)}.$$ The value of $\beta'$ gives the time evolution in the optically thick phase, where the flux is still rising, while $\beta$ gives the flux in the optically thin phase for all the models. \citet{Weiler_1990} also showed (for SN 1986J in particular) that in the dense environment of an SN IIn, the dense shell of gas that forms can mix with the synchrotron-emitting region of the shock and contribute to the free-free absorption. Known as internal free-free absorption, this model can be written as 
\begin{equation}
\begin{split}
F(\nu,t) &= K_{1} \left(\frac{\nu}{5 \, \mathrm{GHz}}\right)^{-\alpha} 
\times \left(\frac{t}{\mathrm{1000 \, days}}\right)^{-\beta} \\
&\quad \times \left(\frac{1 - e^{-\tau_{\mathrm{intFFA}}(\nu,t)}}{\tau_{\mathrm{intFFA}}(\nu,t)}\right),
\end{split}
\end{equation}
where the optical depth $$\tau_{\rm{intFFA}}=K_{3}\left(\mathrm{\frac{\nu}{\mathrm{5\hspace{0.1 cm} GHz}}}\right)^{-2.1}\left(\frac{t}{
\mathrm{1000\hspace{0.1 cm} days}}\right)^{-\delta'}.$$ We fit all three of these models to our combined radio data (at all times and frequencies) and then finally with a combined internal+external FFA model as has been seen for other SNe IIn such as SN 1988Z \citep{VanDyk_1988Z} and SN 2001em \citep{Chandra_2020}.
The fits are done with MCMC, with 10,000 steps, a 4000-step burn-in, and 100 walkers.

\par We see from Table \ref{table:chi_radio} that the internal FFA model is the best fit for the radio data (with $\chi^2_{\nu}=1.3$), and adding the external FFA model (and enforcing a physical value for $\delta = 2.7m$) leads to a worse $\chi^2_{\nu}$. This suggests that the external medium is contributing quite minimally to the absorption.
The better fit is made clear from the low-frequency GMRT data points, which reveal a shallow dropoff that is not accounted for in an SSA or external/external+internal FFA model.

\begin{figure*}[htp!]
    \centering
    \includegraphics[width=18 cm, height=6 cm]{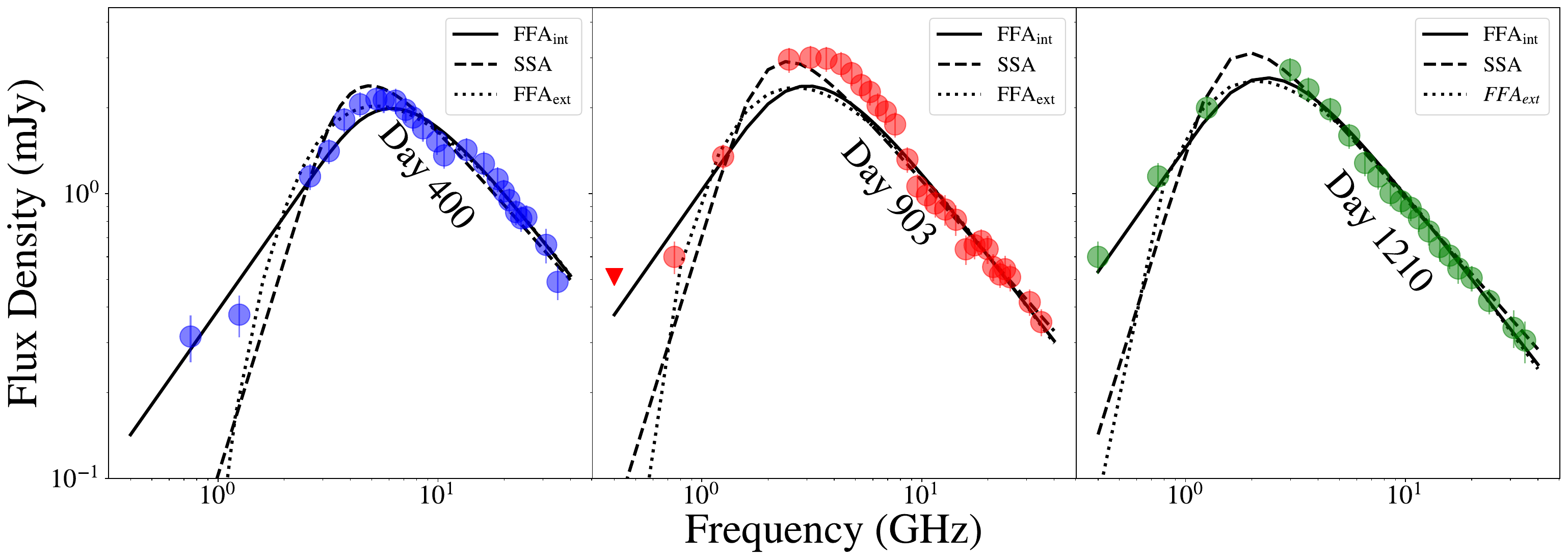}
    %add upper limit figure
    \caption{Radio spectra of SN 2020ywx at the 3 VLA epochs with the best-fit internal free-free absorption (FFA) model (denoted with the solid line as $\mathrm{FFA_{int}}$), the synchrotron self-absorption model (denoted with the dashed line as SSA), and the external FFA model (denoted with the dotted line as $\mathrm{FFA_{ext}}$. The internal FFA model is the best-fitting model as the other models do not account for the elevated flux at low frequencies. However, we note that the data above and below 10 GHz seems to be inconsistent with our model at 400 and 903 days. We attempted two-component external FFA/FFA+SSA fits and did not see an improvement in the $\chi^2$, however, and thus accept the one-component internal FFA as the best fit. We attribute the discrepancies around 10 GHz to some physical mechanism that the models do not fully capture. We interpolate the GMRT data to get a flux density value at the exact VLA epochs given that GMRT data were not always taken coincidentally with the VLA epochs.}
    \label{fig:Radio_spec}
\end{figure*}
\par
\begin{figure*}[htp!]
    \centering
    \includegraphics[width=14 cm, height=8 cm]{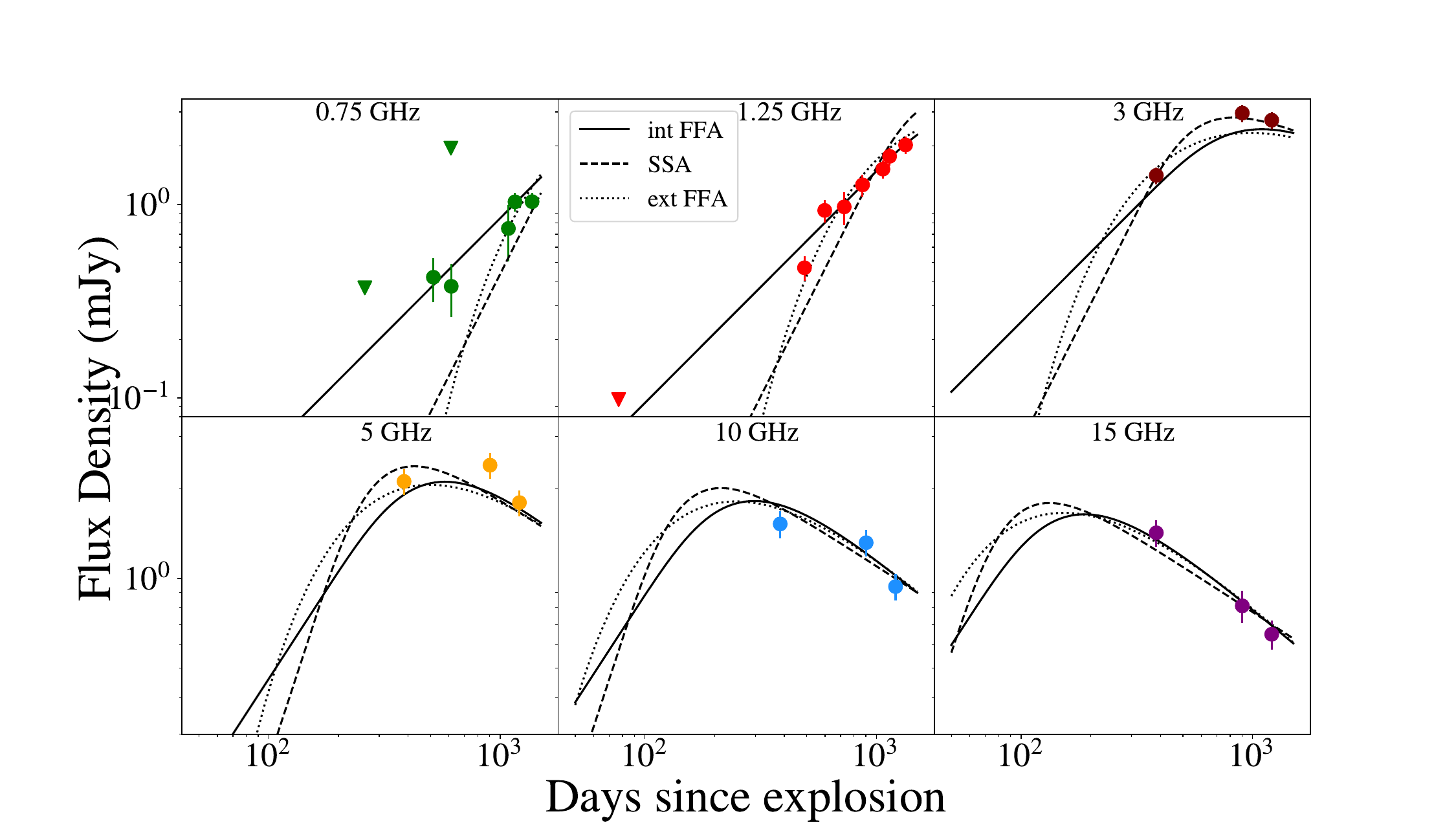}
    %reduce lower y range of lightcurves%change upper limit marker size-change to lower triangles
    \caption{Radio light curves of SN 2020ywx with the associated internal free-free absorption model at 6 frequencies across VLA and GMRT bands. The model generally fits the data well despite the VLA frequencies not being well sampled over time.  }
    \label{fig:Radio_lc}
\end{figure*}

We show our fits to the three VLA+GMRT spectra in Figure \ref{fig:Radio_spec}. There is some discrepancy in flux above/below 10 GHz which is not accounted for even when trying multi-component FFA fits. We attribute this to some microphysics aspects of the emission not captured by the models. Further monitoring of SN 2020ywx at radio wavelengths would be key to understanding this discrepancy.
The model also fits well to the light curves as seen in Figure \ref{fig:Radio_lc}. The light curves reveal the transition from optically thick to thin around 5\,GHz, while the 10/15\,GHz light curves are optically thin throughout our epochs and the 3/1.25\,GHz light curve is still rising due to high optical depth. By constraining the optical depth through our model at 5\,GHz, we are able to make measurements of the radio mass-loss rate in \S~\ref{sec:Interpretations}.

\section{Interpretations}\label{sec:Interpretations}
Having modeled the emission across four wavebands, we now interpret our results to understand the overall picture of SN 2020ywx and its progenitor.

\subsection{X-Ray Interpretation} 
Using our X-ray fitting results, we constrain the detailed evolution of SN 2020ywx in the X-rays and place it in context with other X-ray bright SNe IIn. The high temperature across the fits suggests that the X-ray emission must be coming from the forward shock. While the temperature is not fully constrained by our fits, we can use it to approximate the shock velocity as best as possible. \citet{Chevalier_17} showed that assuming solar abundances (which we also assume given we have no measurement of the metallicity of the SN host),

\begin{equation}
    T_{\rm{FS}}=(1.17\times 10^{5}) v_{4}^2\, {\rm keV},
\end{equation} 
assuming equipartition between electrons and ions and giving the temperature in keV, where FS refers to the forward shock and $\mathrm{V_{4}}$ is the shock speed in units of 10,000\,km\,s$^{-1}$. We assume equipartition given that with the measured forward shock temperature and density $\sim 10^{-14}$  g $cm^{-3}$ (using the mass-loss rate we derive in this section), we find an equipartition time $\sim$ 10 days using equation 12 of \cite{Chevalier_17} along with the shock speed $\sim$ 4000 km/s.
In our last two X-ray epochs at 921 and 1219 days post-explosion, the temperature was roughly constrained at 13.5 and 10.5\,keV, respectively. 
Based on our extrapolated values from the expected temperature evolution assuming $n \approx 6$ (which we justify later in this section), we find that at the first two epochs, the shock speeds were 4100 and 3900 $\pm$ 500\,km\,s$^{-1}$ (with error bars found from extrapolating uncertainties in the temperatures at later times). These are somewhat higher than estimates from the FWHM of the intermediate component in the optical H$\alpha$ fits ($\sim 2500$--3000\,km\,s$^{-1}$), as expected given that the intermediate component traces the dense shell and not the shock directly. We note that these velocities/temperatures are consistent with measurements of the forward-shock temperature made for other SNe IIn around 20\,keV, such as SN 2014C \citep{Margutti_2017, Brethauer_2022,Thomas_2014C} or SN 2010jl \citep{Chandra_2012, Ofek_2010jl}. We adopt 4100\,km\,s$^{-1}$ as the shock velocity at 230 days post-explosion when the first X-ray data were taken. The shock velocity at the subsequent epochs is given by the evolving temperature. We emphasize that the highest velocity components ($\sim 15,000$\,km\,s$^{-1}$) measured in the optical data are from the ejecta, not the shock. This suggests asymmetry as we see the ejecta at fast speeds from early times. 
The shock speeds are relevant for our optical mass-loss measurements as well as the X-rays.
\par
\begin{figure}[htbp]
    \centering
    \includegraphics[width=8 cm,height=12 cm]{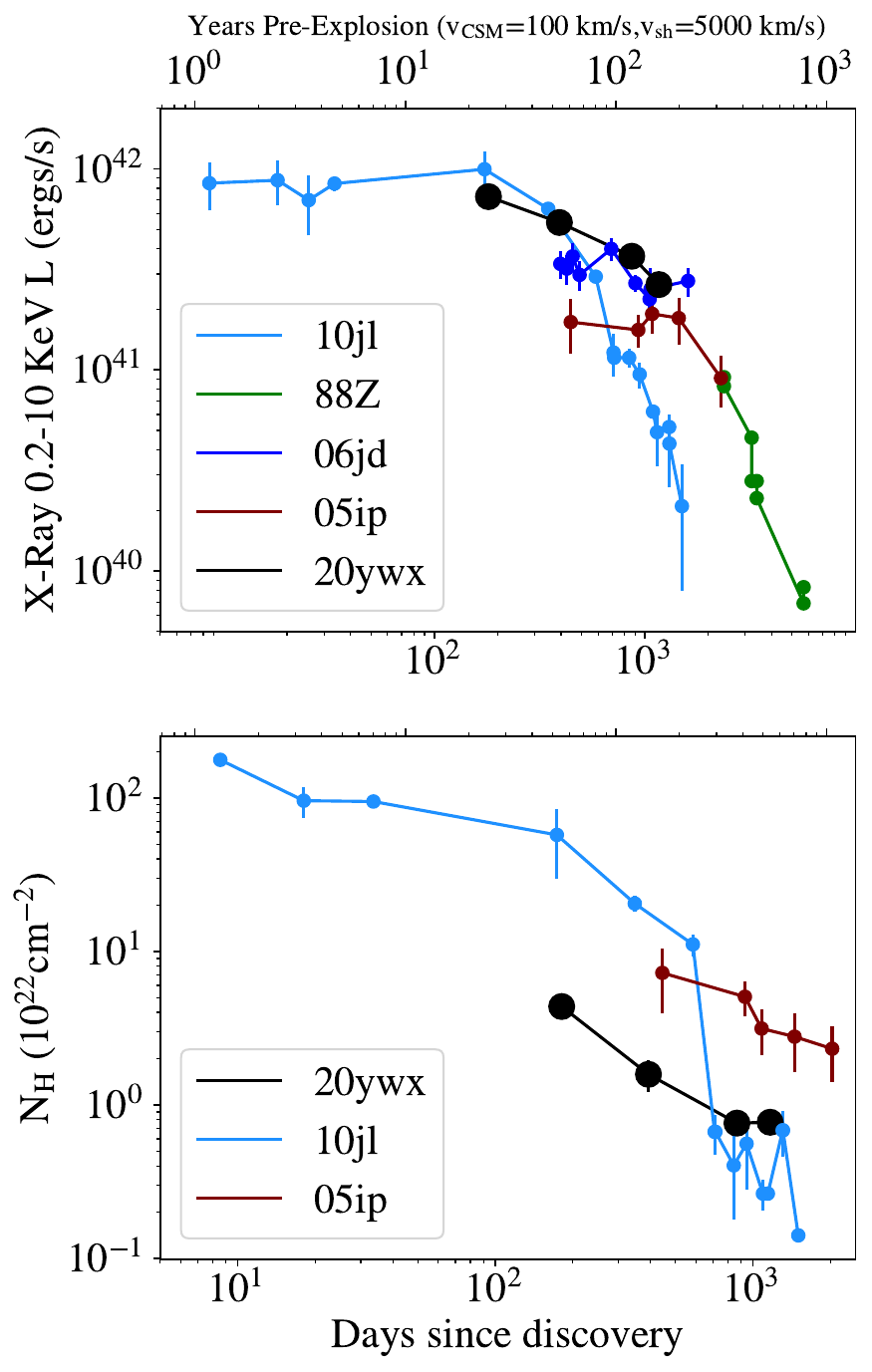}
    \caption{X-ray 0.2--10\,keV luminosity and column density comparison between SN 2020ywx and other SNe IIn. We see that 2020ywx is one of the most luminous IIn Sne of all time, surpassing other SNe at certain epochs. We show the pre-explosion mass-loss timescale corresponding to the data points based on a typical 5000\,km\,s$^{-1}$ shock and 100\,km\,s$^{-1}$ CSM speed. Notably, the plateau in the X-ray light curve of SN 2020ywx is similar to the early stages of the evolution in SN 2005ip and SN 2010jl. We note that varying amounts of the X-ray emission are captured in the early-time 0.2--10\,keV luminosity of these objects given the likely differing temperatures $> 10$\ keV. }
    \label{fig:Xray_L}
\end{figure}

In terms of the measured column density, we find a general decline followed by a similar measurement at the final two epochs (see Figure \ref{fig:Xray_L}). This is similar to the evolution seen in prototypical X-ray SNe IIn (SN 2006jd and SN 2010jl) \citep{Chandra_2015}, with the late-time plateau interesting to consider. The late-time plateau is at a column density ($7 \times 10^{21}\, \mathrm{\mathrm{cm^{-2}}}$) much higher than the expected galactic value ($1.6 \times 10^{20}\, \mathrm{\mathrm{cm^{-2}}}$) found from the IPAC database. We thus conclude that it is unlikely that this column density is from the host H\,I emission given the low column density measured at the site prior to the SN. It is possible that this elevated value of the column density is due to the reverse shock beginning to emerge through the dense shell, or this is simply the column density of the unshocked CSM. We examine this possibility after considering the evolution of the flux and constraining whether the shocks are adiabatic or radiative. Using the unabsorbed flux from our models, we constructed a luminosity curve and compared  with that of other prototypical SNe IIn. 
\par Despite the fact that given the high measured temperatures we are only sampling part of the X-ray spectrum of SN 2020ywx, it is evident from Figure \ref{fig:Xray_L} that SN 2020ywx is the second most luminous type IIn X-ray SN of all time. The best-fit power law to the light curve of SN 2020ywx gives a decaying exponential with a decline rate of $0.77_{-0.06}^{+0.06}$.
This can be broken down to understand the density structure of the CSM and ejecta. As derived by \citet{Frannson_96} and detailed by \citet{Dwarkadas_16}, given an ejecta density profile $\mathrm{\rho_{ej}}\approx v^{-n}t^{-3}$ and CSM density profile $\mathrm{\rho_{CSM}}\approx r^{-s}$ (we assume an unchanging CSM density profile --- which is confirmed from our results across wavelengths, in particular the already discussed constant optical light-curve decline), the X-ray luminosity if we capture all of it (which we attempt to do by simulating the 0.2--50\,keV flux) is
\begin{equation}
    L\propto t^{-\frac{12-7s+2ns-3n}{n-s}}.
\end{equation}
We set our fitted exponent equal to the exponent in the theoretical expression to constrain $s$ and $n$ \citep{Chevalier_17}. We fit for a range of $n=6$--12 as has been seen for many previous SNe \citep{Dwarkadas_16}. We find $s \approx  1.8$--1.9 for our initial range in $n$. We fix $s$ at 1.85, with $n \approx 6$ (taken with some input from the radio-derived $n$) suggesting a shallower ejecta/density  gradient.  It is possible that n $<$ 6 and thus $s \sim 2$ but given the results at radio wavelengths and the discrepancies in CSM densities we eventually measure, we find it more likely that s $<$ 2. We emphasize that this $s$ suggests non-constant mass-loss \citep{Chevalier_17}. 
\par
To calculate the pre-explosion mass-loss rate from here, we must first determine if the shocks are radiative or adiabatic. This can be done using the derived $s$ and $n$.  The reverse shock cooling time is given assuming Solar abundances, our $s=1.85$, and the measured forward-shock temperature $\sim 10^{8}\, \mathrm{K}$ as \citep{Dwarkadas_16}
\begin{equation}
t_{\rm cool,RS} = 3.5 \times 10^{9} \frac{(4-s)(3-s)^{4.34}}{(n-3)^{4.34+s}(n-4)(n-s)^{s}} V_{4}^{3.34+s} C_{*}^{-1} 
\end{equation}
\[
\times \left[\frac{t_{d}}{11.57}\right]^{s}
\]
where $V_{4}$ is the shock speed in $10^{4}$\,km\,s$^{-1}$ and $C_{*}$ is defined as the ratio $\dot{M}_{-5}/v_{w}$, where $\dot{M}_{-5}$ is the mass-loss rate of the progenitor in units of $10^{-5}\, \mathrm{M_{\odot}\,yr^{-1}}$ and $v_{w}$ is the wind speed in units of 10\,km\,s$^{-1}$. We use the derived wind speed of 120\,km\,s$^{-1}$ and the $C_{*} \approx 100$ derived from the radio analysis (independent of X-ray assumptions and detailed in the radio interpretations section) to calculate the cooling times. We use the radio mass-loss rate derived in the radio portion of \S~\ref{sec:Interpretations} as the radio emission is the most independent probe of the mass-loss rate \citep{Chevalier_17}. With all parameters defined as before and assuming an average 3500\,km\,s$^{-1}$ shock velocity, we find that the reverse shock is radiative for our n=6 for more than 1500 days, so it is radiative for all of the evolution thus far. We find the column density of the cool gas at the discontinuity between the forward and reverse shock (not in front of the forward shock) by using another equation slightly modified from \citet{Chevalier_17},
\begin{equation}
    N_{\rm cool}=1.0 \times 10^{21}\frac{n-3}{n-s}(n-4)C_{*}V_{4}^{1-s}\bigg{(}\frac{t}{100 \mathrm{\,days}}\bigg{)}^{1-s} \, \mathrm{\mathrm{cm^{-2}}},
\end{equation}
assuming spherical symmetry. Plugging values in, we get an initial estimate at our first X-ray epoch around $2.4 \times 10^{23}\, \mathrm{cm^{-2}}$, decreasing over time by the final epoch at 1250 days to $2.4 \times 10^{22}\, \mathrm{cm^{-2}}$. Thus, we verify that there is a dense shell between the shock that absorbs the reverse shock throughout the SN's evolution. Since we are seeing emission from the forward shock, it is not surprising that we do not find this high column density from our data, as what we are measuring in our data is the column density of the CSM, not the dense shell. 
\par
We now seek to probe whether the forward shock is radiative or adiabatic to obtain an X-ray measurement of the mass-loss rate. \citet{Chevalier_17} derive for the forward shock that the cooling time adapted for our parameters is 
\begin{equation}
    t_{\rm cool,FS}=\frac{3.5(n-s)^sC_{*}^{-1}}{(n-3)^2}V_{4}^3t_{D}^s \, \mathrm{\,days}
\end{equation}
where $t_{D}$ is the time since the explosion in days. Plugging in our radio values for the mass-loss rate and X-ray shock velocity as we did for the reverse shock as well as our $s$, we find that the shock will be cooling for $<$ 150 days for $n$=6 (and for any other realistic $n$), and thus that the forward shock is no longer radiative but adiabatic in our data as all X-ray data were taken at $> 200$ days.  
\par
Having confirmed that the X-ray emission is coming from an adiabatic forward shock, we now calculate the mass-loss rate this implies for the progenitor of SN 2020ywx. \citet{Frannson_96} derived that an adiabatic forward shock will have spectral luminosity at 1\,keV of 
\begin{equation}
\begin{split}
L_{\rm CS}(1\,\text{ keV}) &= \left(\frac{4.1 \times 10^{37}}{3-s}\right) T_{9}^{0.16} 
\times e^{-0.0116/T_{9}} \xi C_{*}^{2} V_{4}^{3-2s} \\
&\quad \times \left( \frac{t_{d}}{11.57} \right)^{3-2s},
\end{split}
\end{equation}
where $t_{d}$ is the time since explosion, $\xi$ is related to the chemical composition of the gas (around 0.86 for Solar abundance), and $T_{9}$ is the temperature of the shock (which we have constrained along with the shock velocity). We use $s=1.85$ and the X-ray 0.2--10\,keV flux at each epoch to obtain estimates of the luminosity at 1\,keV using PIMMS\footnote{https://cxc.harvard.edu/toolkit/pimms.jsp}. The result is $C_{*} = 210$ at the first X-ray epoch, suggesting a mass-loss rate of $(1.3 \pm 0.7) \times 10^{-2}\, \mathrm{M_{\odot}\, yr^{-1}}$ , where we have used equation 2  of \citep{Frannson_96} (assuming an initial radius $\sim$ $10^{15}$ cm as is usually taken for type IIn SNe) to adjust the rate to the fact that this first observation at 230 days is taken when the shock has already traversed a considerable distance $\sim 6\times 10^{15}$ cm if we take an average 4000 km/s shock velocity. Converting to density using the same equation 2 of \citep{Frannson_96}, this would mean a CSM density at the forward shock (using our s$\sim$ 1.85) of $2.0 
\times 10^{-16} \mathrm{g \hspace{0.2 cm} cm^{-3}}$. 
\par Uncertainties are estimated by propagating errors in the distance, time since explosion, luminosity, and shock speeds (assuming a shock-speed uncertainty of 20\% given the uncertainty in the temperature). This mass-loss rate does not change significantly over the following epochs even when adjusting for the evolution of the shock radius and hovers around 0.01\,$\mathrm{M_{\odot}\,yr^{-1}}$ throughout the evolution (see Figure \ref{fig:Final_ML}). This suggests that the star was losing mass at a near-constant rate with no uptick before the explosion, which is surprising for most potential progenitor mechanisms. 
\par
We can also attempt to constrain the mass-loss rate from the column density for comparison based on our assumption that the X-ray emission comes from the forward shock. The column density of the circumstellar gas in spherical symmetry can be calculated in general as \citep{Dwarkadas_16} 
\begin{equation}
    N({\rm H})_{\rm CS}=\frac{2.1 \times 10^{22}}{s-1}\,C_{*}V_{4}^{1-s}\bigg{(}\frac{t_{d}}{8.9}\bigg{)}^{1-s},
\end{equation} 
assuming $n(\rm He)/n(\rm H)=0.1$ with $V_{4}$ being the shock speed as usual. Using this equation and our measured column densities (Table \ref{table:xray_results}) we find $C_{*}$ values at the earliest X-Ray epoch $\sim$ 15. This is in disagreement with the mass-loss rate measured from the X-ray luminosity as well as what we calculate in the optical and radio later in this section. We attribute this to potential asymmetry effects as was seen in SN 2023ixf \citep{Chandra_23IXF} or line-of-sight effects affecting the column density measurements. 
\par 
Finally, we interpret the presence of the blended highly ionized Fe lines in our earlier spectra. 
We expect that the X-rays produced will ionize the medium to some extent. 
To derive a measure of the ionization parameter, we use from \citet{Dwarkadas_16}
\begin{equation}\chi=(2\times 10^{-38})\,L\xi^{-2}C_{*}^{-1}V_{4}^{s-2}\,(t_{d}/8.9 \hspace{0.1 cm} {\rm d})^{s-2}\end{equation}
This yields an ionization parameter at the first epoch of $\chi= 360$. However, this value is likely underestimated because we are probably still not capturing the full X-ray luminosity at the first epoch even with the 0.2--50\,keV simulation. Thus, the ionization parameter is probably $\sim 1000$ at early times (for at least the first 500 days), consistent with the values expected given strong ionized iron lines \citep{Dwarkadas_16}. This value slowly decreases as the luminosity falls given our $s \approx 1.85$.   It is also worth exploring the presence of what is likely the calcium K$\beta$ 4.0\,keV line in the second epoch. While this line is not seen in interacting SNe as ubiquitously as the ionized Fe line \citep{Chandra_2012}, it has been seen in certain SNe IIn \citep{Dwarkadas_16} and SN remnants \citep{Miceli_2015}, and it is not unexpected given the high temperature and relatively high ionization level.

\subsection{Optical/IR Interpretation}

\begin{figure}
    \centering
    \includegraphics[width=8 cm, height=6 cm]{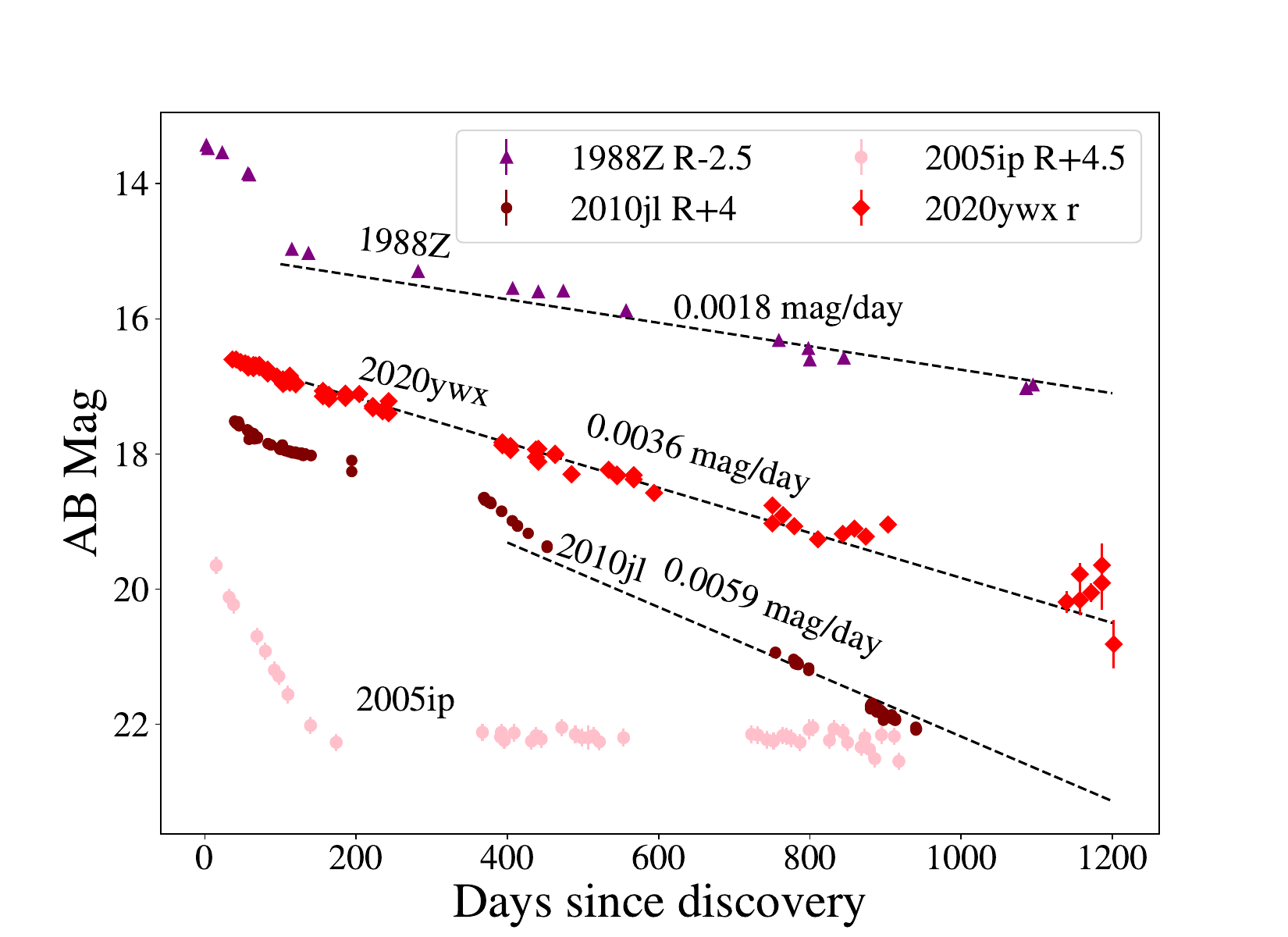}
    %add 2010jl errorbars
    \caption{A comparison of $R/r$-band optical light curves for SN 2020ywx and three prototypical SNe IIn, SNe 1988Z, 2005ip and 2010jl. We note that SN 2020ywx is constant in its decline. The SN 2010jl data are from \citet{Zhang_2010jl,Baerway2024}, the SN 2005ip data are from \citep{Smith_2005ip} and the SN 1988Z data are from \citet{Aretxaga_88z} and \citet{Turatto_88z}. SN 2005ip showed a remarkable plateau in it's lightcurve due to emission from a large number of coronal lines. We note that the SN 1988Z data are from earlier phases than the X-ray data shown in Figure \ref{fig:Xray_L}.}
    \label{fig:Comparison}
\end{figure}

Our optical data reveal a linear decline in the $gri$-band light curves and multicomponent line emission with a dramatic blueshift in the intermediate-width emission lines. The rate of the decline in $r$-band magnitude in particular (0.003\,mag\,day$^{-1}$) is extremely slow relative to radioactive decay (0.01\,mag\,day$^{-1}$), but also even relative to some other strongly interacting SNe IIn such as SN 2010jl \citep{Ofek_2010jl} with a decay rate closer to 0.006\,mag\,day$^{-1}$ at late times. SN 1988Z, however, still has SN 2020ywx beat at a 0.0018\,mag\,day$^{-1}$ decline during the $\sim 100$--1000 days post-explosion \citep{Turatto_88z,Aretxaga_88z}. SN 2010jl and SN 2006jd had distinct regimes in their light curves which revealed different components of the CSM --- shells of mass lost \citep{2006jd_stritzinger,Jencson_2016}; by contrast, SN 2020ywx seems to be undergoing a long phase of continued interaction with a highly dense CSM more similar to that of SN 1988Z or SN~2005ip \citep{Smith_2017}. We thus conclude based on the optical light curve that the mass was lost relatively steadily by the progenitor and there were no varying-density shells of CSM which would potentially produce bumps in the light curve (and would have caused bumps in the radio light curves as well). We compare the $r/R$-band decline rates of SNe 2010jl, 2005ip, 1988Z, and 2020ywx in Figure \ref{fig:Comparison}.
\par
We now interpret the spectral fitting results. First, we note that it is somewhat expected that the line profiles are best fit by Gaussians. The physical rationale for a Lorentzian or Voigt profile at late times is difficult to understand as the optical depth should decrease enough such that electron scattering is insignificant \citep{Smith_2016,smith10}. The Gaussian fits to the H$\alpha$ profiles in the spectra are insightful for understanding the nature and origin of the various lines. 
\par In the earlier spectra, the lines are best fit by a combination of broad, intermediate, and narrow (in velocity) components as seen in Figure \ref{fig:MCMC_fits}. This can be explained through freely expanding ejecta, shocked circumstellar gas, and unshocked photoionized CSM. The intermediate component traces the dense shell between the forward and reverse shocks, and the narrow component comes from the unshocked CSM \citep{Smith_2016}. The narrow H$\alpha$ emission is definitively contaminated by the H\,II region in most of our spectra. However, the fact  that there is narrow H$\alpha$ emission from the SN (i.e., we are not completely mislabeling the H\,II emission as narrow H$\alpha$) is undeniable due to both the presence of the P~Cygni profiles in the higher resolution spectra and the strong radio and X-ray emission which indicate there must be extensive CSM (shocked and unshocked) and thus should be accompanying narrow lines.

\par
The broad component of the emission is blueshifted and prominent at early times as was seen in the prototypical SN IIn 2006jd \citep{2006jd_stritzinger}, which is indicative of asymmetry in the CSM given that we do not necessarily expect to see ``through'' the photosphere to the freely expanding SN ejecta. This broad component is at $\sim 10,000$\,km\,s$^{-1}$ throughout the early evolution, and it fades at later epochs as one would expect if it originates from freely expanding ejecta. Another potential reason for this broad blueshift from the first spectral epoch (as seen in Figure \ref{fig:MCMC_fits}) is optically thick ejecta shrouding the red side of the SN at early times as was suspected in SN 2021foa \citep{Farias_2021foa}. These blueshifted broad lines at early times have been seen in other type II supernovae \citep{Reynolds}-although in other cases the lines are likely linked more directly with the interaction given they do not fade over time. Further evidence for asymmetry comes from the red bump in the intermediate profiles seen at early times (causing the initial intermediate profiles to be centered at redshifted velocities), suggesting a highly asymmetric CSM given that it seems that we are preferentially seeing the back side of the dense shell.
\begin{figure}[htbp]
    \centering
    \includegraphics[width=8 cm, height=6 cm]{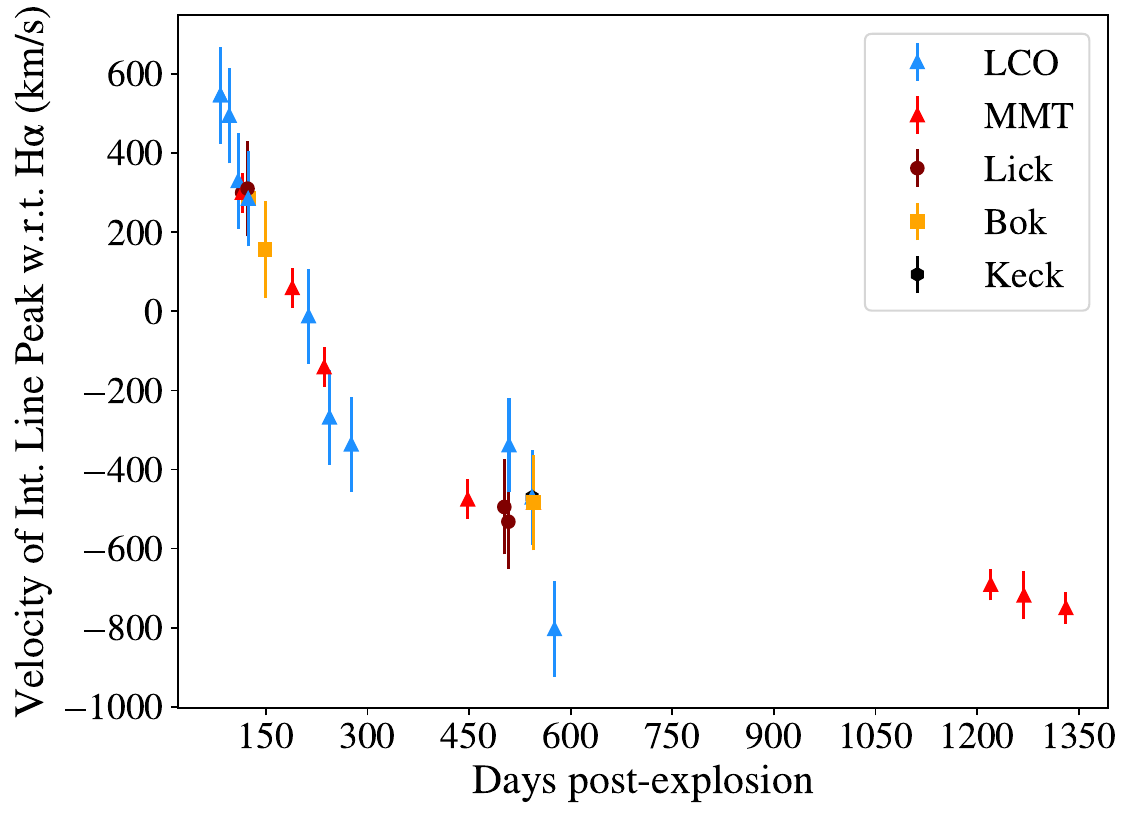}
    %flip marker on MMT points
    %use empty circle for lick
    \caption{Evolution of the central velocity of the intermediate component fit to SN 2020ywx's $\mathrm{H\alpha}$ profiles with points from different telescopes labeled. The uncertainties are added in quadrature from the fitting and the resolution of the telescope (detailed in Table \ref{table:optical_log}). The evolution suggests a strong blueshift which can be explained by invoking dust forming in the dense shell.}
    \label{fig:Blueshift}
\end{figure}

\par The intermediate component also decreases in velocity over time, which is unsurprising as the shock and dense shell decelerate. Additionally, the intermediate lines undergo an intriguing evolution from a redshifted central velocity to a blueshifted central velocity, changing in central velocity by $\sim 1500$\,km\,s$^{-1}$ across the SN evolution as shown in Figure \ref{fig:Blueshift}. For our purposes, we did not fix the broad component, so instead of displaying the red component ``eaten away'' we use the changing central velocity as proof of the blueshift.
\begin{figure}[h!]
    \centering
    \includegraphics[width=8 cm, height= 7cm]{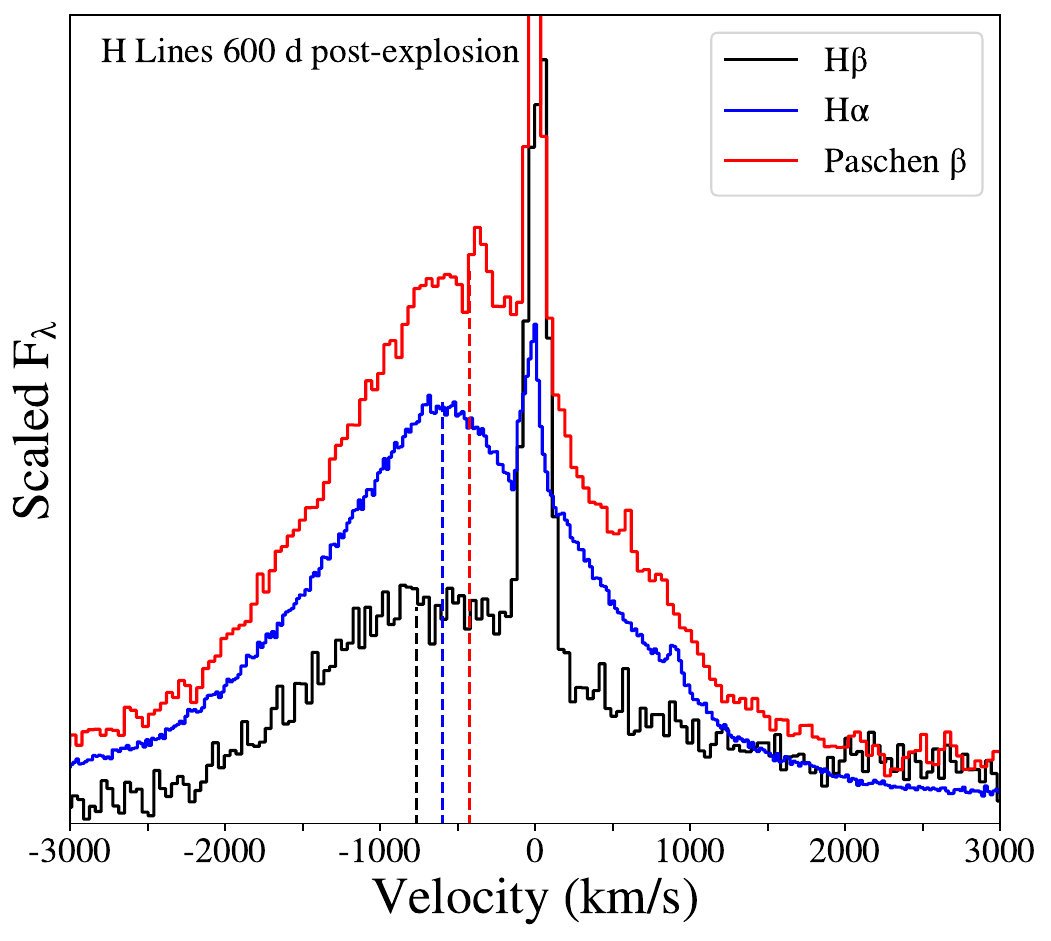}
    \caption{A comparison between optical and NIR hydrogen lines at $\sim 600$ days post-explosion. The dashed lines show the MCMC-fitted center velocity for each line. We note a more prominent blueshift in the bluest line, suggesting dust effects. We find that the flux ratio between these lines exceed the expected nebular case B values of P$\beta$/H$\beta$=0.16 and H$\alpha$/H$\beta$=2.86( see e.g. \cite{84_line_ratio}), suggesting dust effects. We do not plot this ratio due to different resolutions for different instruments not allowing direct comparison in velocity space.}
    \label{fig:H_comp}
\end{figure}
\par
The reason for the blueshift has been contested.
\citet{Fransson_2010jl} favored radiatively accelerated Lorentzians causing the blueshifted profiles, while \citet{Smith_2012},  \citet{Smith_2015da}, and \citet{Sarangi_2018} have argued from the theoretical and observational perspective for dust formation in the dense shell between shocks for prototypical SNe IIn, and also for SNe Ibn \citep{smith08,Mattila_2006jc}. Given the evolution of SN 2020ywx, this seems to be another case in favor of the  \citet{Sarangi_2018} ideas, given that the blueshift only occurs in the intermediate-width lines and continues to evolve over $> 1000$ days.
We also measured the intermediate central velocity over time in H$\beta$, and found a similarly strong blueshift (Figure \ref{fig:Halpha}), suggesting there is dust formation throughout the shell affecting lines both blue and red. We also plot a comparison between hydrogen lines from the optical to the NIR in Figure \ref{fig:H_comp} to show the more pronounced effect in the bluer H$\beta$, and a less pronounced effect in the infrared Pa $\beta$ line.  This wavelength dependence, with a more significant blueshift at shorter wavelengths, is expected for dust extinction and is inconsistent with electron scattering effects.  
\subsubsection{Optical Mass-Loss Rate Calculation}
\begin{figure}
    \centering  \includegraphics[width=8 cm, height =6 cm]{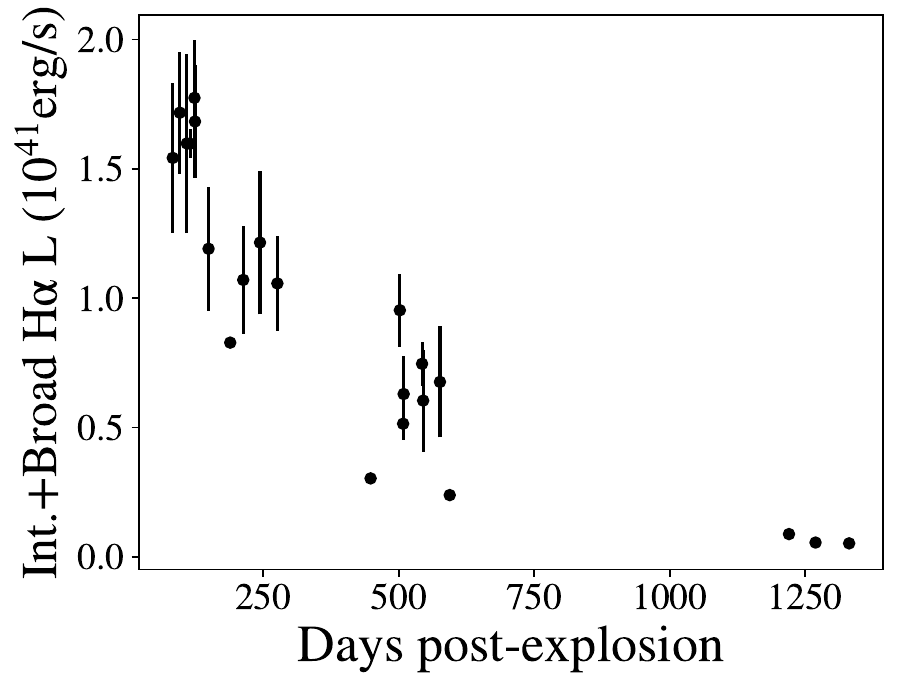}
    \caption{The evolution of the photometerically-calibrated H$\alpha$ luminosity from the broad and intermediate Gaussian components of our fits. These values are averaged at similar epochs and used to calculate the optically derived mass-loss rate. We notice a general decline with the luminosity peaking around $1.5 \times 10^{41}$\,erg\,s$^{-1}$. The error bars vary significantly owing to differing resolutions of different telescopes/instruments.}
    \label{fig:Ha_Lum}
\end{figure}

Now with all the relevant optical parameters calculated, we derive an optical estimate for the mass-loss rate.  Given that we lack sufficient UV/IR data (so we cannot truly constrain the bolometric luminosity), we use the H$\alpha$ luminosity as a proxy for the interaction-driven luminosity contribution for calculating the mass-loss rate. We use the H$\alpha$ luminosity from the intermediate and broad components (whenever the latter is present --- it is mostly weak compared to the intermediate component except at early times) owing to blending between these components in our early-time spectra.  As \citet{Chugai_1991} showed, if the luminosity of the SN is dominated by interaction, the luminosity in the H$\alpha$ line will be proportional to the kinetic energy dissipated per unit time across the shock front and thus the mass-loss rate (without assuming any density profile as we have constraints on the shock speed across time and do not have to assume a constant shock velocity). The mass-loss rate can then be generally written as
\begin{equation}
    \dot{M}=\frac{4 L_{{\rm H}\alpha}v_{\rm{w}}}{\epsilon v_{\rm{s}}^3}.
\end{equation}
The shock speed is calculated from the X-ray measurements given that the intermediate line width is both an underestimate of the shock speed and that the line components are likely blended in many of our earlier low-resolution spectra. We note that the ratio of X-ray shock speed to optical shock speed is $\sim$ 1.3-1.7 across all epochs, which would agree with the ratio of the shell radius to contact discontinuity radius found by \citep{Chevalier_1982} for our values n$\sim$ 6 and s$\sim$ 1.85.
Here $\epsilon$ is related to the efficiency of conversion of kinetic energy to H$\alpha$ luminosity, and we fix this at 0.1 for these first $\sim 4$\,yr post-explosion as is generally expected for a strong shock and dense CSM \citep{Taddia_2013}. We note that this efficiency is not well-constrained and is a significant source of uncertainty. Uncertainties are calculated through the propagation of the luminosity/shock speed errors. The luminosity values are taken from photometrically calibrated (using r band ZTF data) estimates from the Gaussian fits to our spectra when the optical data were coincident with our X-ray data and interpolated between neighboring measurements when data were not coincident. The uncertainties are larger in cases where interpolation was necessary, as we added the standard deviation between points in quadrature with the errors on the individual luminosities and the distance.  The evolution of the H$\alpha$ luminosity is shown in Figure \ref{fig:Ha_Lum}.
\par At 215 days post-explosion (at the first X-ray epoch), we calculate $\dot{M}=(1.1 \pm 0.4) \times 10^{-2}\,\mathrm{M_{\odot}\,yr^{-1}}$.  There is a general decline in mass-loss rate over the first 1300 days post-explosion to $(2.0 \pm 0.6) \times 10^{-3}\, \mathrm{M_{\odot}\,yr^{-1}}$. 
We thus find a non-constant, non-$r^{-2}$ wind density profile for the mass loss; see Figure \ref{fig:Final_ML} for the full evolution. Using the relative CSM (120\,km\,s$^{-1}$) and shock ($\sim 4000$\,km\,s$^{-1}$) speeds, we find that the mass loss must have occurred at these elevated rates for at least 100\,yr pre-explosion. This duration was determined using 
$t_{\rm ML}=(v_{\rm sh}/v_{\rm CSM})t$, where $t_{\rm ML}$ refers to the amount of time over which mass was lost.  We thus determine based on integrating the optical and X-ray mass-loss results over time that there must have been at least $> 1\,{\rm M}_{\odot}$ of mass ejected (given the duration of mass loss) by the progenitor pre-explosion. This is a lower limit considering it is extremely likely (based on the unchanging light curve decline) that the SN will remain luminous in the optical for years to come and thus imply many $\mathrm{M_{\odot}}$ of CSM.
\subsubsection{Confirmation of Dust from the NIR}
Our NIR spectra are well-fit by blackbodies which decrease in temperature over time as seen in Figure \ref{fig:IR_tot}.   When combined with the optical blueshift seen in Figure \ref{fig:Blueshift} this is definitive evidence for dust in the picture.
The blueshift in the optical lines from the shell between the shocks confirms that dust must have formed after the explosion since the location of dust causing the blueshift is internal to the forward shock. This newly formed dust is either in the ejecta or in a dense shell in the post-shock gas. \cite{Arka_2022} argue that in interacting SNe, dust is likely to form in the post-shock gas. Moreover, the blueshift of the optical line profiles is the most prominent if the dust is present close to the line-forming region, in this case the dense shell.
The dust is in all likelihood forming in the dense shell where dust formation can happen quite efficiently and at early times (as the blueshift is notable as early as 230 days post-explosion) \citep{Sarangi_2018}. \cite{Arka_2022} found that dust formed behind the reverse shock (i.e., not in the ejecta) is much more likely to survive and have a notable effect on the SN line emission out to late times. They also modeled the dust mass as a function of mass-loss rate given some total ejecta mass. As we do not have a good constraint on the ejecta mass, we note simply that given the scale of mass loss around 500 days post-explosion ($\sim 10^{-2}\, \mathrm{M_{\odot}}$\,yr), we find that the amount of dust formed would be around $10^{-3}\, \mathrm{M_{\odot}}$ regardless of ejecta mass based on their work. 
\par The temperature of the NIR emission at 1000\,K does not allow us to constrain the origin of the dust grains (in terms of O-rich vs. Si-rich vs. C-rich grains), but we conclude that there is definitive evidence for dust formation. Further probing into the mid-IR and out to later times would be vital to provide more constraints and compare with the extensive modeling being done. Further monitoring could also provide information on the development of the CO overtone lines associated with dust formation which we are unable to resolve in our spectra.

\subsection{Radio Interpretation}
Having outlined the radio fitting and analysis process in the previous section, we now interpret our radio results to obtain a third measure of the mass-loss rate. The fact that the radio emission is best described by internal FFA is expected considering the high density of the CSM measured from the X-ray and optical observations. 
\par While internal FFA is the best fit, the model parameters do not match the expected hydrodynamic evolution (i.e., as was seen in SN 1993J; \citealt{Frannson_96}) in which the post-shock energy density is proportional to the magnetic field density as detailed by \citet{Chevalier_1982}. This hydrodynamic model suggests that $\beta=3+\alpha-3m$. However, this is not possible for our best-fit values given that our best-fit $\beta=0.66$ and $\alpha \approx 1$; $m>1$ is the only way to satisfy this condition, suggesting a nonphysical acceleration of the forward shock. However, \citet{Chevalier_1998} developed another model (his model 4) in which the flux of electrons is proportional to the flux of particles in the shock front. In this parametrization, we have $\beta=2\alpha m/2$.  Given our $\alpha$ and $\beta$, we derive $m=0.68$. This value of $m$ suggests $n \approx 6$ if $s \approx 2$ (which it is likely not based on the overall non-constant mass loss --- but the X-ray results suggest a shallower $s$ which would only imply a shallower $n$). This is interesting to note given that it suggests a slightly less dramatic evolution of the shock than seen in other radio SNe, but it is not surprising for SNe IIn with dense CSM \citep{Weiler_1990}. 
\par The $\alpha$ value ($\approx$ 1) measured from the fits suggests a relatively soft nonthermal electron index. In terms of $\delta'$, the secondary time exponent in the internal FFA model, \citet{Weiler_1990} showed that $\delta' = 5m$ given the internal absorbing gas. However, we let this be a free parameter as we consider other cases.  The $\delta'$ value found does not line up with the preferred hydrodynamic model, but we attribute this to variations in the medium or additional complexities in the microparameters. 
\par 
While radio SNe have been notorious for their differences, the early-time radio evolution of SN 2020ywx is consistent with that seen for prototypical radio SNe IIn such as SNe 1988Z and 1986J \citep{Weiler_1990, VanDyk_1988Z}. The 5\,GHz peak time of SN 2020ywx ($\sim 900$ days) and peak 5\,GHz luminosity ($3 \times 10^{28}\, \mathrm{erg \hspace{0.1 cm}\,cm^{-2}\,s^{-1}Hz^{-1}}$) line up almost precisely with those two objects. SNe 1988Z and 1986J were both modeled with a combined internal+external FFA fit. SN 1986J had an increase in integrated flux at low frequencies due to an unexplained central component\citep{Bietenholtz_2017}, which has not been seen thus far in SN 2020ywx. 1986J also showed clear evidence for asymmetry in the CSM given the shell hotspot seen in VLBI data. Full VLBI monitoring of SN 1993J has also revealed that the Rayleigh-Taylor instability likely amplifies the magnetic field and that the brightness distribution peaks in the region near the contact discontinuity \citep{Vidal_2024}, suggesting there may be high-density regions in the emission that contribute to the ``internal" FFA. These effects could well be occurring for SN 2020ywx, but without VLBI/higher-resolution radio data, it is not possible to disentangle emission components. 
\par
\begin{figure*}
    \centering
    \includegraphics[width=10 cm, height= 8cm]{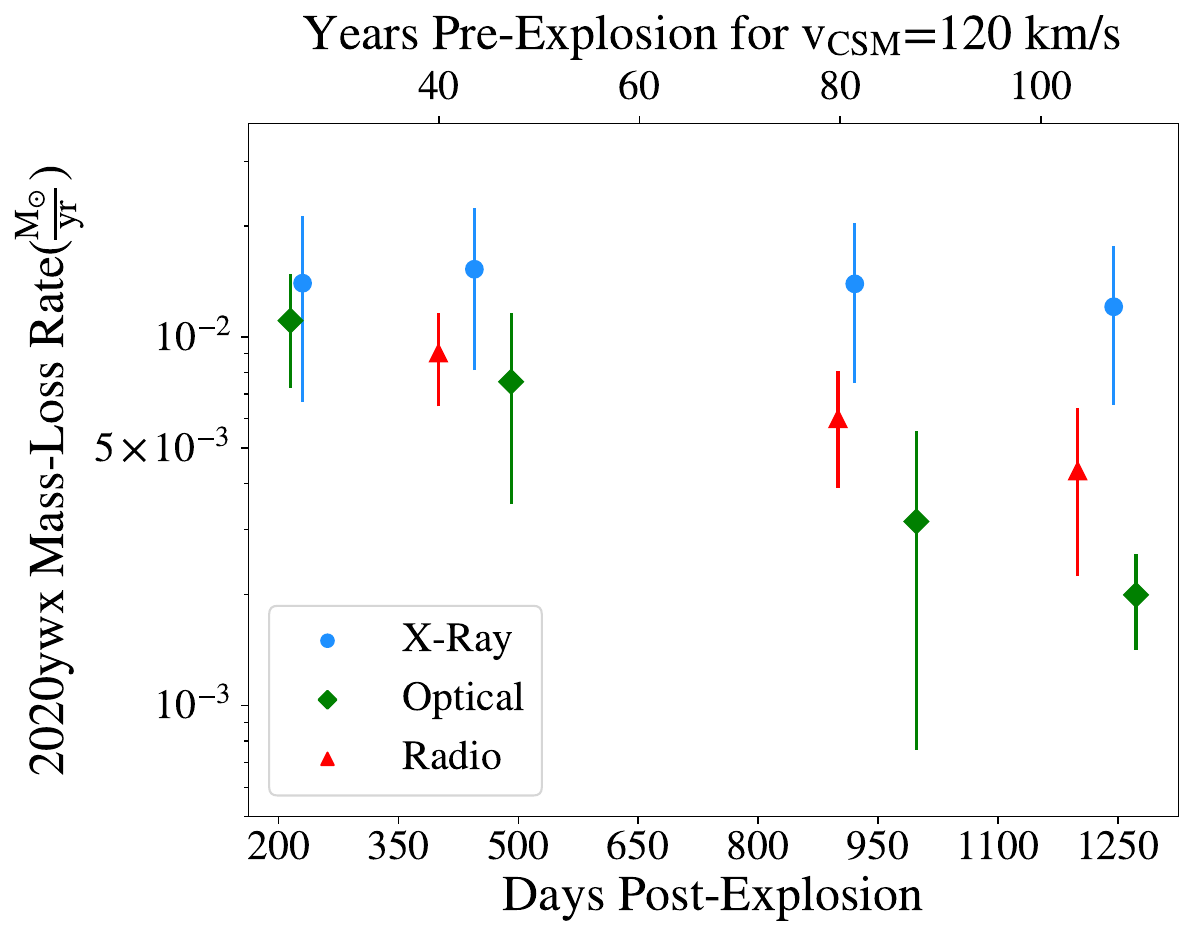}
    \caption{Mass-loss rate constrained across wavelengths at epochs for which we have data in each band. The mass loss is consistently high at $\sim 0.01\,\mathrm{M_{\odot}\,yr^{-1}}$ and persists near this rate for close to 100\,yr (with this time frame measured through the relative speed of the shock to the CSM) at each wavelength. We note the discrepancy at late times between the X-rays and the radio/optical results, which we ascribe to asymmetries in the CSM.}
    \label{fig:Final_ML}
\end{figure*}

To find the radio mass-loss rate, we consider the implications of our internal FFA modeling.  \citet{Weiler_1986} found an expression for the mass-loss rate which we modify for our case considering the early ejecta velocities (found from our earliest optical spectra at 83 days),
\begin{equation}
\begin{split}
    \dot{M} = (2.40 \times 10^{-5})\,\tau_{5}^{0.5}\bigg{(}\frac{v_{\rm{w}}}{10\,\rm{km\,s^{-1}}}\bigg{)}
    \bigg{(}\frac{t_{d}}{83\,{\rm d}}\bigg{)} \\
    \times \bigg{(}\frac{T_{e}}{10^4\,{\rm K}}\bigg{)}^{0.68}\, \rm{M_{\odot}\,yr^{-1}},
\end{split}
\end{equation}
where $T_{e}$ is the electron temperature in the CSM assuming equipartition between ions and electrons, and $\tau_{5}$ is the 5\,GHz optical depth as found from the internal FFA model. We note that given a fit of internal+external FFA, we find $\dot{M}$ at values $\sim 1.3$ times higher if we take this optical depth. We adopt the internal FFA values given the better fit. The actual value of the CSM temperature $T_{e}$ has been debated, and the contention comes down to assumptions about the ionization state of the wind/CSM, whether the wind is preionized by the progenitor or ionized by the X-rays themselves. In our case, given the unequivocally high measured mass loss, we assume that the CSM is not pre-ionized and thus that the initial CSM temperature must be higher at $10^{5}$\,K. At later epochs, we assume that the wind temperature decreases to around $10^{4}$\,K.  We ran a series of quick CLOUDY \citep{Cloudy_2013} steady-state models and found that this is a reasonable assumption given the CSM conditions derived in the optial/X-rays. See \citet{Lundqvist_1991,Lundqvist_88,Lundqvist_2013} and \citet{Frannson_96} for details on these considerations. At the initial radio epoch at 400 days, we derive a mass-loss rate of $(8.9 \pm 2.5) \times 10^{-3}\, \mathrm{M_{\odot}\,yr^{-1}}$. Under the assumption that the electron temperature decreases linearly over time, the mass-loss rate was gradually increasing in the decades before the explosion and was at $(4.3 \pm 2.0) \times 10^{-3} \, \mathrm{M_{\odot}\,yr^{-1}}$ in the final epoch of radio observations which trace 100\,yr before the explosion. This suggests a nonconstant $s \neq 2$ evolution of the density profile, similar to the optical and X-ray results, although the decline differs from the X-ray results. The full radio mass-loss results are shown in Figure \ref{fig:Final_ML}.

\section{Discussion}\label{sec:Disc}
The three wavelengths of data across epochs paint a consistent picture for SN 2020ywx: sustained mass loss around $10^{-3}$--$10^{-2}$\,M$_{\odot}$\,yr$^{-1}$ over at least 100\, yrs pre-explosion. However, there are differences across wavelengths, particularly between the X-rays and other wavelengths, which point to asymmetries. In more detail, the results support an asymmetric CSM with extensive CSM/clumps along certain regions around the star and lesser CSM density in other areas. We may be receiving X-ray emission from certain denser inner regions with radio emission from outer faster-moving regions. We note that given the shock and wind speed and average mass-loss rate measured, we are in a regime when our data were taken when the wind optical depth $\tau_{w} \approx 1$ \citep{Chevalier_2012}, and thus in a regime where the X-ray emission should be maximized and not suffer heavily from photoabsorption and other cooling effects. As we lack VLBI data, we cannot confirm this picture for the radio emission coming from faster-moving less-dense material, but this is a scenario which was seen in SN 2014C with similar elevated X-ray luminosity \citep{Brethauer_2022}.
\par This conclusion is supported by the X-ray plateau in the luminosity and hence mass-loss rate. While the X-ray-derived mass-loss rate stays relatively constant as the emission is theoretically stemming from some clumpier and denser region, the radio and optically derived rates decrease over time.  The assumption of a decreasing $10^{5-4}$\,K electron temperature in the radio analysis does cause the radio mass-loss evolution to be steeper than assuming a constant temperature. Still, even this assumption does not account for the entire discrepancy as this does not explain why the discrepancy increases over time. Furthermore, in the optical, the efficiency of conversion to H$\alpha$ luminosity is not well constrained. However, even correcting by a factor of 1.5 would not resolve the magnitude or evolution discrepancy. These differences across wavelengths could also be due to the X-ray emission coming from both the forward and reverse shocks at later times (and thus our assumption of reverse shock absorption may be slightly incorrect), and thus the measured mass-loss rate may be slightly overestimated.
\par More detailed modeling beyond the scope of this paper is needed to model clumps/asymmetries to fully understand the mass-loss rate. Regardless of the specific reason, this discrepancy points to a nonuniform CSM. Distinguishing between clumps and other kinds of asymmetry is difficult based on the current set of models and the lack of full coverage, particularly at early epochs for this SN.
\par 
\par 
SN 2020ywx in general most closely resembles the prototypical class of SNe IIn which includes SNe 1986J, 1988Z, 2006jd, and 2010jl across wavelengths. With this in mind, we now consider various progenitor possibilities for SN 2020ywx.  Given that the SN has only been around for $\sim 4$\,yrs, the deduced amount of hydrogen-rich mass ejected ($> 1\,{\rm M}_{\odot}$) is not enough to constrain the progenitor in a meaningful way. 
\par 
It would not be surprising, based on the evolution so far, if this SN were to remain bright for years to come and thus suggest a very high-mass progenitor. The mass-loss rate derived for SN 2020ywx, as was the case for SN 2010jl and many other SNe IIn, defies the conventional values found for wind-driven steady mass loss which reaches its limits around $10^{-4}\, \mathrm{M_{\odot}\,yr^{-1}}$ \citep{Smith_2014}.
\par We consider the possibility of eruptive mass loss from an LBV-like star or cool supergiant. For an LBV-like star, the eruptive event is expected to last a few years to a decade as seen observationally  \citep[e.g.,][]{Smith11lbv}, while for a cool hypergiant, it can last for hundreds of years \citep{Smith_2014}. The evolution of the optical light curve in particular does not lend credence to the idea of repeated outbursts from a single star given that if there were shells with varying densities, there would be less of a constant decline in the light curve. We note, however, that LBVs can eject bipolar/aspherical shells with a wide range of speeds and CSM at a large range of radii, even from a single event. A hypergiant could sustain high mass loss for more extended periods \citep{Smith_2014}. However, the derived CSM speed  ($\sim 100$\,km\,s$^{-1}$) exceeds the boundaries of the expected wind speed for these red supergiants or yellow hypergiants \citep{Goldman_2017}. Thus, yellow hypergiant mass loss seems unlikely as the explanation for the progenitor of SN 2020ywx. Other proposed channels include wave-driven mass loss and burning instabilities \citep{Quataert_2012,sa:14}  as well as enhanced red supergiant mass loss \citep{Yoon_Cantiello}. However, the timescales for wave driving are too short (on the scale of around a year) to explain the long-lasting ($> 100$\,yr) mass loss in SN 2020ywx. For other nuclear-burning instabilities, the presence of strong Hydrogen in nearby CSM disfavors this hypothesis. For enhanced red supergiant winds, the measured wind speed is still much too high. 
\par 
We thus consider the possibility of binary-driven mass loss as perhaps the only mechanism that can explain the quantity and duration of mass loss derived across all wavelengths \citep{sa:14}. First, the non-uniformity ($s \neq 2$) suggested by our mass-loss measurements and the presence of broad ejecta from early times in the spectra suggests a binary progenitor mechanism, as single stars would not generally be expected to lose mass in such a non-geometrically-uniform way %\citep{Woosley_1993} 
(although there may be exceptions for rapid rotators). In other words, there must be some pockets of low optical depth that allow us to see ``through'' the highly dense CSM which are difficult to explain under single-star mass loss. It has become clear that binarity has a far more important role to play than previously thought for massive stars \citep{Sana_2012}, and binarity leading to a merger (which is also plausible here) has been suggested as a possible progenitor mechanism for other SNe IIn such as SN 2015da \citep{Smith_2015da}, SN 2014C \citep{Thomas_2014C}, or SN 2001em given common-envelope evolution \citep{Chandra_2020}. In this context, it is worth remembering that LBV eruptions, often invoked to account for the CSM of SNe IIn, may themselves be a binary interaction phenomenon.  The best example is the massive LBV star $\eta$ Carinae, for which a binary merger is the leading explanation for its eruption \citep{smith18}. Binary evolution in the form of Roche-lobe overflow could explain the very dense H-rich CSM formed while also accounting for the 100\,km\,s$^{-1}$ speed at which the material was lost and the timescale.
\par Distinguishing between binary models is difficult without the full picture of the SN evolution at early or later times. The lack of deviations from the standard model in the radio evolution prevents a definitive confirmation of the binary.

\section{Conclusions}\label{sec:Conclusion}
%add portion about low frequency GMRT points
Through an exploration of the radio, NIR, optical, and X-ray emission of the Type IIn SN 2020ywx, we obtained three independent estimates of the mass-loss rate for the progenitor of SN 2020ywx. This was made possible by combining high-resolution spectral measurements in the optical with data in the radio and X-rays. All three values were found to be relatively consistent and suggestive of extreme mass loss (at $\sim 10^{-2}\,{\rm M}_{\odot}$\,yr$^{-1}$) for at least a century preceding the explosion. However, the differences in mass-loss calculated across wavelengths suggest asymmetries in the CSM. In more detail, this work has provided a variety of insights regarding SN 2020ywx, as follows:
\begin{enumerate}
    \item SN 2020ywx is the second most luminous X-ray IIn SN ever discovered.  We extrapolated an initial temperature of the forward shock of 20\,keV using results from the late-time X-ray data and use this number to deduce that SN 2020ywx had a forward shock speed $\sim 4000$\,km\,s$^{-1}$ which declined over time.  
    \item From the X-ray luminosity evolution, we calculate values for the mass-loss rate across time. There is a plateau in the X-ray mass-loss rate, which does not match results at other wavelengths. We thus suggest that the X-ray emission originates in a dense region of CSM. 
    \item At optical wavelengths, we find multi-component line emission in hydrogen and helium suggestive of three components of the system: the ejecta, shocked material, and unshocked CSM.
    \item The FWHM of the line components originating from the shock and ejecta decline over time as expected for a decelerating shock and ejecta. The intermediate-width components in all emission lines experience a dramatic blueshift, shifting bluewards by $>$ 1000 km/s, suggesting dust effects. 
    \item The high-resolution optical spectra show P~Cygni profiles in the narrow lines, which directly constrain the CSM speed at $120 \pm 22$\,km\,s$^{-1}$.
    \item Using the H$\alpha$ luminosity, we derive a declining optical mass-loss rate over 100\,years pre-explosion for the progenitor of SN 2020ywx. This disagrees with the X-ray results while also implying the CSM density profile exponent $s \neq 2$,  suggesting asymmetries.
    \item The NIR spectra show additional emission lines of hydrogen and helium. The continuum is well fit by a 1000\,K blackbody which is indicative of formation of new dust and its evolution.
    \item The optical $gri$-band light curve has been declining at a nearly constant rate for $> 1000$ days, suggesting a continuous (not shell-like) CSM. The radio and X-ray light curves also indicate a continuous CSM.
    \item  At radio wavelengths, we find synchrotron emission attenuated by free-free absorption from the emitting electrons (internal FFA due to mixing of cool gas in the synchrotron-emitting region).
    \item The derived radio optical depth shows a mass-loss rate declining over time, assuming the expected decrease in electron temperature. This again suggests $s \neq 2$ and confirms asymmetric mass loss. 
\end{enumerate}    We explore different progenitor channels and conclude that binary interaction is the most plausible channel that can explain the overall mass-loss picture of SN 2020ywx. This is another case that adds to the consensus that SN IIn progenitors are diverse and complex. We emphasize the importance of obtaining both low-frequency radio data, and high-resolution optical/X-ray spectra for IIn SNe to fully constrain the mass-loss history. We also emphasize the need for 3D modeling of the wide variety of potential binary interactions that are likely leading to SNe IIn. 

All optical and NIR photometry and spectra will be made public via the Weizmann
Interactive Supernova Data REPository (WISeREP) \citep{Wiserep}.
The code used to perform all of the MCMC fits in this work is provided for the case of optical emission line spectra on github here: \href{https://github.com/18rway/Emission_fitting_MCMC}{Emission Fitting-github}.
Continued observational monitoring of SN 2020ywx would be beneficial to further explore various channels of IIn formation, including the potential presence of a binary companion. We also emphasize that the dust as yet is relatively unconstrained and further high-resolution observational monitoring particularly in the mid-infrared would provide great insight.

%% Also note that the akcnowlodgment environment does not support long amounts of text. If you have a lot of people and institutions to acknowledge, do not use this command. Instead, create a new \section{Acknowledgments}.
\section{acknowledgments}

R.B.-W. and P.C. acknowledge NASA award GO3-24056X. R.B.-W. acknowledges support from a VSGC Fellowship. M.M. and the METAL group at UVa acknowledge support in part from ADAP program grant  80NSSC22K0486, from NSF grant AST-2206657, and from {\it HST} program GO-16656. 
Time-domain research by the University of Arizona team and D.J.S. is supported by NSF grants AST-2108032, AST-2308181, AST-2407566, and AST-2432036, and by the Heising-Simons Foundation under grant \#2020-1864. Observations reported here were obtained at the MMT Observatory, a joint facility of the University of Arizona and the Smithsonian Institution. K.M. acknowledges
support from JSPS KAKENHI grants JP24KK0070, JP24H01810, and JP20H00174.
A.V.F.’s research group at UC Berkeley acknowledges financial assistance from the Christopher R. Redlich  
Fund, Gary and Cynthia Bengier, Clark and Sharon Winslow, Alan Eustace (W.Z. is a Bengier-Winslow-Eustace Specialist in Astronomy), William Draper, Timothy and Melissa Draper, Briggs and Kathleen Wood, Sanford Robertson (T.G.B. is a Draper-Wood-Robertson Specialist in Astronomy), and numerous other donors.   

The Very Large Array is operated by the National Radio Astronomy Observatory, a facility of the U.S. National Science Foundation (NSF) operated under cooperative agreement by Associated Universities, Inc. 
%We thank the staff of the GMRT that made the GMRT observations possible. 
GMRT is run by the National Centre for Radio Astrophysics of the Tata Institute of Fundamental Research. 

This work contains observations from the Las Cumbres Observatory (LCO) Global Telescope Network; the LCO group is supported by NSF grants AST-1911225 and AST-1911151.  It also contains
observations obtained with the Samuel Oschin 48-
inch Telescope at the Palomar Observatory as part of
the Zwicky Transient Facility project. ZTF is supported
by NSF grants 
AST-1440341 and AST-2034437, and a collaboration including Caltech, IPAC, the Weizmann Institute for Science, the Oskar Klein Center at Stockholm
University, the University of Maryland, the University
of Washington, Deutsches Elektronen-Synchrotron and
Humboldt University, Los Alamos National Laboratories, the TANGO Consortium of Taiwan, the University
of Wisconsin at Milwaukee, and Lawrence Berkeley National Laboratories, Trinity College Dublin, and IN2P3,
France. Operations are conducted by COO, IPAC, and
UW. 

Some of the data presented herein were obtained at
the W. M. Keck Observatory, which is operated as a scientific
partnership among the California Institute of Technology, the
University of California, and NASA. The Observatory was
made possible by the generous financial support of the W. M.
Keck Foundation. The authors wish to recognize and acknowledge
the very significant cultural role and reverence that the
summit of Maunakea has always had within the indigenous
Hawaiian community. We are most fortunate to have the
opportunity to conduct observations from this mountain. 
A major upgrade of the Kast spectrograph on the        Shane 3\,m telescope at Lick Observatory, led by Brad 
Holden, was made possible through generous gifts from 
the Heising-Simons Foundation, William and Marina  Kast, and the University of California Observatories.  
Research at Lick Observatory is partially supported by a generous gift from Google.  

We thank the staffs of the various observatories used
to obtain data for their expert assistance.

\facilities{GMRT, VLA, LBT, MMT, Lick Observatory, Keck, LCO, Chandra, Swift (XRT)}
\\
\textit{Software:} \texttt{Astropy} \citep{Astropy_2013,Astropy_2022,Astropy_2018},CASA \citep{Casa_desc},\texttt{xspec}\citep{xspec},CIAO\citep{Ciao_2006}
\appendix
\setcounter{table}{0}
\renewcommand{\thetable}{A\arabic{table}}

We provide tables of all data taken in the radio and X-rays and all optical+IR spectra. We also list some of the optical photometry of SN 2020ywx.

\begin{deluxetable*}{cccccc}[h!]
\tablecaption{X-Ray Observations Log \label{table:xray_log}}
\tablehead{
\colhead{Date of Observation} & 
\colhead{Mission} & 
\colhead{SN Age (days)} & 
\colhead{Instrument} & 
\colhead{Obs ID} & 
\colhead{Exp Time (ks)}
} 
\startdata
2021 March 17.58 & \textit{Swift} & 185  & XRT    & 00014168001 &  4.53 \\
2021 March 25.62 & \textit{Swift} & 193  & XRT    & 00014168002 &  4.37 \\
2021 March 26.02 & \textit{Swift} & 194  & XRT    & 00014168003 &  2.74 \\
2021 March 29.47 & \textit{Swift} & 198  & XRT    & 00014168004 &  1.34 \\
2021 May 2.48   & \textit{Chandra} & 231 & ACIS-S & 23581       & 40.00 \\
2021 December 2.32 & \textit{Chandra} & 445 & ACIS-S & 23582  & 40.00 \\
2023 March 20.99 & \textit{Chandra} & 921 & ACIS-S & 26660       & 15.00 \\
2023 March 23.90 & \textit{Chandra} & 921 & ACIS-S & 27755       & 15.00 \\
2024 January 15.35 & \textit{Chandra} & 1219 & ACIS-S & 26661    & 50.00 \\
\enddata
\end{deluxetable*}
\begin{deluxetable*}{ccccccc}

\tablecaption{ Optical+IR Spectra Log \label{table:optical_log}}

\tablehead{\colhead{Date of Observations}& \colhead{Faclity/Inst} & \colhead{Exp. Time (s)}& \colhead{Slit Width ($''$)}& \colhead{Resolution (\AA)}& \colhead{SN Age (days)}  & \colhead{Wavelength Range (\AA)} } 

\startdata
2020-12-04.58 & LCO/FLOYDS-N &2700 & 2 &10-11 &83 & 3500-10000 \\
2020-12-17.58 & LCO/FLOYDS-N &2700 & 2& 8-9&96 & 3500-10000 \\
2020-12-30.63 & LCO/FLOYDS-N &2700 & 2 &14-15 &109 & 3500-10000 \\
2021-01-06.42 & MMT/BlueChannel &1200& 1 & 1-2&116 & 5598-6867 \\
2021-01-13 & Lick/Kast & 2400 & 2 & 7-8&123 & 3622-10712 \\
2021-01-14.48 & LCO/FLOYDS-N &2700 & 2 &8-9 &124 & 3500-10000 \\
2021-02-08.42 & Bok/B$\&$C spec &3600 & 1.5 & 8-9&149 & 4000-8000 \\
2021-03-21.89 & MMT/Binospec &3600 & 1.66 & 1-2&189 & 5688-7209 \\
2021-03-23.44 & LCO/FLOYDS-N &2700 & 2 & 11-12&191 & 3500-10000 \\
2021-04-14.23 & LCO/FLOYDS-N &2700 & 2 &10-11 &213 & 3500-10000 \\
2021-05-07.13 & MMT/Binospec &1800& 1.66 & 1-2&236 & 5688-7209 \\
2021-05-15.41 & LCO/FLOYDS-N &3600 & 2 &13-14 &244 & 3500-10000 \\
2021-06-14.39 & LCO/FLOYDS-S &3600 & 2 & 8-9&276 & 3500-10000 \\
2021-12-05.53 & MMT/BlueChannel &1200& 1 & 1-2&448 & 5598-6867 \\
2022-01-27 & Lick/Kast & 3600 & 2 & 6-7&505 & 3622-10766 \\
2022-02-02 & Lick/Kast & 3600 & 2 & 6-7&508 & 3622-10680 \\
2022-02-03.42 & LCO/FLOYDS-N &3600 & 2 & 11-12&509 & 3500-10000 \\
2022-03-10.30 & Bok/B$\&$C spec &1800 & 1.5 & 6-7&544 & 3936-7934 \\
2022-03-11.57 & LCO/FLOYDS-S &3600 & 2 &17-18 &545 & 3500-10000 \\
2022-04-11.56 & LCO/FLOYDS-S &3600 & 2 &17-18 &576 & 3500-10000 \\
2022-04-29 & Keck/LRIS &1200 & 1 &2-3 &594 & 3100-6890 \\
2022-05-05.27 & Bok/B$\&$C spec &2700 & 1.5 &2-3 &599 & 4870-6020 \\
2022-05-14.36& Keck/NIRES &3600&0.55& 2-3&608 & 9657-24669 \\
2024-01-14.47 & MMT/Binospec &3600& 1.66 &2-3 &1220 & 3824-9197 \\
2024-02-06.26& Magellan FIRE&126.8&0.6 &11-12&1240 &7678-25596 \\
2024-03-05.35 & MMT/Binospec &4800& 1.66 &1-2 &1269 & 5255-7752 \\
2024-05-12.02& Magellan FIRE&158.6&0.6 &9-10&1336 &7678-25596 \\
2024-06-05.15 & MMT/Binospec &3600& 1.66 &3-4 &1361 & 3825-9197 \\
\enddata
%\tablecomments{}
\end{deluxetable*}
\clearpage
\startlongtable
\begin{deluxetable*}{ccccccc}

\tablecaption{ Radio Data Log \label{table:Radio_log}}
\tablehead{\colhead{Tel.} & \colhead{UTC Date of Obs.} & \colhead{Days Since Expl.} & \colhead{Rep Freq (GHz)} & \colhead{Flux Density (mJy)}& \colhead{$1\sigma$ Fit uncertainty ($\mu$Jy)}&\colhead{RMS ($\mu Jy$})}
\startdata
GMRT & 2020 November 29 & 77 & 1.25& 0.098(UL) &N/A &33\\
GMRT & 2020 December 14 & 93 & 1.25& 0.051(UL) & &17\\
GMRT & 2021 May 27 & 256 & 1.25& 0.13 & 50&30\\
GMRT & 2021 May 31 & 260 & 0.75& 0.37(UL) &N/A &123\\
VLA & 2021 October 4 & 386 & 2.62 & 1.15 & 33& 26 \\
VLA & 2021 October 4 & 386 & 3.19 & 1.41 & 17& 39\\
VLA & 2021 October 4 & 386 & 3.75 & 1.82 & 13& 21 \\
VLA & 2021 October 4 & 386 & 4.42 & 2.06 & 20&18 \\
VLA & 2021 October 4 & 386 & 5.25 & 2.15 & 20&21 \\
VLA & 2021 October 4 & 386 & 5.67 & 2.13 & 34 &22\\
VLA & 2021 October 4 & 386 & 6.42 & 2.12 & 31& 15 \\
VLA & 2021 October 4 & 386 & 7.13 & 1.97 & 28&22 \\
VLA & 2021 October 4 & 386 & 7.7 & 1.85 & 35 &19\\
VLA & 2021 October 2 & 384 & 8.55 & 1.696 & 29 &35\\
VLA & 2021 October 2 & 384 & 9.9 & 1.526 & 45& 26 \\
VLA & 2021 October 2 & 384 & 10.67 & 1.362 & 24&28 \\
VLA & 2021 October 8 & 390 & 13.49 & 1.425 & 21&115\\
VLA & 2021 October 8 & 390 & 16.19 & 1.274 & 37 &101\\
VLA & 2021 October 8 & 412 & 18.68 & 1.125 & 15 &18\\
VLA & 2021 October 30 & 412 & 19.93 & 1.018 & 10 &21\\
VLA & 2021 October 30 & 412 & 21.12 & 0.948 & 17 &24\\
VLA & 2021 October 30 & 412 & 22.67 & 0.861 & 15 &23\\
VLA & 2021 October 30 & 412 & 23.93 & 0.819 & 18 &22\\
VLA & 2021 October 30 & 414 & 25.25 & 0.827 & 12 &21\\
VLA & 2021 November 1 & 414 & 31.06 & 0.661 & 64 &17\\
VLA & 2021 November 1 & 414 & 34.99 & 0.49 & 47 &19\\
GMRT & 2022 February 02 & 512 & 0.47 & 50&100&39\\
GMRT & 2022 February 03 & 514 & 1.4& 0.93& 100&78\\
GMRT & 2022 May 18 & 612 & 0.40 & 1.95(UL)&N/A&650\\
GMRT & 2022 May 20 & 614 & 1.25& 0.93& 80&58\\
GMRT & 2022 May 20 & 614 & 0.75& 0.377& 110&133\\
GMRT & 2022 September 27& 744 & 0.40& 0.8(UL)&N/A& 267\\
GMRT & 2022 September 27 & 744 & 0.75& 0.6(UL)&N/A &200\\
GMRT & 2022 September 27 & 744 & 1.25& 0.97& 160&76\\
GMRT & 2023 February 18 & 888 & 0.75& 0.255(UL)&N/A &85\\
GMRT & 2023 February 19 & 888 & 1.25& 1.26& 11&90\\
VLA & 2023 March 21 & 919 & 2.49 & 2.957 & 36& 17 \\
VLA & 2023 March 21 & 919 & 3.12 & 3.004 & 48&61 \\
VLA & 2023 March 21 & 919 & 3.69 & 2.98 & 32& 55 \\
VLA & 2023 March 21 & 919 & 4.29 & 2.852 & 3&43 \\
VLA & 2023 March 21 & 919 & 4.8 & 2.648 & 21&53 \\
VLA & 2023 March 21 & 919 & 5.31 & 2.404 & 25&45 \\
VLA & 2023 March 21 & 919 & 5.81 & 2.275 & 43&47 \\
VLA & 2023 March 21 & 919 & 6.3 & 2.04 & 39&43 \\
VLA & 2023 March 21 & 919 & 6.87 & 1.94 & 21&36\\
VLA & 2023 March 21 & 919 & 7.57 & 1.75 & 22&36 \\
VLA & 2023 March 21 & 919 & 8.61 & 1.32 & 25&38 \\
VLA & 2023 March 21 & 919 & 9.55 & 1.06 & 31&45 \\
VLA & 2023 March 21 & 919 & 10.55 & 0.987 & 38&43 \\
VLA & 2023 March 21 & 919 & 11.51 & 0.924 & 36&44\\
VLA & 2023 March 20 & 918 & 12.78 & 0.88 & 63 &35\\
VLA & 2023 March 20 & 918 & 14.31 & 0.81 & 60&31 \\
VLA & 2023 March 20 & 918 & 15.85 & 0.638 & 42&31 \\
VLA & 2023 March 20 & 918 & 17.39 & 0.659 & 28&38 \\
VLA & 2023 March 23 & 921 & 18.67 & 0.681 & 16 &16\\
VLA & 2023 March 23 & 921 & 19.93 & 0.639 & 34&17 \\
VLA & 2023 March 23 & 921 & 21.13 & 0.553 & 22&99 \\
VLA & 2023 March 23 & 921 & 22.67 & 0.52 & 15 &23\\
VLA & 2023 March 23 & 921 & 23.93 & 0.543 & 32&21 \\
VLA & 2023 March 23 & 921 & 25.221 & 0.509 & 24 &18\\
VLA & 2023 March 23 & 921 & 30.99 & 0.416 & 16&55 \\
VLA & 2023 March 23 & 921 & 34.95 & 0.354 & 15&66 \\
GMRT & 2023 September 2& 744 & 0.40& 0.51(UL)& N/A&140\\
GMRT & 2023 September 2 & 888 & 0.75& 0.75& 23&247\\
GMRT & 2023 September 3 & 888 & 1.25& 1.52& 50&62\\
GMRT & 2023 November 15 & 888 & 0.4& 0.5&50&153\\
GMRT & 2023 November 17 & 888 & 0.75& 1.03& 5&94\\
GMRT & 2023 November 18 & 888 & 1.25& 1.77& 5&88\\
VLA & 2024 January 1 & 1205 & 2.99 & 2.72 & 50&171 \\
VLA & 2024 January 1 & 1205 & 3.62 & 2.324 & 25&98 \\
VLA & 2024 January 1 & 1205 & 4.55 & 1.976 & 17&175 \\
VLA & 2024 January 1 & 1205 & 5.55 & 1.6 & 17&70 \\
VLA & 2024 January 1 & 1205 & 6.55 & 1.282 & 22&30 \\
VLA & 2024 January 1 & 1205 & 7.51 & 1.148 & 16&27 \\
VLA & 2024 January 1 & 1205 & 8.55 & 1.005 & 17 &21\\
VLA & 2024 January 1 & 1205 & 9.55 & 0.941 & 09 &24\\
VLA & 2024 January 1 & 1205 & 10.55 & 0.892 & 14&20 \\
VLA & 2024 January 1 & 1205 & 11.51 & 0.818 & 16&23 \\
VLA & 2024 January 1 & 1205 & 12.77 & 0.737 & 34&24 \\
VLA & 2024 January 1 & 1205 & 14.31 & 0.65 & 34&28 \\
VLA & 2024 January 1 & 1205 & 15.85 & 0.606 & 18&27\\
VLA & 2024 January 1 & 1205 & 17.38 & 0.546 & 34&26 \\
VLA & 2024 January 1 & 1205 & 19.99 & 0.506 & 13&13 \\
VLA & 2024 January 1 & 1205 & 23.99 & 0.42 & 8&12 \\
VLA & 2024 January 1 & 1205 & 30.99 & 0.338 & 38&72 \\
VLA & 2024 January 1 & 1205 & 34.99 & 0.305 & 41&35 \\
GMRT & 2024 May 12 & 1337 & 1.25 & 2.026& 88 &35 \\
GMRT & 2024 June 16  & 1372 & 0.75 & 1.032& 58 &118 \\
\enddata
\tablecomments{Fit uncertainties are from CASA \textit{imfit}, not including systematic uncertainties. UL denotes 3$\sigma$ upper limits from the image rms.}
\end{deluxetable*}

\begin{deluxetable}{cccc}
\tablecaption{Photometric Observations Log 
\label{table:photometry_log}}
\tablehead{
\colhead{MJD} & 
\colhead{Magnitude (AB)} & 
\colhead{$\sigma_{\mathrm{mag}}$ (AB)} & 
\colhead{Instrument/Filter}
}
\startdata
59157.61995 & 16.499 & 0.027 & ATLAS o \\
59157.62549 & 16.479 & 0.025 & ATLAS o \\
59157.64256 & 16.534 & 0.029 & ATLAS o \\
59157.64992 & 16.420 & 0.059 & ATLAS o \\
59166.48980 & 16.990 & 0.050 & ZTF g \\
59166.53544 & 16.530 & 0.040 & ZTF r \\
59168.50731 & 16.990 & 0.050 & ZTF g \\
59168.53863 & 16.530 & 0.040 & ZTF r \\
59170.49674 & 17.020 & 0.040 & ZTF g \\
59170.55676 & 16.510 & 0.040 & ZTF r \\
59173.48462 & 17.040 & 0.050 & ZTF g \\
59173.52400 & 16.550 & 0.040 & ZTF r \\
59176.51689 & 17.050 & 0.040 & ZTF g \\
59176.55172 & 16.570 & 0.030 & ZTF r \\
59179.52704 & 16.590 & 0.040 & ZTF r \\
59181.49191 & 16.600 & 0.040 & ZTF r \\
59181.52844 & 17.060 & 0.050 & ZTF g \\
59181.57916 & 16.620 & 0.036 & ATLAS o \\
59181.58469 & 16.722 & 0.032 & ATLAS o \\
59181.58837 & 16.682 & 0.031 & ATLAS o \\
59181.60404 & 16.658 & 0.029 & ATLAS o \\
59183.58891 & 16.637 & 0.039 & ATLAS o \\
59183.59305 & 16.668 & 0.035 & ATLAS o \\
59183.59860 & 16.699 & 0.036 & ATLAS o \\
59183.60915 & 16.608 & 0.032 & ATLAS o \\
59183.66102 & 16.628 & 0.036 & ATLAS o \\
\enddata
\tablecomments{Early optical ZTF/ATLAS photometry of SN 2020ywx. The full photometry will be provided in the online version of this work.} 
\end{deluxetable}

\vspace{5mm}




\let\cleardoublepage\clearpage

\

\bibliography{ms}{}

\begin{thebibliography}{}
\expandafter\ifx\csname natexlab\endcsname\relax\def\natexlab#1{#1}\fi
\providecommand{\url}[1]{\href{#1}{#1}}
\providecommand{\dodoi}[1]{doi:~\href{http://doi.org/#1}{\nolinkurl{#1}}}
\providecommand{\doeprint}[1]{\href{http://ascl.net/#1}{\nolinkurl{http://ascl.net/#1}}}
\providecommand{\doarXiv}[1]{\href{https://arxiv.org/abs/#1}{\nolinkurl{https://arxiv.org/abs/#1}}}

\bibitem[{{Aghakhanloo} {et~al.}(2023){Aghakhanloo}, {Smith}, {Milne},
  {Andrews}, {Van Dyk}, {Filippenko}, {Jencson}, {Lau}, {Sand}, {Wyatt}, \&
  {Zheng}}]{Aghakhanloo_2023}
{Aghakhanloo}, M., {Smith}, N., {Milne}, P., {et~al.} 2023, \mnras, 526, 456,
  \dodoi{10.1093/mnras/stad2702}

\bibitem[{{Anderson} {et~al.}(2014){Anderson}, {Gonz{\'a}lez-Gait{\'a}n},
  {Hamuy}, {Guti{\'e}rrez}, {Stritzinger}, {Olivares E.}, {Phillips},
  {Schulze}, {Antezana}, {Bolt}, {Campillay}, {Castell{\'o}n}, {Contreras}, {de
  Jaeger}, {Folatelli}, {F{\"o}rster}, {Freedman}, {Gonz{\'a}lez}, {Hsiao},
  {Krzemi{\'n}ski}, {Krisciunas}, {Maza}, {McCarthy}, {Morrell}, {Persson},
  {Roth}, {Salgado}, {Suntzeff}, \& {Thomas-Osip}}]{Anderson-2014}
{Anderson}, J.~P., {Gonz{\'a}lez-Gait{\'a}n}, S., {Hamuy}, M., {et~al.} 2014,
  \apj, 786, 67, \dodoi{10.1088/0004-637X/786/1/67}

\bibitem[{{Andrews} {et~al.}(2011){Andrews}, {Clayton}, {Wesson}, {Sugerman},
  {Barlow}, {Clem}, {Ercolano}, {Fabbri}, {Gallagher}, {Landolt}, {Meixner},
  {Otsuka}, {Riebel}, \& {Welch}}]{Andrews_2010jl}
{Andrews}, J.~E., {Clayton}, G.~C., {Wesson}, R., {et~al.} 2011, \aj, 142, 45,
  \dodoi{10.1088/0004-6256/142/2/45}

\bibitem[{{Angel} {et~al.}(1979){Angel}, {Hilliard}, \& {Weymann}}]{Angel_79}
{Angel}, J.~R.~P., {Hilliard}, R.~L., \& {Weymann}, R.~J. 1979, in The MMT and
  the Future of Ground-Based Astronomy, ed. T.~C. {Weekes}, Vol. 385, 87

\bibitem[{{Aretxaga} {et~al.}(1999){Aretxaga}, {Benetti}, {Terlevich},
  {Fabian}, {Cappellaro}, {Turatto}, \& {della Valle}}]{Aretxaga_88z}
{Aretxaga}, I., {Benetti}, S., {Terlevich}, R.~J., {et~al.} 1999, \mnras, 309,
  343, \dodoi{10.1046/j.1365-8711.1999.02830.x}

\bibitem[{{Arnaud}(1996)}]{xspec}
{Arnaud}, K.~A. 1996, in Astronomical Society of the Pacific Conference Series,
  Vol. 101, Astronomical Data Analysis Software and Systems V, ed. G.~H.
  {Jacoby} \& J.~{Barnes}, 17

\bibitem[{{Astropy Collaboration} {et~al.}(2013){Astropy Collaboration},
  {Robitaille}, {Tollerud}, {Greenfield}, {Droettboom}, {Bray}, {Aldcroft},
  {Davis}, {Ginsburg}, {Price-Whelan}, {Kerzendorf}, {Conley}, {Crighton},
  {Barbary}, {Muna}, {Ferguson}, {Grollier}, {Parikh}, {Nair}, {Unther},
  {Deil}, {Woillez}, {Conseil}, {Kramer}, {Turner}, {Singer}, {Fox}, {Weaver},
  {Zabalza}, {Edwards}, {Azalee Bostroem}, {Burke}, {Casey}, {Crawford},
  {Dencheva}, {Ely}, {Jenness}, {Labrie}, {Lim}, {Pierfederici}, {Pontzen},
  {Ptak}, {Refsdal}, {Servillat}, \& {Streicher}}]{Astropy_2013}
{Astropy Collaboration}, {Robitaille}, T.~P., {Tollerud}, E.~J., {et~al.} 2013,
  \aap, 558, A33, \dodoi{10.1051/0004-6361/201322068}

\bibitem[{{Astropy Collaboration} {et~al.}(2018){Astropy Collaboration},
  {Price-Whelan}, {Sip{\H{o}}cz}, {G{\"u}nther}, {Lim}, {Crawford}, {Conseil},
  {Shupe}, {Craig}, {Dencheva}, {Ginsburg}, {VanderPlas}, {Bradley},
  {P{\'e}rez-Su{\'a}rez}, {de Val-Borro}, {Aldcroft}, {Cruz}, {Robitaille},
  {Tollerud}, {Ardelean}, {Babej}, {Bach}, {Bachetti}, {Bakanov}, {Bamford},
  {Barentsen}, {Barmby}, {Baumbach}, {Berry}, {Biscani}, {Boquien}, {Bostroem},
  {Bouma}, {Brammer}, {Bray}, {Breytenbach}, {Buddelmeijer}, {Burke},
  {Calderone}, {Cano Rodr{\'\i}guez}, {Cara}, {Cardoso}, {Cheedella}, {Copin},
  {Corrales}, {Crichton}, {D'Avella}, {Deil}, {Depagne}, {Dietrich}, {Donath},
  {Droettboom}, {Earl}, {Erben}, {Fabbro}, {Ferreira}, {Finethy}, {Fox},
  {Garrison}, {Gibbons}, {Goldstein}, {Gommers}, {Greco}, {Greenfield},
  {Groener}, {Grollier}, {Hagen}, {Hirst}, {Homeier}, {Horton}, {Hosseinzadeh},
  {Hu}, {Hunkeler}, {Ivezi{\'c}}, {Jain}, {Jenness}, {Kanarek}, {Kendrew},
  {Kern}, {Kerzendorf}, {Khvalko}, {King}, {Kirkby}, {Kulkarni}, {Kumar},
  {Lee}, {Lenz}, {Littlefair}, {Ma}, {Macleod}, {Mastropietro}, {McCully},
  {Montagnac}, {Morris}, {Mueller}, {Mumford}, {Muna}, {Murphy}, {Nelson},
  {Nguyen}, {Ninan}, {N{\"o}the}, {Ogaz}, {Oh}, {Parejko}, {Parley}, {Pascual},
  {Patil}, {Patil}, {Plunkett}, {Prochaska}, {Rastogi}, {Reddy Janga},
  {Sabater}, {Sakurikar}, {Seifert}, {Sherbert}, {Sherwood-Taylor}, {Shih},
  {Sick}, {Silbiger}, {Singanamalla}, {Singer}, {Sladen}, {Sooley},
  {Sornarajah}, {Streicher}, {Teuben}, {Thomas}, {Tremblay}, {Turner},
  {Terr{\'o}n}, {van Kerkwijk}, {de la Vega}, {Watkins}, {Weaver}, {Whitmore},
  {Woillez}, {Zabalza}, \& {Astropy Contributors}}]{Astropy_2018}
{Astropy Collaboration}, {Price-Whelan}, A.~M., {Sip{\H{o}}cz}, B.~M., {et~al.}
  2018, \aj, 156, 123, \dodoi{10.3847/1538-3881/aabc4f}

\bibitem[{{Astropy Collaboration} {et~al.}(2022){Astropy Collaboration},
  {Price-Whelan}, {Lim}, {Earl}, {Starkman}, {Bradley}, {Shupe}, {Patil},
  {Corrales}, {Brasseur}, {N{\"o}the}, {Donath}, {Tollerud}, {Morris},
  {Ginsburg}, {Vaher}, {Weaver}, {Tocknell}, {Jamieson}, {van Kerkwijk},
  {Robitaille}, {Merry}, {Bachetti}, {G{\"u}nther}, {Aldcroft},
  {Alvarado-Montes}, {Archibald}, {B{\'o}di}, {Bapat}, {Barentsen},
  {Baz{\'a}n}, {Biswas}, {Boquien}, {Burke}, {Cara}, {Cara}, {Conroy},
  {Conseil}, {Craig}, {Cross}, {Cruz}, {D'Eugenio}, {Dencheva}, {Devillepoix},
  {Dietrich}, {Eigenbrot}, {Erben}, {Ferreira}, {Foreman-Mackey}, {Fox},
  {Freij}, {Garg}, {Geda}, {Glattly}, {Gondhalekar}, {Gordon}, {Grant},
  {Greenfield}, {Groener}, {Guest}, {Gurovich}, {Handberg}, {Hart},
  {Hatfield-Dodds}, {Homeier}, {Hosseinzadeh}, {Jenness}, {Jones}, {Joseph},
  {Kalmbach}, {Karamehmetoglu}, {Ka{\l}uszy{\'n}ski}, {Kelley}, {Kern},
  {Kerzendorf}, {Koch}, {Kulumani}, {Lee}, {Ly}, {Ma}, {MacBride}, {Maljaars},
  {Muna}, {Murphy}, {Norman}, {O'Steen}, {Oman}, {Pacifici}, {Pascual},
  {Pascual-Granado}, {Patil}, {Perren}, {Pickering}, {Rastogi}, {Roulston},
  {Ryan}, {Rykoff}, {Sabater}, {Sakurikar}, {Salgado}, {Sanghi}, {Saunders},
  {Savchenko}, {Schwardt}, {Seifert-Eckert}, {Shih}, {Jain}, {Shukla}, {Sick},
  {Simpson}, {Singanamalla}, {Singer}, {Singhal}, {Sinha}, {Sip{\H{o}}cz},
  {Spitler}, {Stansby}, {Streicher}, {{\v{S}}umak}, {Swinbank}, {Taranu},
  {Tewary}, {Tremblay}, {de Val-Borro}, {Van Kooten}, {Vasovi{\'c}}, {Verma},
  {de Miranda Cardoso}, {Williams}, {Wilson}, {Winkel}, {Wood-Vasey}, {Xue},
  {Yoachim}, {Zhang}, {Zonca}, \& {Astropy Project
  Contributors}}]{Astropy_2022}
{Astropy Collaboration}, {Price-Whelan}, A.~M., {Lim}, P.~L., {et~al.} 2022,
  \apj, 935, 167, \dodoi{10.3847/1538-4357/ac7c74}

\bibitem[{{Baer-Way} {et~al.}(2024){Baer-Way}, {DeGraw}, {Zheng}, {Van Dyk},
  {Filippenko}, {Fox}, {Brink}, {Kelly}, {Smith}, {Vasylyev}, {de Jaeger},
  {Zhang}, {Stegman}, {Ross}, \& {Yunus}}]{Baerway2024}
{Baer-Way}, R., {DeGraw}, A., {Zheng}, W., {et~al.} 2024, \apj, 964, 172,
  \dodoi{10.3847/1538-4357/ad2175}

\bibitem[{{Becker}(2015)}]{HOTPANTS}
{Becker}, A. 2015, {HOTPANTS: High Order Transform of PSF ANd Template
  Subtraction}, Astrophysics Source Code Library, record ascl:1504.004

\bibitem[{{Bellm} {et~al.}(2019){Bellm}, {Kulkarni}, {Graham}, {Dekany},
  {Smith}, {Riddle}, {Masci}, {Helou}, {Prince}, {Adams}, {Barbarino},
  {Barlow}, {Bauer}, {Beck}, {Belicki}, {Biswas}, {Blagorodnova}, {Bodewits},
  {Bolin}, {Brinnel}, {Brooke}, {Bue}, {Bulla}, {Burruss}, {Cenko}, {Chang},
  {Connolly}, {Coughlin}, {Cromer}, {Cunningham}, {De}, {Delacroix}, {Desai},
  {Duev}, {Eadie}, {Farnham}, {Feeney}, {Feindt}, {Flynn}, {Franckowiak},
  {Frederick}, {Fremling}, {Gal-Yam}, {Gezari}, {Giomi}, {Goldstein},
  {Golkhou}, {Goobar}, {Groom}, {Hacopians}, {Hale}, {Henning}, {Ho}, {Hover},
  {Howell}, {Hung}, {Huppenkothen}, {Imel}, {Ip}, {Ivezi{\'c}}, {Jackson},
  {Jones}, {Juric}, {Kasliwal}, {Kaspi}, {Kaye}, {Kelley}, {Kowalski},
  {Kramer}, {Kupfer}, {Landry}, {Laher}, {Lee}, {Lin}, {Lin}, {Lunnan},
  {Giomi}, {Mahabal}, {Mao}, {Miller}, {Monkewitz}, {Murphy}, {Ngeow},
  {Nordin}, {Nugent}, {Ofek}, {Patterson}, {Penprase}, {Porter}, {Rauch},
  {Rebbapragada}, {Reiley}, {Rigault}, {Rodriguez}, {van Roestel}, {Rusholme},
  {van Santen}, {Schulze}, {Shupe}, {Singer}, {Soumagnac}, {Stein}, {Surace},
  {Sollerman}, {Szkody}, {Taddia}, {Terek}, {Van Sistine}, {van Velzen},
  {Vestrand}, {Walters}, {Ward}, {Ye}, {Yu}, {Yan}, \&
  {Zolkower}}]{ztf_overview}
{Bellm}, E.~C., {Kulkarni}, S.~R., {Graham}, M.~J., {et~al.} 2019, \pasp, 131,
  018002, \dodoi{10.1088/1538-3873/aaecbe}

\bibitem[{{Bietenholz} \& {Bartel}(2017)}]{Bietenholtz_2017}
{Bietenholz}, M.~F., \& {Bartel}, N. 2017, \apj, 851, 7,
  \dodoi{10.3847/1538-4357/aa960b}

\bibitem[{{Bietenholz} {et~al.}(2021){Bietenholz}, {Bartel}, {Argo}, {Dua},
  {Ryder}, \& {Soderberg}}]{Bietenholz_2021}
{Bietenholz}, M.~F., {Bartel}, N., {Argo}, M., {et~al.} 2021, \apj, 908, 75,
  \dodoi{10.3847/1538-4357/abccd9}

\bibitem[{{Bietenholz} {et~al.}(2012){Bietenholz}, {Brunthaler}, {Soderberg},
  {Krauss}, {Zauderer}, {Bartel}, {Chomiuk}, \& {Rupen}}]{Bietenholz_2012}
{Bietenholz}, M.~F., {Brunthaler}, A., {Soderberg}, A.~M., {et~al.} 2012, \apj,
  751, 125, \dodoi{10.1088/0004-637X/751/2/125}

\bibitem[{{Bilinski} {et~al.}(2024){Bilinski}, {Smith}, {Williams}, {Smith},
  {Leonard}, {Hoffman}, {Andrews}, \& {Milne}}]{Bilinksi}
{Bilinski}, C., {Smith}, N., {Williams}, G.~G., {et~al.} 2024, \mnras, 529,
  1104, \dodoi{10.1093/mnras/stae380}

\bibitem[{{Brethauer} {et~al.}(2022){Brethauer}, {Margutti}, {Milisavljevic},
  {Bietenholz}, {Chornock}, {Coppejans}, {De Colle}, {Hajela}, {Terreran},
  {Vargas}, {DeMarchi}, {Harris}, {Jacobson-Gal{\'a}n}, {Kamble}, {Patnaude},
  \& {Stroh}}]{Brethauer_2022}
{Brethauer}, D., {Margutti}, R., {Milisavljevic}, D., {et~al.} 2022, \apj, 939,
  105, \dodoi{10.3847/1538-4357/ac8b14}

\bibitem[{{Chandra}(2018)}]{Chandra_2018}
{Chandra}, P. 2018, \ssr, 214, 27, \dodoi{10.1007/s11214-017-0461-6}

\bibitem[{{Chandra} {et~al.}(2012){Chandra}, {Chevalier}, {Chugai}, {Fransson},
  {Irwin}, {Soderberg}, {Chakraborti}, \& {Immler}}]{Chandra_2012}
{Chandra}, P., {Chevalier}, R.~A., {Chugai}, N., {et~al.} 2012, \apj, 755, 110,
  \dodoi{10.1088/0004-637X/755/2/110}

\bibitem[{{Chandra} {et~al.}(2015){Chandra}, {Chevalier}, {Chugai}, {Fransson},
  \& {Soderberg}}]{Chandra_2015}
{Chandra}, P., {Chevalier}, R.~A., {Chugai}, N., {Fransson}, C., \&
  {Soderberg}, A.~M. 2015, \apj, 810, 32, \dodoi{10.1088/0004-637X/810/1/32}

\bibitem[{{Chandra} {et~al.}(2020){Chandra}, {Chevalier}, {Chugai},
  {Milisavljevic}, \& {Fransson}}]{Chandra_2020}
{Chandra}, P., {Chevalier}, R.~A., {Chugai}, N., {Milisavljevic}, D., \&
  {Fransson}, C. 2020, \apj, 902, 55, \dodoi{10.3847/1538-4357/abb460}

\bibitem[{{Chandra} {et~al.}(2024){Chandra}, {Chevalier}, {Maeda}, {Ray}, \&
  {Nayana}}]{Chandra_23IXF}
{Chandra}, P., {Chevalier}, R.~A., {Maeda}, K., {Ray}, A.~K., \& {Nayana},
  A.~J. 2024, \apjl, 963, L4, \dodoi{10.3847/2041-8213/ad275d}

\bibitem[{{Chandra} \& {Kanekar}(2017)}]{Chandra_Kanekar_2017}
{Chandra}, P., \& {Kanekar}, N. 2017, \apj, 846, 111,
  \dodoi{10.3847/1538-4357/aa85a2}

\bibitem[{{Chevalier}(1982)}]{Chevalier_1982}
{Chevalier}, R.~A. 1982, \apj, 259, 302, \dodoi{10.1086/160167}

\bibitem[{{Chevalier}(1998)}]{Chevalier_1998}
---. 1998, \apj, 499, 810, \dodoi{10.1086/305676}

\bibitem[{{Chevalier} \& {Fransson}(2017)}]{Chevalier_17}
{Chevalier}, R.~A., \& {Fransson}, C. 2017, in Handbook of Supernovae, ed.
  A.~W. {Alsabti} \& P.~{Murdin} (Springer), 875,
  \dodoi{10.1007/978-3-319-21846-5_34}

\bibitem[{{Chevalier} \& {Irwin}(2012)}]{Chevalier_2012}
{Chevalier}, R.~A., \& {Irwin}, C.~M. 2012, \apjl, 747, L17,
  \dodoi{10.1088/2041-8205/747/1/L17}

\bibitem[{{Chugai}(1991)}]{Chugai_1991}
{Chugai}, N.~N. 1991, \mnras, 250, 513, \dodoi{10.1093/mnras/250.3.513}

\bibitem[{{Chugai} \& {Danziger}(1994)}]{Chugai_Danziger}
{Chugai}, N.~N., \& {Danziger}, I.~J. 1994, \mnras, 268, 173,
  \dodoi{10.1093/mnras/268.1.173}

\bibitem[{{Cold} \& {Hjorth}(2023)}]{Cold23-SNIInrate-PTFZTF}
{Cold}, C., \& {Hjorth}, J. 2023, \aap, 670, A48,
  \dodoi{10.1051/0004-6361/202244867}

\bibitem[{{Cushing} {et~al.}(2004){Cushing}, {Vacca}, \& {Rayner}}]{Cushing04}
{Cushing}, M.~C., {Vacca}, W.~D., \& {Rayner}, J.~T. 2004, \pasp, 116, 362,
  \dodoi{10.1086/382907}

\bibitem[{{Dessart} \& {Hillier}(2022)}]{Dessart_2022}
{Dessart}, L., \& {Hillier}, D.~J. 2022, \aap, 660, L9,
  \dodoi{10.1051/0004-6361/202243372}

\bibitem[{{Dwarkadas} \& {Gruszko}(2012)}]{Dwarkadas_2012}
{Dwarkadas}, V.~V., \& {Gruszko}, J. 2012, \mnras, 419, 1515,
  \dodoi{10.1111/j.1365-2966.2011.19808.x}

\bibitem[{{Dwarkadas} {et~al.}(2016){Dwarkadas}, {Romero-Ca{\~n}izales},
  {Reddy}, \& {Bauer}}]{Dwarkadas_16}
{Dwarkadas}, V.~V., {Romero-Ca{\~n}izales}, C., {Reddy}, R., \& {Bauer}, F.~E.
  2016, \mnras, 462, 1101, \dodoi{10.1093/mnras/stw1717}

\bibitem[{{Evans} {et~al.}(2007){Evans}, {Beardmore}, {Page}, {Tyler},
  {Osborne}, {Goad}, {O'Brien}, {Vetere}, {Racusin}, {Morris}, {Burrows},
  {Capalbi}, {Perri}, {Gehrels}, \& {Romano}}]{Swift_tools}
{Evans}, P.~A., {Beardmore}, A.~P., {Page}, K.~L., {et~al.} 2007, \aap, 469,
  379, \dodoi{10.1051/0004-6361:20077530}

\bibitem[{{Fabricant} {et~al.}(2019){Fabricant}, {Fata}, {Epps}, {Gauron},
  {Mueller}, {Zajac}, {Amato}, {Barberis}, {Bergner}, {Brennan}, {Brown},
  {Chilingarian}, {Geary}, {Kradinov}, {McLeod}, {Smith}, \&
  {Woods}}]{Binospec_2019}
{Fabricant}, D., {Fata}, R., {Epps}, H., {et~al.} 2019, \pasp, 131, 075004,
  \dodoi{10.1088/1538-3873/ab1d78}

\bibitem[{{Farias} {et~al.}(2024){Farias}, {Gall}, {Narayan}, {Rest}, {Villar},
  {Angus}, {Auchettl}, {Davis}, {Foley}, {Gagliano}, {Hjorth}, {Izzo},
  {Kilpatrick}, {Perkins}, {Ramirez-Ruiz}, {Ransome}, {Sarangi.}, {Yarza},
  {Coulter}, {Jones}, {Khetan}, {Rest}, {Siebert}, {Swift}, {Taggart},
  {Tinyanont}, {Wrubel}, {de Boer}, {Clever}, {Dhara}, {Gao}, \&
  {Lin}}]{Farias_2021foa}
{Farias}, D., {Gall}, C., {Narayan}, G., {et~al.} 2024, arXiv e-prints,
  arXiv:2409.01359, \dodoi{10.48550/arXiv.2409.01359}

\bibitem[{{Ferland} {et~al.}(2013){Ferland}, {Porter}, {van Hoof}, {Williams},
  {Abel}, {Lykins}, {Shaw}, {Henney}, \& {Stancil}}]{Cloudy_2013}
{Ferland}, G.~J., {Porter}, R.~L., {van Hoof}, P.~A.~M., {et~al.} 2013, \rmxaa,
  49, 137, \dodoi{10.48550/arXiv.1302.4485}

\bibitem[{{Filippenko}(1982)}]{Filippenko_1982}
{Filippenko}, A.~V. 1982, \pasp, 94, 715, \dodoi{10.1086/131052}

\bibitem[{{Filippenko}(1997)}]{Filippenko_1997}
---. 1997, \araa, 35, 309, \dodoi{10.1146/annurev.astro.35.1.309}

\bibitem[{{Foley} {et~al.}(2007){Foley}, {Smith}, {Ganeshalingam}, {Li},
  {Chornock}, \& {Filippenko}}]{Foley_2006jc}
{Foley}, R.~J., {Smith}, N., {Ganeshalingam}, M., {et~al.} 2007, \apjl, 657,
  L105, \dodoi{10.1086/513145}

\bibitem[{{Foreman-Mackey} {et~al.}(2013){Foreman-Mackey}, {Hogg}, {Lang}, \&
  {Goodman}}]{Foreman_2013}
{Foreman-Mackey}, D., {Hogg}, D.~W., {Lang}, D., \& {Goodman}, J. 2013, \pasp,
  125, 306, \dodoi{10.1086/670067}

\bibitem[{{Fox} {et~al.}(2011){Fox}, {Chevalier}, {Skrutskie}, {Soderberg},
  {Filippenko}, {Ganeshalingam}, {Silverman}, {Smith}, \& {Steele}}]{Fox_2011}
{Fox}, O.~D., {Chevalier}, R.~A., {Skrutskie}, M.~F., {et~al.} 2011, \apj, 741,
  7, \dodoi{10.1088/0004-637X/741/1/7}

\bibitem[{{Fox} {et~al.}(2015){Fox}, {Silverman}, {Filippenko}, {Mauerhan},
  {Becker}, {Borish}, {Cenko}, {Clubb}, {Graham}, {Hsiao}, {Kelly}, {Lee},
  {Marion}, {Milisavljevic}, {Parrent}, {Shivvers}, {Skrutskie}, {Smith},
  {Wilson}, \& {Zheng}}]{Fox_2015}
{Fox}, O.~D., {Silverman}, J.~M., {Filippenko}, A.~V., {et~al.} 2015, \mnras,
  447, 772, \dodoi{10.1093/mnras/stu2435}

\bibitem[{{Fransson} {et~al.}(1996){Fransson}, {Lundqvist}, \&
  {Chevalier}}]{Frannson_96}
{Fransson}, C., {Lundqvist}, P., \& {Chevalier}, R.~A. 1996, \apj, 461, 993,
  \dodoi{10.1086/177119}

\bibitem[{{Fransson} {et~al.}(2014){Fransson}, {Ergon}, {Challis}, {Chevalier},
  {France}, {Kirshner}, {Marion}, {Milisavljevic}, {Smith}, {Bufano},
  {Friedman}, {Kangas}, {Larsson}, {Mattila}, {Benetti}, {Chornock}, {Czekala},
  {Soderberg}, \& {Sollerman}}]{Fransson_2010jl}
{Fransson}, C., {Ergon}, M., {Challis}, P.~J., {et~al.} 2014, \apj, 797, 118,
  \dodoi{10.1088/0004-637X/797/2/118}

\bibitem[{{Fruscione} {et~al.}(2006){Fruscione}, {McDowell}, {Allen},
  {Brickhouse}, {Burke}, {Davis}, {Durham}, {Elvis}, {Galle}, {Harris},
  {Huenemoerder}, {Houck}, {Ishibashi}, {Karovska}, {Nicastro}, {Noble},
  {Nowak}, {Primini}, {Siemiginowska}, {Smith}, \& {Wise}}]{Ciao_2006}
{Fruscione}, A., {McDowell}, J.~C., {Allen}, G.~E., {et~al.} 2006, in Society
  of Photo-Optical Instrumentation Engineers (SPIE) Conference Series, Vol.
  6270, Observatory Operations: Strategies, Processes, and Systems, ed. D.~R.
  {Silva} \& R.~E. {Doxsey}, 62701V, \dodoi{10.1117/12.671760}

\bibitem[{{Gal-Yam} {et~al.}(2021){Gal-Yam}, {Yaron}, {Pastorello},
  {Taubenberger}, {Fraser}, \& {Perley}}]{Galyam_2017}
{Gal-Yam}, A., {Yaron}, O., {Pastorello}, A., {et~al.} 2021, Transient Name
  Server AstroNote, 76, 1

\bibitem[{{Gal-Yam} {et~al.}(2007){Gal-Yam}, {Leonard}, {Fox}, {Cenko},
  {Soderberg}, {Moon}, {Sand}, {Caltech Core Collapse Program}, {Li},
  {Filippenko}, {Aldering}, \& {Copin}}]{GalYam2007}
{Gal-Yam}, A., {Leonard}, D.~C., {Fox}, D.~B., {et~al.} 2007, \apj, 656, 372,
  \dodoi{10.1086/510523}

\bibitem[{{Gall} {et~al.}(2014){Gall}, {Hjorth}, {Watson}, {Dwek}, {Maund},
  {Fox}, {Leloudas}, {Malesani}, \& {Day-Jones}}]{Gall_2010jl}
{Gall}, C., {Hjorth}, J., {Watson}, D., {et~al.} 2014, \nat, 511, 326,
  \dodoi{10.1038/nature13558}

\bibitem[{{Gaskell} \& {Ferland}(1984)}]{84_line_ratio}
{Gaskell}, C.~M., \& {Ferland}, G.~J. 1984, \pasp, 96, 393,
  \dodoi{10.1086/131352}

\bibitem[{{Goldman} {et~al.}(2017){Goldman}, {van Loon}, {Zijlstra}, {Green},
  {Wood}, {Nanni}, {Imai}, {Whitelock}, {Matsuura}, {Groenewegen}, \&
  {G{\'o}mez}}]{Goldman_2017}
{Goldman}, S.~R., {van Loon}, J.~T., {Zijlstra}, A.~A., {et~al.} 2017, \mnras,
  465, 403, \dodoi{10.1093/mnras/stw2708}

\bibitem[{{Heger} {et~al.}(2003){Heger}, {Fryer}, {Woosley}, {Langer}, \&
  {Hartmann}}]{heger03}
{Heger}, A., {Fryer}, C.~L., {Woosley}, S.~E., {Langer}, N., \& {Hartmann},
  D.~H. 2003, \apj, 591, 288, \dodoi{10.1086/375341}

\bibitem[{{Jencson} {et~al.}(2016){Jencson}, {Prieto}, {Kochanek}, {Shappee},
  {Stanek}, \& {Pogge}}]{Jencson_2016}
{Jencson}, J.~E., {Prieto}, J.~L., {Kochanek}, C.~S., {et~al.} 2016, \mnras,
  456, 2622, \dodoi{10.1093/mnras/stv2795}

\bibitem[{{Kale} \& {Ishwara-Chandra}(2021)}]{GMRT_pipeline}
{Kale}, R., \& {Ishwara-Chandra}, C.~H. 2021, Experimental Astronomy, 51, 95,
  \dodoi{10.1007/s10686-020-09677-6}

\bibitem[{{Kundu} {et~al.}(2019){Kundu}, {Lundqvist}, {Sorokina},
  {P{\'e}rez-Torres}, {Blinnikov}, {O'Connor}, {Ergon}, {Chandra}, \&
  {Das}}]{Kundu_2019}
{Kundu}, E., {Lundqvist}, P., {Sorokina}, E., {et~al.} 2019, \apj, 875, 17,
  \dodoi{10.3847/1538-4357/ab0d81}

\bibitem[{{Liu} {et~al.}(2016){Liu}, {Modjaz}, {Bianco}, \& {Graur}}]{Liu_2016}
{Liu}, Y.-Q., {Modjaz}, M., {Bianco}, F.~B., \& {Graur}, O. 2016, \apj, 827,
  90, \dodoi{10.3847/0004-637X/827/2/90}

\bibitem[{{Lundqvist} \& {Fransson}(1988)}]{Lundqvist_88}
{Lundqvist}, P., \& {Fransson}, C. 1988, \aap, 192, 221

\bibitem[{{Lundqvist} \& {Fransson}(1991)}]{Lundqvist_1991}
---. 1991, \apj, 380, 575, \dodoi{10.1086/170615}

\bibitem[{{Lundqvist} {et~al.}(2013){Lundqvist}, {Mattila}, {Sollerman},
  {Kozma}, {Baron}, {Cox}, {Fransson}, {Leibundgut}, \&
  {Spyromilio}}]{Lundqvist_2013}
{Lundqvist}, P., {Mattila}, S., {Sollerman}, J., {et~al.} 2013, \mnras, 435,
  329, \dodoi{10.1093/mnras/stt1303}

\bibitem[{{Maeda} {et~al.}(2013){Maeda}, {Nozawa}, {Sahu}, {Minowa},
  {Motohara}, {Ueno}, {Folatelli}, {Pyo}, {Kitagawa}, {Kawabata}, {Anupama},
  {Kozasa}, {Moriya}, {Yamanaka}, {Nomoto}, {Bersten}, {Quimby}, \&
  {Iye}}]{Maeda_2010jl}
{Maeda}, K., {Nozawa}, T., {Sahu}, D.~K., {et~al.} 2013, \apj, 776, 5,
  \dodoi{10.1088/0004-637X/776/1/5}

\bibitem[{{Margalit} {et~al.}(2022){Margalit}, {Quataert}, \&
  {Ho}}]{Margalit_2022}
{Margalit}, B., {Quataert}, E., \& {Ho}, A. Y.~Q. 2022, \apj, 928, 122,
  \dodoi{10.3847/1538-4357/ac53b0}

\bibitem[{{Margutti} {et~al.}(2017){Margutti}, {Kamble}, {Milisavljevic},
  {Zapartas}, {de Mink}, {Drout}, {Chornock}, {Risaliti}, {Zauderer},
  {Bietenholz}, {Cantiello}, {Chakraborti}, {Chomiuk}, {Fong}, {Grefenstette},
  {Guidorzi}, {Kirshner}, {Parrent}, {Patnaude}, {Soderberg}, {Gehrels}, \&
  {Harrison}}]{Margutti_2017}
{Margutti}, R., {Kamble}, A., {Milisavljevic}, D., {et~al.} 2017, \apj, 835,
  140, \dodoi{10.3847/1538-4357/835/2/140}

\bibitem[{{Mart{\'\i}-Vidal} {et~al.}(2024){Mart{\'\i}-Vidal}, {Bj{\"o}rnsson},
  {P{\'e}rez-Torres}, {Lundqvist}, \& {Marcaide}}]{Vidal_2024}
{Mart{\'\i}-Vidal}, I., {Bj{\"o}rnsson}, C.~I., {P{\'e}rez-Torres}, M.~A.,
  {Lundqvist}, P., \& {Marcaide}, J.~M. 2024, \aap, 691, A171,
  \dodoi{10.1051/0004-6361/202450329}

\bibitem[{{Masci} {et~al.}(2019){Masci}, {Laher}, {Rusholme}, {Shupe}, {Groom},
  {Surace}, {Jackson}, {Monkewitz}, {Beck}, {Flynn}, {Terek}, {Landry},
  {Hacopians}, {Desai}, {Howell}, {Brooke}, {Imel}, {Wachter}, {Ye}, {Lin},
  {Cenko}, {Cunningham}, {Rebbapragada}, {Bue}, {Miller}, {Mahabal}, {Bellm},
  {Patterson}, {Juri{\'c}}, {Golkhou}, {Ofek}, {Walters}, {Graham}, {Kasliwal},
  {Dekany}, {Kupfer}, {Burdge}, {Cannella}, {Barlow}, {Van Sistine}, {Giomi},
  {Fremling}, {Blagorodnova}, {Levitan}, {Riddle}, {Smith}, {Helou}, {Prince},
  \& {Kulkarni}}]{ztf_reduction}
{Masci}, F.~J., {Laher}, R.~R., {Rusholme}, B., {et~al.} 2019, \pasp, 131,
  018003, \dodoi{10.1088/1538-3873/aae8ac}

\bibitem[{{Mattila} {et~al.}(2008){Mattila}, {Meikle}, {Lundqvist},
  {Pastorello}, {Kotak}, {Eldridge}, {Smartt}, {Adamson}, {Gerardy}, {Rizzi},
  {Stephens}, \& {van Dyk}}]{Mattila_2006jc}
{Mattila}, S., {Meikle}, W.~P.~S., {Lundqvist}, P., {et~al.} 2008, \mnras, 389,
  141, \dodoi{10.1111/j.1365-2966.2008.13516.x}

\bibitem[{{Mauerhan} {et~al.}(2013){Mauerhan}, {Smith}, {Filippenko},
  {Blanchard}, {Blanchard}, {Casper}, {Cenko}, {Clubb}, {Cohen}, {Fuller},
  {Li}, \& {Silverman}}]{Mauerhan_2012}
{Mauerhan}, J.~C., {Smith}, N., {Filippenko}, A.~V., {et~al.} 2013, \mnras,
  430, 1801, \dodoi{10.1093/mnras/stt009}

\bibitem[{{McMullin} {et~al.}(2007){McMullin}, {Waters}, {Schiebel}, {Young},
  \& {Golap}}]{Casa_desc}
{McMullin}, J.~P., {Waters}, B., {Schiebel}, D., {Young}, W., \& {Golap}, K.
  2007, in Astronomical Society of the Pacific Conference Series, Vol. 376,
  Astronomical Data Analysis Software and Systems XVI, ed. R.~A. {Shaw},
  F.~{Hill}, \& D.~J. {Bell}, 127

\bibitem[{{Miceli} {et~al.}(2015){Miceli}, {Sciortino}, {Troja}, \&
  {Orlando}}]{Miceli_2015}
{Miceli}, M., {Sciortino}, S., {Troja}, E., \& {Orlando}, S. 2015, \apj, 805,
  120, \dodoi{10.1088/0004-637X/805/2/120}

\bibitem[{Miller \& Stone(1993)}]{Kast_1993}
Miller, J., \& Stone, R. 1993, Lick Obs. Tech. Rep. 66, Tech. rep., Lick
  Obs.,Santa Cruz

\bibitem[{{Modjaz} {et~al.}(2019){Modjaz}, {Guti{\'e}rrez}, \&
  {Arcavi}}]{Modjaz_2019}
{Modjaz}, M., {Guti{\'e}rrez}, C.~P., \& {Arcavi}, I. 2019, Nature Astronomy,
  3, 717, \dodoi{10.1038/s41550-019-0856-2}

\bibitem[{{Ofek} {et~al.}(2014){Ofek}, {Zoglauer}, {Boggs}, {Barri{\'e}re},
  {Reynolds}, {Fryer}, {Harrison}, {Cenko}, {Kulkarni}, {Gal-Yam}, {Arcavi},
  {Bellm}, {Bloom}, {Christensen}, {Craig}, {Even}, {Filippenko},
  {Grefenstette}, {Hailey}, {Laher}, {Madsen}, {Nakar}, {Nugent}, {Stern},
  {Sullivan}, {Surace}, \& {Zhang}}]{Ofek_2010jl}
{Ofek}, E.~O., {Zoglauer}, A., {Boggs}, S.~E., {et~al.} 2014, \apj, 781, 42,
  \dodoi{10.1088/0004-637X/781/1/42}

\bibitem[{{Oke} \& {Gunn}(1983)}]{Oke_Gunn}
{Oke}, J.~B., \& {Gunn}, J.~E. 1983, \apj, 266, 713, \dodoi{10.1086/160817}

\bibitem[{{Oke} {et~al.}(1995){Oke}, {Cohen}, {Carr}, {Cromer}, {Dingizian},
  {Harris}, {Labrecque}, {Lucinio}, {Schaal}, {Epps}, \& {Miller}}]{LRIS_Keck}
{Oke}, J.~B., {Cohen}, J.~G., {Carr}, M., {et~al.} 1995, \pasp, 107, 375,
  \dodoi{10.1086/133562}

\bibitem[{{Pessi} {et~al.}(2020){Pessi}, {Anderson}, {Gutierrez}, \&
  {Irani}}]{2020ywx_class}
{Pessi}, P., {Anderson}, J., {Gutierrez}, C., \& {Irani}, I. 2020, Transient
  Name Server Classification Report, 2020-3822, 1

\bibitem[{{Quataert} \& {Shiode}(2012)}]{Quataert_2012}
{Quataert}, E., \& {Shiode}, J. 2012, \mnras, 423, L92,
  \dodoi{10.1111/j.1745-3933.2012.01264.x}

\bibitem[{{Ransome} {et~al.}(2021){Ransome}, {Habergham-Mawson}, {Darnley},
  {James}, {Filippenko}, \& {Schlegel}}]{Ransome_2021}
{Ransome}, C.~L., {Habergham-Mawson}, S.~M., {Darnley}, M.~J., {et~al.} 2021,
  \mnras, 506, 4715, \dodoi{10.1093/mnras/stab1938}

\bibitem[{{Ransome} \& {Villar}(2024)}]{Ransome_2024}
{Ransome}, C.~L., \& {Villar}, V.~A. 2024, arXiv e-prints, arXiv:2409.10596.
\newblock \doarXiv{2409.10596}

\bibitem[{{Reguitti} {et~al.}(2024){Reguitti}, {Pignata}, {Pastorello},
  {Dastidar}, {Reichart}, {Haislip}, \& {Kouprianov}}]{Reguitti_2024}
{Reguitti}, A., {Pignata}, G., {Pastorello}, A., {et~al.} 2024, arXiv e-prints,
  arXiv:2403.10398, \dodoi{10.48550/arXiv.2403.10398}

\bibitem[{{Reynolds} {et~al.}(2025){Reynolds}, {Nagao}, {Gottumukkala},
  {Guti{\'e}rrez}, {Kangas}, {Kravtsov}, {Kuncarayakti}, {Maeda}, {Elias-Rosa},
  {Fraser}, {Kotak}, {Mattila}, {Pastorello}, {Pessi}, {Cai}, {Fynbo},
  {Kawabata}, {Lundqvist}, {Matilainen}, {Moran}, {Reguitti}, {Taguchi}, \&
  {Yamanaka}}]{Reynolds}
{Reynolds}, T.~M., {Nagao}, T., {Gottumukkala}, R., {et~al.} 2025, arXiv
  e-prints, arXiv:2501.13619, \dodoi{10.48550/arXiv.2501.13619}

\bibitem[{{Ryder} {et~al.}(1993){Ryder}, {Staveley-Smith}, {Dopita}, {Petre},
  {Colbert}, {Malin}, \& {Schlegel}}]{Ryder}
{Ryder}, S., {Staveley-Smith}, L., {Dopita}, M., {et~al.} 1993, \apj, 416, 167,
  \dodoi{10.1086/173223}

\bibitem[{{Sana} {et~al.}(2012){Sana}, {de Mink}, {de Koter}, {Langer},
  {Evans}, {Gieles}, {Gosset}, {Izzard}, {Le Bouquin}, \&
  {Schneider}}]{Sana_2012}
{Sana}, H., {de Mink}, S.~E., {de Koter}, A., {et~al.} 2012, Science, 337, 444,
  \dodoi{10.1126/science.1223344}

\bibitem[{{Sarangi} {et~al.}(2018){Sarangi}, {Dwek}, \&
  {Arendt}}]{Sarangi_2018}
{Sarangi}, A., {Dwek}, E., \& {Arendt}, R.~G. 2018, \apj, 859, 66,
  \dodoi{10.3847/1538-4357/aabfc3}

\bibitem[{{Sarangi} \& {Slavin}(2022)}]{Arka_2022}
{Sarangi}, A., \& {Slavin}, J.~D. 2022, \apj, 933, 89,
  \dodoi{10.3847/1538-4357/ac713d}

\bibitem[{{Schlafly} \& {Finkbeiner}(2011)}]{2011_reddening}
{Schlafly}, E.~F., \& {Finkbeiner}, D.~P. 2011, \apj, 737, 103,
  \dodoi{10.1088/0004-637X/737/2/103}

\bibitem[{{Schlegel}(1990)}]{Schlegel_1990}
{Schlegel}, E.~M. 1990, \mnras, 244, 269

\bibitem[{{Smith} {et~al.}(2002){Smith}, {Tucker}, {Kent}, {Richmond},
  {Fukugita}, {Ichikawa}, {Ichikawa}, {Jorgensen}, {Uomoto}, {Gunn}, {Hamabe},
  {Watanabe}, {Tolea}, {Henden}, {Annis}, {Pier}, {McKay}, {Brinkmann}, {Chen},
  {Holtzman}, {Shimasaku}, \& {York}}]{Smith_AB}
{Smith}, J.~A., {Tucker}, D.~L., {Kent}, S., {et~al.} 2002, \aj, 123, 2121,
  \dodoi{10.1086/339311}

\bibitem[{{Smith} {et~al.}(2020){Smith}, {Smartt}, {Young}, {Tonry}, {Denneau},
  {Flewelling}, {Heinze}, {Weiland}, {Stalder}, {Rest}, {Stubbs}, {Anderson},
  {Chen}, {Clark}, {Do}, {F{\"o}rster}, {Fulton}, {Gillanders}, {McBrien},
  {O'Neill}, {Srivastav}, \& {Wright}}]{atlas_server}
{Smith}, K.~W., {Smartt}, S.~J., {Young}, D.~R., {et~al.} 2020, \pasp, 132,
  085002, \dodoi{10.1088/1538-3873/ab936e}

\bibitem[{{Smith}(2014)}]{Smith_2014}
{Smith}, N. 2014, \araa, 52, 487, \dodoi{10.1146/annurev-astro-081913-040025}

\bibitem[{Smith(2017)}]{Smith_2016}
Smith, N. 2017, Interacting Supernovae: Types IIn and Ibn (Springer
  International Publishing), 403–429, \dodoi{10.1007/978-3-319-21846-5_38}

\bibitem[{{Smith} \& {Andrews}(2020)}]{smith20}
{Smith}, N., \& {Andrews}, J.~E. 2020, \mnras, 499, 3544,
  \dodoi{10.1093/mnras/staa3047}

\bibitem[{{Smith} {et~al.}(2023){Smith}, {Andrews}, {Milne}, {Filippenko},
  {Brink}, {Kelly}, {Yuk}, \& {Jencson}}]{Smith_2015da}
{Smith}, N., {Andrews}, J.~E., {Milne}, P., {et~al.} 2023, arXiv e-prints,
  arXiv:2312.00253, \dodoi{10.48550/arXiv.2312.00253}

\bibitem[{{Smith} \& {Arnett}(2014)}]{sa:14}
{Smith}, N., \& {Arnett}, W.~D. 2014, \apj, 785, 82,
  \dodoi{10.1088/0004-637X/785/2/82}

\bibitem[{{Smith} {et~al.}(2010){Smith}, {Chornock}, {Silverman}, {Filippenko},
  \& {Foley}}]{smith10}
{Smith}, N., {Chornock}, R., {Silverman}, J.~M., {Filippenko}, A.~V., \&
  {Foley}, R.~J. 2010, \apj, 709, 856, \dodoi{10.1088/0004-637X/709/2/856}

\bibitem[{{Smith} {et~al.}(2008){Smith}, {Foley}, \& {Filippenko}}]{smith08}
{Smith}, N., {Foley}, R.~J., \& {Filippenko}, A.~V. 2008, \apj, 680, 568,
  \dodoi{10.1086/587860}

\bibitem[{{Smith} {et~al.}(2011{\natexlab{a}}){Smith}, {Li}, {Filippenko}, \&
  {Chornock}}]{Smith_2011}
{Smith}, N., {Li}, W., {Filippenko}, A.~V., \& {Chornock}, R.
  2011{\natexlab{a}}, \mnras, 412, 1522,
  \dodoi{10.1111/j.1365-2966.2011.17229.x}

\bibitem[{{Smith} {et~al.}(2011{\natexlab{b}}){Smith}, {Li}, {Silverman},
  {Ganeshalingam}, \& {Filippenko}}]{Smith11lbv}
{Smith}, N., {Li}, W., {Silverman}, J.~M., {Ganeshalingam}, M., \&
  {Filippenko}, A.~V. 2011{\natexlab{b}}, \mnras, 415, 773,
  \dodoi{10.1111/j.1365-2966.2011.18763.x}

\bibitem[{{Smith} \& {Owocki}(2006)}]{so:06}
{Smith}, N., \& {Owocki}, S.~P. 2006, \apjl, 645, L45, \dodoi{10.1086/506523}

\bibitem[{{Smith} {et~al.}(2012){Smith}, {Silverman}, {Filippenko}, {Cooper},
  {Matheson}, {Bian}, {Weiner}, \& {Comerford}}]{Smith_2012}
{Smith}, N., {Silverman}, J.~M., {Filippenko}, A.~V., {et~al.} 2012, \aj, 143,
  17, \dodoi{10.1088/0004-6256/143/1/17}

\bibitem[{{Smith} \& {Tombleson}(2015)}]{st:15}
{Smith}, N., \& {Tombleson}, R. 2015, \mnras, 447, 598,
  \dodoi{10.1093/mnras/stu2430}

\bibitem[{{Smith} {et~al.}(2007){Smith}, {Li}, {Foley}, {Wheeler}, {Pooley},
  {Chornock}, {Filippenko}, {Silverman}, {Quimby}, {Bloom}, \&
  {Hansen}}]{Smith_2007}
{Smith}, N., {Li}, W., {Foley}, R.~J., {et~al.} 2007, \apj, 666, 1116,
  \dodoi{10.1086/519949}

\bibitem[{{Smith} {et~al.}(2009){Smith}, {Silverman}, {Chornock}, {Filippenko},
  {Wang}, {Li}, {Ganeshalingam}, {Foley}, {Rex}, \& {Steele}}]{Smith_2005ip}
{Smith}, N., {Silverman}, J.~M., {Chornock}, R., {et~al.} 2009, \apj, 695,
  1334, \dodoi{10.1088/0004-637X/695/2/1334}

\bibitem[{{Smith} {et~al.}(2017){Smith}, {Kilpatrick}, {Mauerhan}, {Andrews},
  {Margutti}, {Fong}, {Graham}, {Zheng}, {Kelly}, {Filippenko}, \&
  {Fox}}]{Smith_2017}
{Smith}, N., {Kilpatrick}, C.~D., {Mauerhan}, J.~C., {et~al.} 2017, \mnras,
  466, 3021, \dodoi{10.1093/mnras/stw3204}

\bibitem[{{Smith} {et~al.}(2018){Smith}, {Andrews}, {Rest}, {Bianco}, {Prieto},
  {Matheson}, {James}, {Smith}, {Strampelli}, \& {Zenteno}}]{smith18}
{Smith}, N., {Andrews}, J.~E., {Rest}, A., {et~al.} 2018, \mnras, 480, 1466,
  \dodoi{10.1093/mnras/sty1500}

\bibitem[{{Srivastav} {et~al.}(2020){Srivastav}, {Smith}, {McBrien}, {Smartt},
  {Gillanders}, {Fulton}, {Young}, {Shingles}, {McCollum}, {Chen}, {Anderson},
  {Denneau}, {Heinze}, {Tonry}, {Weiland}, {Stalder}, {Rest}, \&
  {Wright}}]{2020ywx_disc}
{Srivastav}, S., {Smith}, K.~W., {McBrien}, O., {et~al.} 2020, Transient Name
  Server AstroNote, 215, 1

\bibitem[{{Stetson}(1987)}]{Steson_daophot}
{Stetson}, P.~B. 1987, \pasp, 99, 191, \dodoi{10.1086/131977}

\bibitem[{{Stritzinger} {et~al.}(2012){Stritzinger}, {Taddia}, {Fransson},
  {Fox}, {Morrell}, {Phillips}, {Sollerman}, {Anderson}, {Boldt}, {Brown},
  {Campillay}, {Castellon}, {Contreras}, {Folatelli}, {Habergham}, {Hamuy},
  {Hjorth}, {James}, {Krzeminski}, {Mattila}, {Persson}, \&
  {Roth}}]{2006jd_stritzinger}
{Stritzinger}, M., {Taddia}, F., {Fransson}, C., {et~al.} 2012, \apj, 756, 173,
  \dodoi{10.1088/0004-637X/756/2/173}

\bibitem[{{Taddia} {et~al.}(2013){Taddia}, {Stritzinger}, {Sollerman},
  {Phillips}, {Anderson}, {Boldt}, {Campillay}, {Castell{\'o}n}, {Contreras},
  {Folatelli}, {Hamuy}, {Heinrich-Josties}, {Krzeminski}, {Morrell}, {Burns},
  {Freedman}, {Madore}, {Persson}, \& {Suntzeff}}]{Taddia_2013}
{Taddia}, F., {Stritzinger}, M.~D., {Sollerman}, J., {et~al.} 2013, \aap, 555,
  A10, \dodoi{10.1051/0004-6361/201321180}

\bibitem[{{Tartaglia} {et~al.}(2020){Tartaglia}, {Pastorello}, {Sollerman},
  {Fransson}, {Mattila}, {Fraser}, {Taddia}, {Tomasella}, {Turatto},
  {Morales-Garoffolo}, {Elias-Rosa}, {Lundqvist}, {Harmanen}, {Reynolds},
  {Cappellaro}, {Barbarino}, {Nyholm}, {Kool}, {Ofek}, {Gao}, {Jin}, {Tan},
  {Sand}, {Ciabattari}, {Wang}, {Zhang}, {Huang}, {Li}, {Mo}, {Rui}, {Xiang},
  {Zhang}, {Hosseinzadeh}, {Howell}, {McCully}, {Valenti}, {Benetti}, {Callis},
  {Carracedo}, {Fremling}, {Kangas}, {Rubin}, {Somero}, \&
  {Terreran}}]{Tartaglia_2020}
{Tartaglia}, L., {Pastorello}, A., {Sollerman}, J., {et~al.} 2020, \aap, 635,
  A39, \dodoi{10.1051/0004-6361/201936553}

\bibitem[{{Thomas} {et~al.}(2022){Thomas}, {Wheeler}, {Dwarkadas}, {Stockdale},
  {Vink{\'o}}, {Pooley}, {Xu}, {Zeimann}, \& {MacQueen}}]{Thomas_2014C}
{Thomas}, B.~P., {Wheeler}, J.~C., {Dwarkadas}, V.~V., {et~al.} 2022, \apj,
  930, 57, \dodoi{10.3847/1538-4357/ac5fa6}

\bibitem[{{Tody}(1986)}]{iraf_86}
{Tody}, D. 1986, in Society of Photo-Optical Instrumentation Engineers (SPIE)
  Conference Series, Vol. 627, Instrumentation in astronomy VI, ed. D.~L.
  {Crawford}, 733, \dodoi{10.1117/12.968154}

\bibitem[{{Tody}(1993)}]{iraf_93}
{Tody}, D. 1993, in Astronomical Society of the Pacific Conference Series,
  Vol.~52, Astronomical Data Analysis Software and Systems II, ed. R.~J.
  {Hanisch}, R.~J.~V. {Brissenden}, \& J.~{Barnes}, 173

\bibitem[{{Tonry} {et~al.}(2018){Tonry}, {Denneau}, {Heinze}, {Stalder},
  {Smith}, {Smartt}, {Stubbs}, {Weiland}, \& {Rest}}]{Tonry_atlas}
{Tonry}, J.~L., {Denneau}, L., {Heinze}, A.~N., {et~al.} 2018, \pasp, 130,
  064505, \dodoi{10.1088/1538-3873/aabadf}

\bibitem[{{Turatto} {et~al.}(1993){Turatto}, {Cappellaro}, {Danziger},
  {Benetti}, {Gouiffes}, \& {della Valle}}]{Turatto_88z}
{Turatto}, M., {Cappellaro}, E., {Danziger}, I.~J., {et~al.} 1993, \mnras, 262,
  128, \dodoi{10.1093/mnras/262.1.128}

\bibitem[{{Valenti} {et~al.}(2016){Valenti}, {Howell}, {Stritzinger}, {Graham},
  {Hosseinzadeh}, {Arcavi}, {Bildsten}, {Jerkstrand}, {McCully}, {Pastorello},
  {Piro}, {Sand}, {Smartt}, {Terreran}, {Baltay}, {Benetti}, {Brown},
  {Filippenko}, {Fraser}, {Rabinowitz}, {Sullivan}, \&
  {Yuan}}]{Valenti_LCOpipe}
{Valenti}, S., {Howell}, D.~A., {Stritzinger}, M.~D., {et~al.} 2016, \mnras,
  459, 3939, \dodoi{10.1093/mnras/stw870}

\bibitem[{{van Dyk} {et~al.}(1993){van Dyk}, {Weiler}, {Sramek}, \&
  {Panagia}}]{VanDyk_1988Z}
{van Dyk}, S.~D., {Weiler}, K.~W., {Sramek}, R.~A., \& {Panagia}, N. 1993,
  \apjl, 419, L69, \dodoi{10.1086/187139}

\bibitem[{{Weiler} {et~al.}(1990){Weiler}, {Panagia}, \&
  {Sramek}}]{Weiler_1990}
{Weiler}, K.~W., {Panagia}, N., \& {Sramek}, R.~A. 1990, \apj, 364, 611,
  \dodoi{10.1086/169444}

\bibitem[{{Weiler} {et~al.}(1986){Weiler}, {Sramek}, {Panagia}, {van der
  Hulst}, \& {Salvati}}]{Weiler_1986}
{Weiler}, K.~W., {Sramek}, R.~A., {Panagia}, N., {van der Hulst}, J.~M., \&
  {Salvati}, M. 1986, \apj, 301, 790, \dodoi{10.1086/163944}

\bibitem[{{Wilson} {et~al.}(2004){Wilson}, {Henderson}, {Herter}, {Matthews},
  {Skrutskie}, {Adams}, {Moon}, {Smith}, {Gautier}, {Ressler}, {Soifer}, {Lin},
  {Howard}, {LaMarr}, {Stolberg}, \& {Zink}}]{Wilson04}
{Wilson}, J.~C., {Henderson}, C.~P., {Herter}, T.~L., {et~al.} 2004, in Society
  of Photo-Optical Instrumentation Engineers (SPIE) Conference Series, Vol.
  5492, Ground-based Instrumentation for Astronomy, ed. A.~F.~M. {Moorwood} \&
  M.~{Iye}, 1295--1305, \dodoi{10.1117/12.550925}

\bibitem[{{Woosley}(2017)}]{woosley17}
{Woosley}, S.~E. 2017, \apj, 836, 244, \dodoi{10.3847/1538-4357/836/2/244}

\bibitem[{{Woosley} \& {Smith}(2022)}]{ws22}
{Woosley}, S.~E., \& {Smith}, N. 2022, \apj, 938, 57,
  \dodoi{10.3847/1538-4357/ac8eb3}

\bibitem[{{Wu} \& {Fuller}(2021)}]{wf21}
{Wu}, S., \& {Fuller}, J. 2021, \apj, 906, 3, \dodoi{10.3847/1538-4357/abc87c}

\bibitem[{{Yaron} \& {Gal-Yam}(2012)}]{Wiserep}
{Yaron}, O., \& {Gal-Yam}, A. 2012, \pasp, 124, 668, \dodoi{10.1086/666656}

\bibitem[{{Yesmin} {et~al.}(2024){Yesmin}, {Pellegrino}, {Modjaz}, {Baer-Way},
  {Howell}, {Arcavi}, {Farah}, {Hiramatsu}, {Hosseinzadeh}, {McCully},
  {Newsome}, {Padilla Gonzalez}, {Terreran}, \& {Jha}}]{Yesmin_2021ukt}
{Yesmin}, N., {Pellegrino}, C., {Modjaz}, M., {et~al.} 2024, arXiv e-prints,
  arXiv:2409.04522, \dodoi{10.48550/arXiv.2409.04522}

\bibitem[{{Yoon} \& {Cantiello}(2010)}]{Yoon_Cantiello}
{Yoon}, S.-C., \& {Cantiello}, M. 2010, \apjl, 717, L62,
  \dodoi{10.1088/2041-8205/717/1/L62}

\bibitem[{{Zhang} {et~al.}(2012){Zhang}, {Wang}, {Wu}, {Chen}, {Chen}, {Liu},
  {Huang}, {Liang}, {Zhao}, {Lin}, {Wang}, {Dennefeld}, {Zhang}, {Zhai}, {Wu},
  {Fan}, {Zou}, {Zhou}, \& {Ma}}]{Zhang_2010jl}
{Zhang}, T., {Wang}, X., {Wu}, C., {et~al.} 2012, \aj, 144, 131,
  \dodoi{10.1088/0004-6256/144/5/131}

\end{thebibliography}
\bibliographystyle{aasjournal}



\end{document}